\documentclass[12pt]{article}
\usepackage{graphicx}
\usepackage{subfigure}
\usepackage{cite}
\usepackage{amsmath}
\usepackage{capt-of}

\begin{document}

\title{\vspace{-2cm} Spontaneous Symmetry Breaking, \\ Group
Decision Making and Beyond \\  2.  Distorted Polarization and Vulnerability}

\author{ Serge Galam\thanks{serge.galam@sciencespo.fr} \\
CEVIPOF - Centre for Political Research, Sciences Po and CNRS,\\
1, Place Saint Thomas d'Aquin, Paris 75007, France}

\date{September 5, 2025}

\maketitle

\begin{abstract}

This paper extends previous work on echo chambers modeled by an Ising-like system at zero temperature (1. Echo Chambers and random Polarization, Symmetry 2024, 16(12), 1566). There, polarization emerged as a spontaneous symmetry-breaking process with a randomly selected direction. Here using a mean-field analysis and Monte Carlo simulations I show that this mechanism is highly vulnerable to minimal distortions. An external symmetry-breaking field, even vanishingly small, suffices to impose a global direction and suppress opposite domains, producing distorted full polarization. In contrast, a handful of quenched local fields with zero average do not erase polarization but reorganize it into opposing domains. Remarkably, as few as two opposed fields, if placed at tipping sites, can redirect the entire system. These fragile sites, indistinguishable from others, act as hidden tipping points that amplify microscopic biases into macroscopic outcomes. Difference in local field proportions is found to be instrumental to guarantee a winning majority. The results highlight how minimal, strategically placed interventions can override autonomous self-organization. The results could, if applicable on social media platforms, question their presumed democratic nature of consensus.

\end{abstract}


\section{Introduction}

In condensed matter systems, long range order can arise either spontaneously, emerging from microscopic fluctuations without directional bias, or as a state imposed by external or internal fields. Spontaneous  long range order reflects the intrinsic capacity of a system to break symmetry under stochastic initial conditions, whereas pre-defined long range order results from constraints that bias the dynamics toward specific ordered configurations. The distinction between these two regimes, intrinsic versus imposed, highlights both the generative power and the fragility of symmetry breaking  \cite{p1, p2, p3, p4}. 

Beyond physical systems, such mechanisms resonate with broader classes of emergent behavior under constraint \cite{p5}. Indeed, in a previous paper entitled ``1. Echo Chambers and random Polarization'' \cite{p6}, I have built a model of opinion dynamics using an Ising-like system at zero temperature. There, polarization is the substitute to  long range order, and similarly emerges as a spontaneous symmetry-breaking process with a randomly selected direction for the final unanimous collective choice. 

However, I have emphasized major differences with physical systems, claiming that with regard to social system, the history, the initial conditions as well the actual scheme used for the updating, do matter and are instrumental relevant features in the making of the final stable outcome. While ergodicity is a requisite in equilibrium statistical mechanics ensuring that time averages measured in experiments coincide with ensemble averages calculated in theory, I claimed that  ergodicity does not apply to social systems.  

As one of the model outcomes, echo chambers were found to emerge as the outcome of a dynamical opinion update process rather than from preferential attachment as usually assumed  \cite{p7}.

Here I show that the random selection of the final outcome is fragile and can be easily distorted. The application of a very small external symmetry-breaking pressure at the early stage of the process suffices to impose the collective ordering and erase the formation of opposite domains. Polarization is thus easily distorted by even minimal external pressures on the community.

In contrast, I find that a few quenched local fields lead to qualitatively different outcomes. Instead of suppressing domains with one unique polarization, they stabilize the coexistence of domains with opposite orientations. Strikingly, as few as two opposed fields are sufficient to redirect the entire system, provided they are located at specific tipping sites. 

Monte Carlo simulations \cite{mc1, mc2, mc3} uncover these tipping sites as hidden weak points. They are indistinguishable from others, yet capable of steering the global evolution when activated. Mean-field analysis confirms these tendencies but missed a variety of outcomes uncovered by Monte Carlo simulations, whose outcomes depend on the microscopic details of the initial conditions.

Altogether, the results demonstrate how both external fields and localized quenched biases provide low-cost paths to distort polarization. They reveal also the disproportionate impact of some tipping sites, and more broadly, the vulnerability of symmetry-breaking processes to minimal but strategically positioned interventions. 

These findings, while obtained in a two-dimensional $30 \times 30$ Ising-like model, could have direct implications for opinion dynamics in social platforms, where structural fragilities may silently convert weak local biases into macroscopic consensus. Indeed, if applicable on social media platforms, the findings question their presumed democratic nature of consensus.

This work relies on earlier work about group decision making \cite{p8, p9} and subscribes to the new \cite{n1, n2} and very active field of sociophysics \cite{a1, a2, a3, a4, a5, a6, a7, a8, a9, z1, z2, z3,z4, z5, z6, z7, z8, t1, t2, t3, t4, t5, t6, t7, t8, t9, t10, t11, t12, t13, t14, t15, t16, t17, t18, u1, u2, u3, u4, u5, u6, u7}. The Ising-like model has inspired many related papers \cite{i1, i2, i3, i4, i5, i6, i7}.

The rest of the paper is organized as follow: Section 2 reviews the model of opinion dynamics used in the first paper of this series \cite{p6}. I apply small external pressure in Section 3. An homogeneous external pressure is studied in Subsection 3.1 while diluted heterogeneous individual pressures are considered in Subsection 3.2. Sub-cases with one single pressure and two opposite individual pressures are included. Having both external uniform pressure and individual pressures is investigated in Subsection 3.3. A series of Monte Carlo simulations are performed in Section 4. Three sub-cases are analyzed with symmetric initial configuration and local fields, asymmetric initial configuration and local fields, and a single local field. Last Section contains the conclusion where results are discussed.

\section{Model used in the first paper}

In the first paper of this series \cite{p6} I considered a collection of $N$ symmetrical equiprobable discrete bimodal individual choices ($c_i=\pm 1$ with $i=1,2,...,N$) who interact by pairs.  Given a configuration of individual choices  $\{c_1, c_2, \dots, c_N \}$, I defined the magnitude of the group utility function as, 
\begin{equation}
U \equiv J \sum_{<i,j>}c_ic_j ,
\label{U}
\end{equation}
where I have assumed a constant amplitude $J$ for all pairs of connected agents. The term $<i,j>$ represents all interacting pairs in the group.  Each agent does not interact with all the others. 

The associated quantity,
\begin{equation}
C=  \sum_{i=1}^{N}c_i ,
\label{C}
\end{equation}
measures the amplitude of the actual social impact of the group aggregate choices with  $-N\leq C \leq N$. In parallel,
\begin{equation}
c= \frac{1}{N} \sum_{i=1}^{N}c_i ,
\label{c}
\end{equation}
measures the related degree of the symmetry breaking with $-1\leq c \leq 1$.

In addition I have introduced $N_+$ and $ N_-$ to count the respective numbers of agents having chosen $+1$ and $-1$. Actual social and cultural impacts are then proportional to $N_+$ and $ N_-$.  Proportions $p=N_+/N$ and $(1-p)$ with $0\leq p \leq 1$ measure the degree of diversity of the $N$-person group.

Invoking the effect of individual anticipation I have performed a mean-field treatment of Eq.(\ref{U}) to get the anticipated group utility function, 
\begin{equation}
U_a=P_g\sum_{i=1}^{N}c_i-\delta ,
\label{Ua}
\end{equation}
where 
\begin{equation}
P_g \equiv\delta  \frac{C}{N} ,
\label{Pg}
\end{equation}
defines the group pressure on each person for an aggregated choice $C$. The parameter $\delta  \equiv\frac{k J N}{2(N-1)}$ is a constant independent of the group choice and thus does not affect  the expected collective choice $C$. The parameter $k$ is the number of persons individual i interacts with by pairs. it is identical for all the group members. The superscript $a$ signals that the anticipating process has been applied.

To get another focus I recast Eq.(\ref{Ua}) as,
\begin{equation}
U_a=P_g\sum_{i=1}^{N}c_i-\gamma N,
\label{Uaa}
\end{equation}
where 
\begin{equation}
P_g \equiv\gamma C,
\label{Pgg}
\end{equation}
and $\gamma  \equiv\frac{k J}{2(N-1)}$. Last term is a constant cost for having an interacting group linear. This cost is a constant independent of the extent of the symmetry breaking, which increase linearly with the  number $N$ of agents in the community.

Assuming that each person seeks to optimize their utility, Eq.(\ref{Uaa}) indicates that  $P_g >0$ favors $+1$ individual choices   while $P_g <0$ favors $-1$ individual choices. Rewriting $\sum_{i=1}^{N}c_i= C$ turns Eq.(\ref{Uaa}) to,
\begin{equation}
U_a=\gamma ( C^2 - N) ,
\label{UaC}
\end{equation}
which is maximum for $C^2=N ^2\rightarrow C=\pm N$. 

The anticipation process is thus found to produce a symmetry breaking along either +1 or -1. It is of importance to emphasize that both possibilities are equirobable. While all pair interactions lead to an identical choice polarization upon optimization of individual utilities, the choice itself is randomly selected. That is a major result, which demonstrates that the symmetry breaking is spontaneous and random with no a priori. 

This unexpected result has been confirmed by Monte Carlo simulations except for the emergence of domains of opposite polarizations, which are found in the simulations but are absent from the mean-filed treatment yielding $U_a$. The domain formation obtained from Monte Carlo simulations was shown to be a function of the history of the system, its size and the algorithm used for individual updates \cite{p6}.

\section{Applying small external pressures}

While the direction of the spontaneous symmetry breaking is random for a pair interacting homogeneous population, applying a small external symmetry breaking uniform pressure may in some case imposes the choice along which the spontaneous polarization is  achieved. 

\subsection{Applying a small external pressure prior to the interacting dynamics}

When the external pressure is applied prior to the launching of the update dynamics, the utility function writes,
\begin{equation}
U_{P} \equiv J \sum_{<i,j>}c_ic_j +  P \sum_{i=1}^{N}c_i ,
\label{UP}
\end{equation}
where $P$ is the external pressure.

Activating the anticipation effect, Eq.(\ref{UP}) becomes,
\begin{equation}
U_{Pa}=(P+P_g)\sum_{i=1}^{N}c_i-\gamma N ,
\label{UaP}
\end{equation}
leading to,
\begin{equation}
U_{Pa}=\gamma( C^2- N) +P C,
\label{UaPC}
\end{equation}
which is no longer maximum for $C^2=N ^2$ due to the term $PC$. Now $P>0 \Rightarrow C>0$ making $C=N$ for maximum and $C=-N$ when $P<0$, independently of the actual magnitude of $P$. Even an infinitesimal external pressure is sufficient to select the final collective state.

\subsection{Applying diluted heterogeneous individual pressures}

Applying an infinitesimal external pressure was shown to have a tremendous impact on the process of spontaneous symmetry breaking driven by pair interactions, by selecting the choice along which the group ends up fully polarized. The underlying condition being that the same infinitesimal external pressure must be applied to each member of the group. Although the cost of activating an infinitesimal external pressure may be low, the implementation of applying it to all group members can be quite demanding. On this basis, I now investigate the impact of having a highly diluted external pressure.

\subsubsection{A single individual pressure}

Applying a single individual pressure $P_k>0$ to agent $m$ yields,
\begin{equation}
U_{m} \equiv J \sum_{<i,j>}c_ic_j +   P_m c_m ,
\label{Um}
\end{equation}
and
\begin{equation}
U_{ma}=\gamma (C^2-N) +   P_m c_m  .
\label{Uma}
\end{equation}

Eq.(\ref{Uma}) shows that an infinitesimal single individual pressure is sufficient to distort the spontaneous symmetry breaking along a pre-selected choice. It is a low cost lever to impose the choice of the group. However, I must stress that this observation raises from the mean-field treatment,  which has de facto waived out all fluctuations. 
The utility gain is then $P_m$ with a cost  $-P_m$ when the alignement of the group is not along the single individual pressure. It is a relatively small value with respect to  $C^2$. I thus expect the Monte Carlo simulation to temper this low cost alignement with the formation of domain barriers hindering a unique full polarization (see below the related Section).

\subsubsection{Two opposite individual pressures}

Adding a second opposite individual pressure $-P_n$ with $P_n>0$ to agent $n$ yields,
\begin{equation}
U_{mn} \equiv J \sum_{<i,j>}c_ic_j +  P_m c_m - P_n c_n ,
\label{Umn}
\end{equation}
and
\begin{equation}
U_{mna}=\gamma \{(C'+c_m+c_n)^2-N\} +  P_m c_m - P_n c_n  ,
\label{Umna}
\end{equation}
with $-(N-2) \leq C'\leq N-2$. 

Maximizing Eq.(\ref{Umna}) yields always $C'=\pm (N-2)$ with the values of $(c_m, c_n)$ to be evaluated. Four configurations are then available with,

\begin{enumerate}

\item $c_m = c_n= +1\Rightarrow  \gamma \{N^2-N\}+ P_m  - P_n$ ,

\item  $c_m = c_n= -1\Rightarrow \gamma (N^2-N) -P_m  + P_n$ ,

\item  $c_m = -c_n= +1\Rightarrow \gamma \{(N-2)^2-N\}+ P_m  + P_n$  ,

\item  $c_m = -c_n= -1\Rightarrow  \gamma \{(N-2)^2-N\} -P_m  - P_n$ ,

\end{enumerate}
which can be recast as, 

\begin{enumerate}

\item $c_m = c_n= +1\Rightarrow U_1\equiv P_m  - P_n$ ,

\item  $c_m = c_n= -1\Rightarrow U_2\equiv -P_m  + P_n$ ,

\item  $c_m = -c_n= +1\Rightarrow U_3\equiv -4\gamma (N-1)+ P_m  + P_n$  ,

\item  $c_m = -c_n= -1\Rightarrow  U_4\equiv -4\gamma (N-1) -P_m  - P_n$ ,

\end{enumerate}
where the constant  $\gamma (N^2-N)$ has been dropped from the four quantities $(U_1, U_2, U_3, U_4)$, which still have to be compared to identify the various maximums as a function of the magnitude of the two individual pressures. The last case $c_m = -c_n= -1$ can be dropped at once with all three terms of $U_4$ being negative. 

To compare the three other cases I start with the condition $P_m>P_n$, which eliminates case 2 with $U_2<0$. Then, $U_1$ must be compared to $U_3$. To reach $U_1>U_3$ requires the condition $P_m  -P_n > -4\gamma (N-1)+ P_m  + P_n$, which is equivalent to $P_n<2\gamma (N-1)$ leading to $P_n<kJ$. In contrast, $P_n>kJ$ makes case 3 the maximum.

Considering $P_m<P_n$ eliminates case 1 with $U_1<0$, leaving cases 2 and 3 to be compared. To have $U_2>U_3$ requires the condition $-P_m  +P_n > -4\gamma (N-1)+ P_m  + P_n$, which is equivalent to $P_m<2\gamma (N-1)$ leading to $P_m<kJ$. In contrast, $P_m>kJ$ makes case 3 the maximum.

To sum up, the maximization process proceeds in three successive steps. First, the symmetry breaking amplitude $C$ is set equal to $\pm(N-2)+c_m+c_n$ with $(c_m, c_n)$ to be determined. Then the larger individual pressure is satisfied, i.e., $c_m=1$ if $P_m>P_n$ or $c_n=-1$ if $P_m<P_n$. That leads respectively to $C=(N-2)+1+c_n=N-1+c_n$ or $C=-(N-2)+c_m-1=-N+1+c_m$.

Finally, the second smaller individual pressure $P_n$ ($P_m$) is satisfied provided $P_n>kJ$ ($P_m>kJ$) leading to  $C=N-1-1=N-2$ ($C=-N+1+1=-N+2$). Otherwise, when $P_n<kJ$ ($P_m<kJ$) yielding $C=N-1+1=N$ ($C=-N-1+1=-N$).

When $P_m=P_n  \equiv P_0$ the four utilities become $U_1=U_2 \equiv U_0=0$, $U_3=-2kJ+2P_0$ and $U_4=-2kJ-2P_0$.. Therefore $U_3>U_0$ when $P_0>kJ$ yielding $c_m=1, c_n=-1, C= \pm (N-2)$. For  $P_0<kJ$ yielding $c_m=c_n= \pm 1, C= \pm N$.  The random breaking of the symmetry is thus recovered.

Accordingly, while one singe individual pressure allows to select the symmetry breaking at a low cost, adding a second opposite one turns the strategy more uncertain. Indeed, to select the symmetry breaking requires an individual pressure larger than the other one, which in turn can lead to a headlong rush.

Adding more individual fields will extend above three-step process to a larger hierarchical step process. In addition, grouping similar local pressures may trigger the formation of domains. This feature is studied in the next Section using Monte Carlo simulations.

\subsection{The full pressure case}

Having both external uniform pressure and individual pressures lead to the complete utilities function,
\begin{eqnarray}
\label{UPei} 
U_{cei} &=& J \sum_{<i,j>}c_ic_j +   \sum_{i=1}^{N}P_i c_i  +  P \sum_{i=1}^{N}c_i , \\ 
&=& J \sum_{<i,j>}c_ic_j +   \sum_{i=1}^{N}(P+P_i) c_i   , \nonumber 
\end{eqnarray}
which is identical to the  statistical physics Hamiltonian of a Random Field Ising Model \cite{} and where each $P+P_i$ is either positive or negative with $P_i$ being positive, negative or null. 

Applying local individual pressures is thus a way to oppose an external uniform pressure provided $sign (P+P_i))=-sign(P)$.

At this stage it is also worth stressing that while both uniform and individual external pressures are produced by som extra-individual body, internal pressures could have an additional origin. Indeed, the individual pressures can stem from the individuals who are being subject to them. In another model I have developed for opinion dynamics, I defined a person $i$ subjected to a local field $h_i$ as inflexible \cite{s1} or stubborn \cite{s2}. In both cases, the effect is identical having this person not shifting opinion keeping always the choice aligned with the local field in the present case.

\section{Monte Carlo simulations to investigate the formation of domains}

In the first paper of this series \cite{p6}, I have shown how the history of a given group of agents does influence the final outcome of the process of decision making. In addition, the choice of the scheme used to implement the update of opinion was also shown to be instrumental in the making of the final outcomes as well as having or not Periodic Boundary Conditions to mimic very large samples. 

Here, I run the simulations using only the sequential update with no boundary conditions as indicated in the Figures with ``BC: Open". The update is implemented via the Metropolis scheme, which is identical to the Glauber scheme at $T=0$.

My goal is to investigate the drastic effect triggered by the presence of small proportions of local fields in the making of the final outcome of the collective choice. I thus run a series of MC simulations. All of them use a grid lattice of size $30 \times 30 =900$ agents with sequential update and fixed parameters $J=1, H=0, T=0$. Amplitudes of the positive and negative local fields are given respectively by $(u, v)$ and their proportions by $(a, b)$. The term ``void" is used to indicate the absence of local field. The proportion of voids is given in by$1-a-b$. To ensure that a person follows their local field when present, I set the field amplitudes at $u=-v=5$, which is larger than 4, the maximum contribution from nearest-neighbor interactions.

For each set of a simulation I show several graphs including the evolution of the magnetization (numbers of +1 minus number of -1 divided by 900) as a function the number of MC steps, the distribution of local fields used for the simulation and a distribution of individual choices after the completion of a certain number of MC. Red color is used for +1 and blue color for -1. The parameter Seed is a parameter allowing to reproduce the simulations with the same sequence of random process. Simulations are performed with local fields generated with both random and manual distributions. In the Figures, this feature is indicated with ``Fields: Random" or ``Fields: Manual". The parameter Seed indicated in the figures is there to allow reproducing the simulations with the same random sequences.

I first start running a series of simulations with equal proportions $a=b$ of positive and negative local fields to preserve the $p=0.50$ symmetry of the initial configuration. In a second part, I consider asymmetric proportions of local fields $a\neq b$ with both symmetric ($p=0.50$) and asymmetric ($p \neq 0.50$) initial configurations.

\subsection{Symmetric initial configuration and local fields}

\subsubsection{Figure (\ref{aa})}

The first upper row of Fig. (\ref{aa}) shows a simulation with an initial configuration with half +1 and half -1 ($p=0.50$, subpart b)) without local fields. The dynamics is shown as a function of MC steps in subpart (a). While the +1 choice gets an increased support till about 10 MC steps, it then starts loosing support before shrinking to zero within only a few MC steps. Subpart (c) shows the existence of two domains with A being majority (485 +1 versus 415 -1) after a completion of 15 MC steps. However few more MC steps leads to an unexpected total victory (spontaneous symmetry breaking) of B as seen in subpart (d). 

The second row of the Figure shows the same simulation, except that it has two local opposite fields shown in subpart (f). These two local fields enlarged the number of MC with A being majority as illustrated in subpart (g) with 539 +1. The final outcome keeps now a small A domain (subpart (h)).

The third row keeps the second row setting with a shift in the locations of the red and blue local fields (subpart (j)).The results are similar to the above ones with now two separate red domains separated by a blue domain (subpart (k)). The final small A domain is in the lower right part of the grid (subpart (l)).

However, the fourth row shows an unexpected outcome. The unique difference with the third row is the swapping of the field colors. During the first 15 MC steps, the dynamics is similar to subpart (i). However, after, instead of a red shrinking, the red is boosted at once to invade the full sample.

\begin{figure}
\centering
\vspace{-3cm}
\subfigure[]{\includegraphics[width=0.3\textwidth]{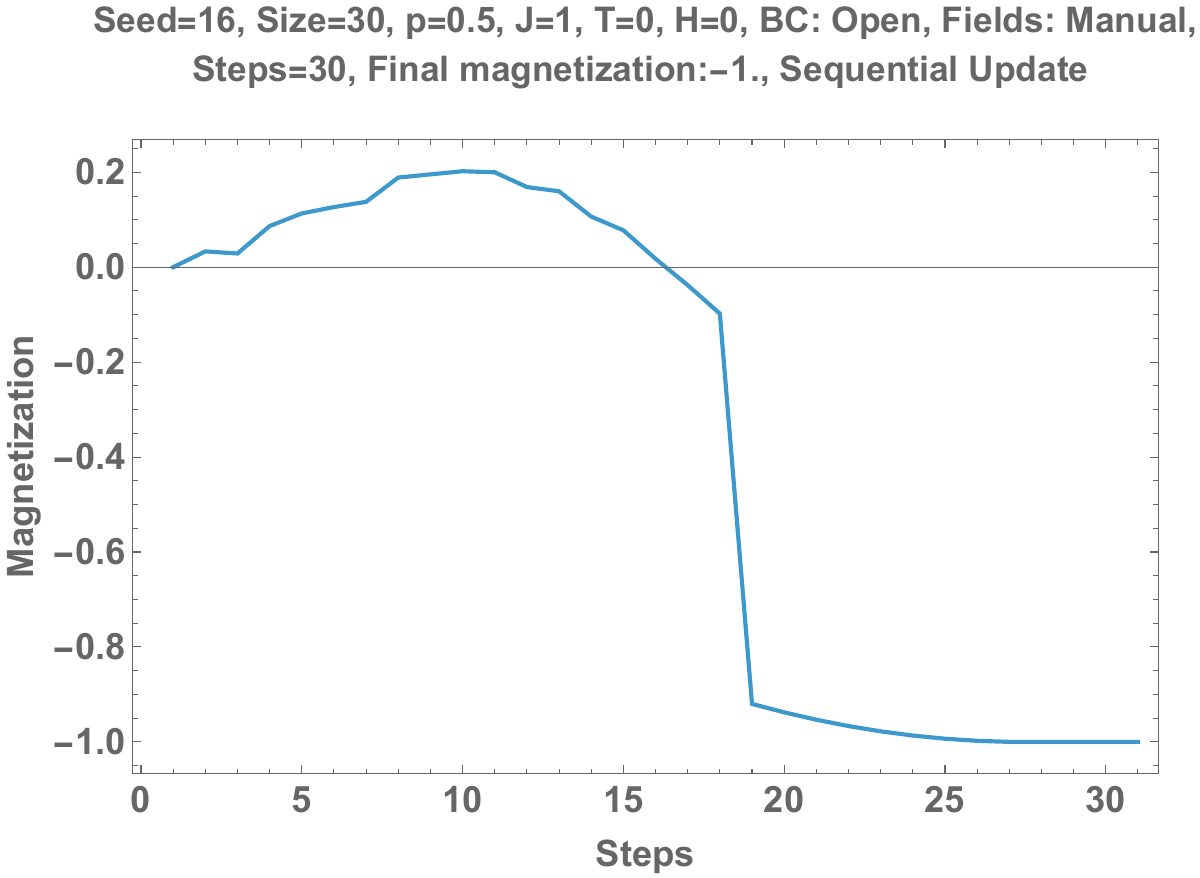}} 
\subfigure[]{\includegraphics[width=0.22\textwidth]{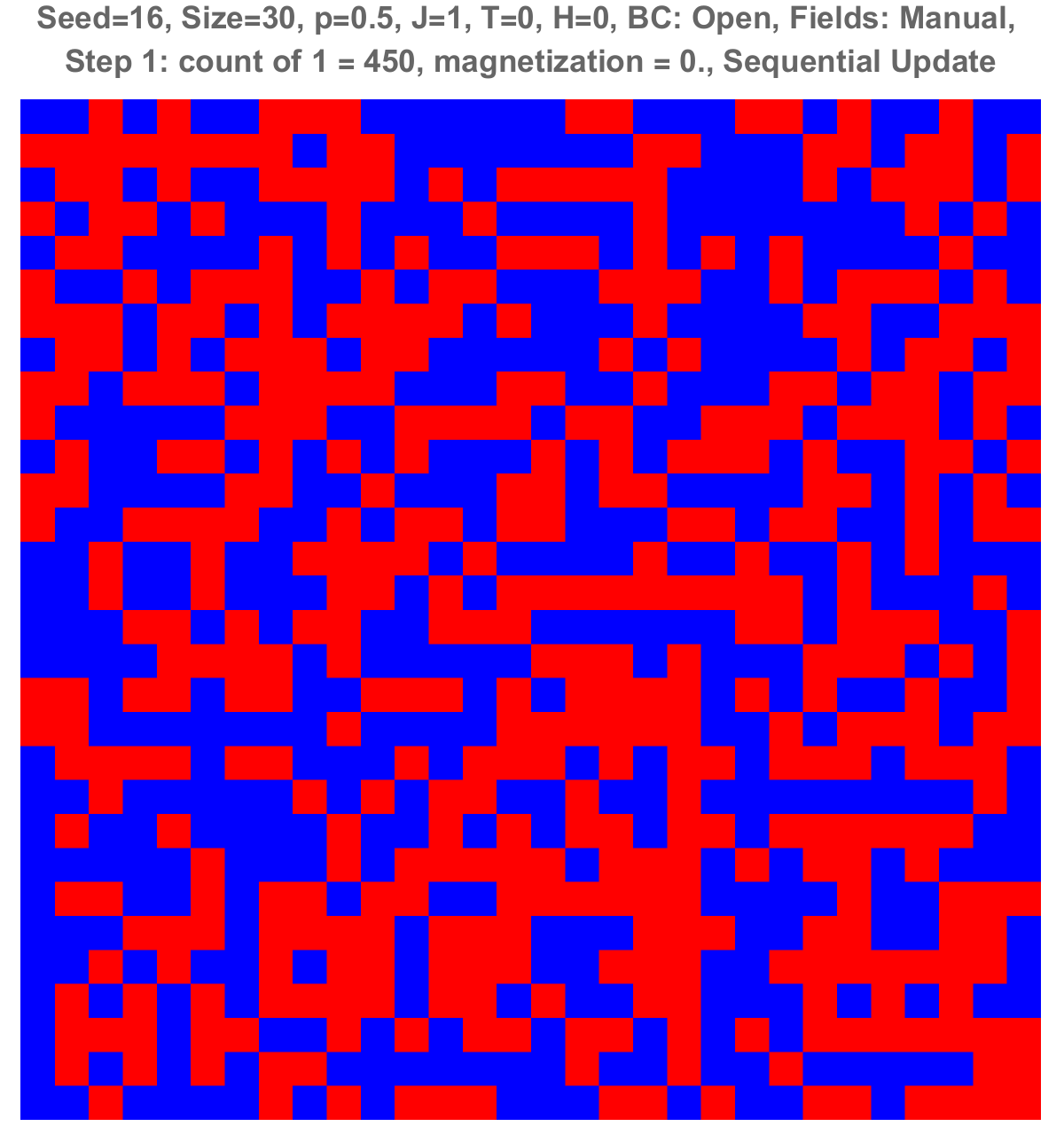}}
\subfigure[]{\includegraphics[width=0.22\textwidth]{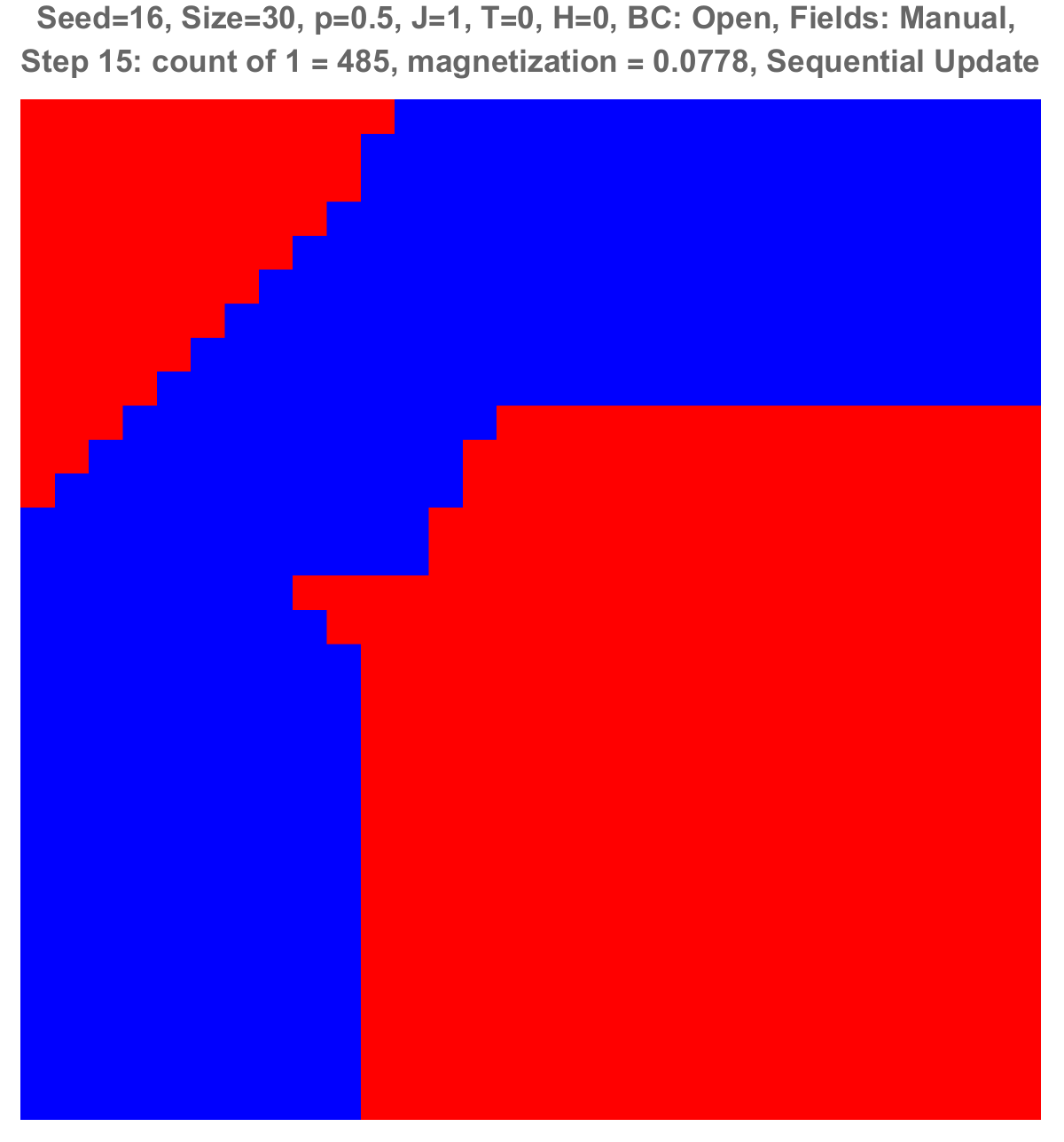}} 
\subfigure[]{\includegraphics[width=0.22\textwidth]{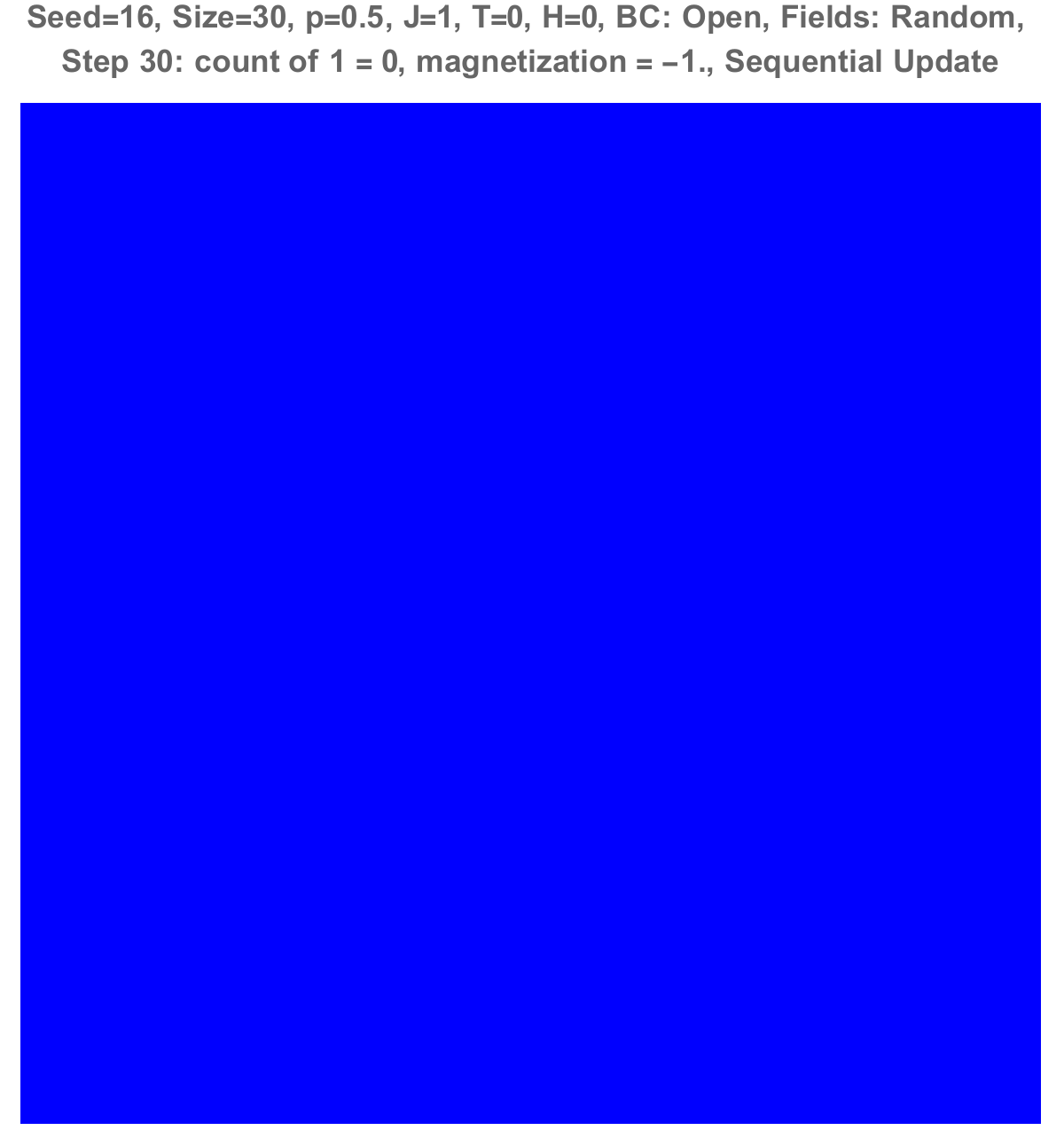}}
\\
\subfigure[]{\includegraphics[width=0.3\textwidth]{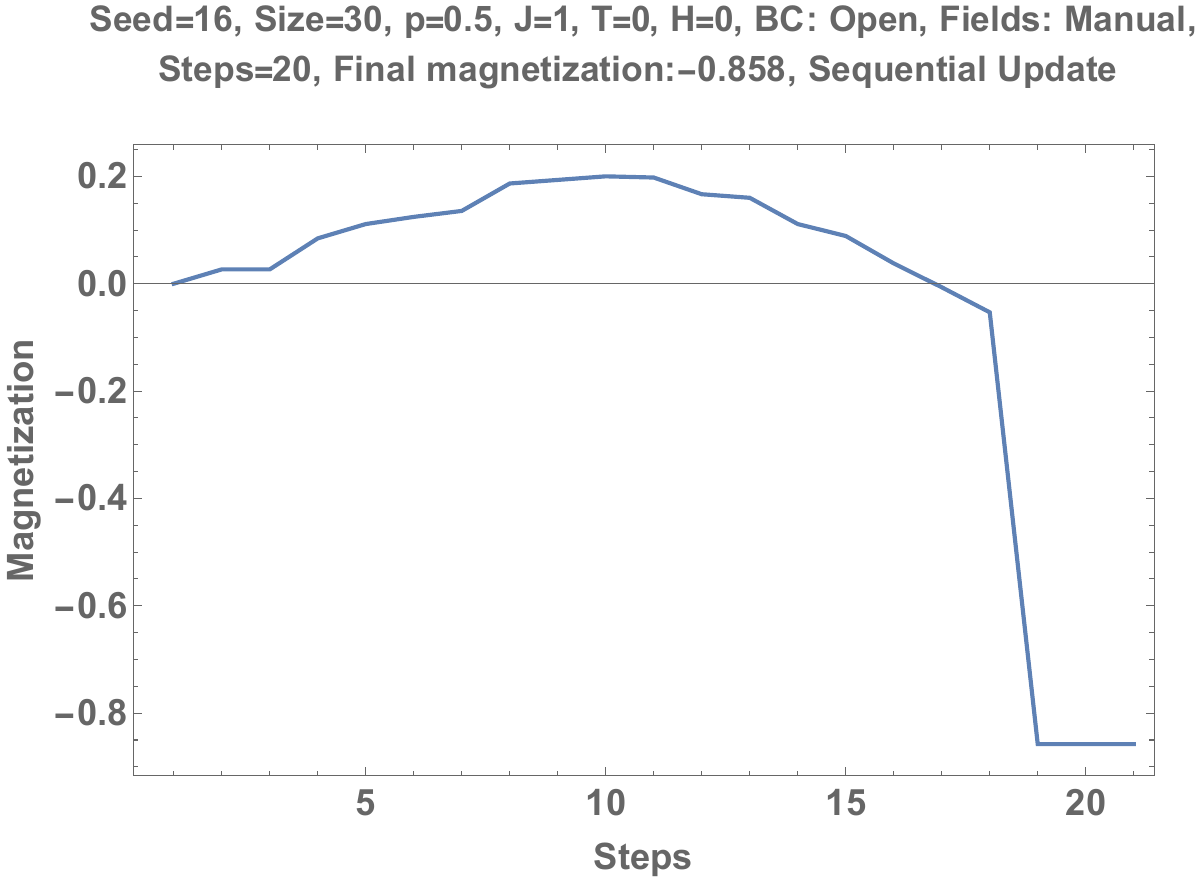}} 
\subfigure[]{\includegraphics[width=0.22\textwidth]{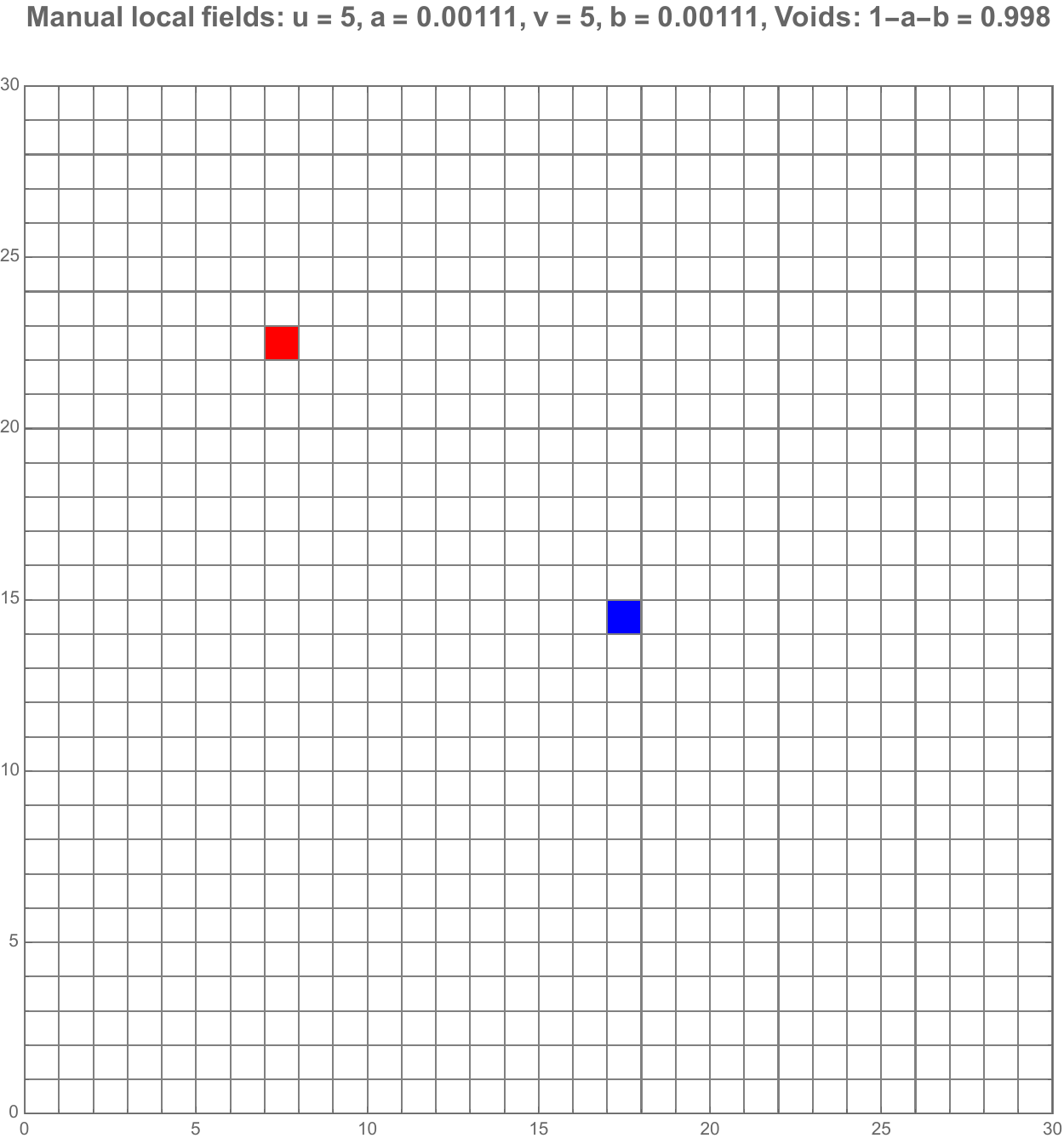}}
\subfigure[]{\includegraphics[width=0.22\textwidth]{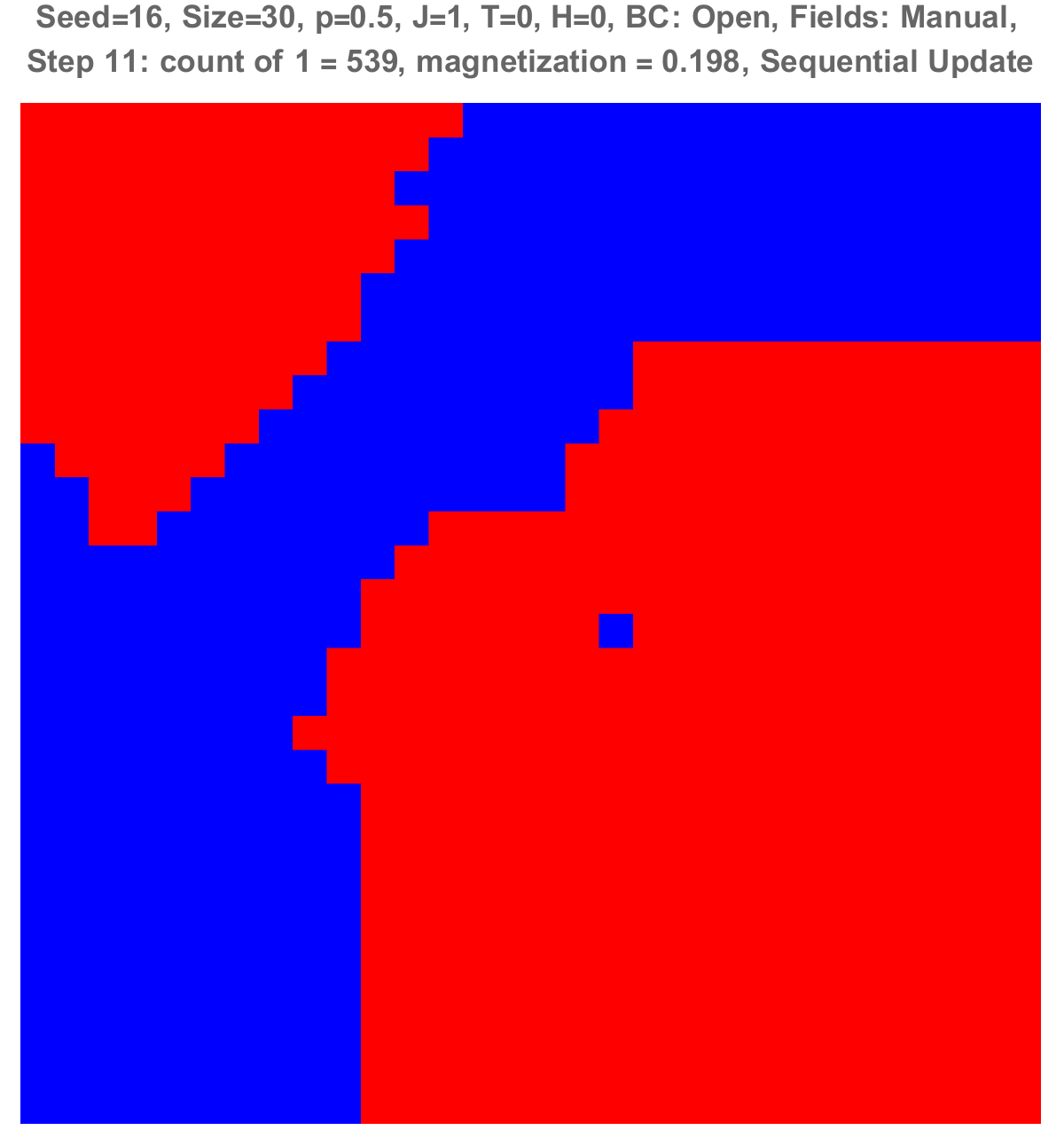}}
\subfigure[]{\includegraphics[width=0.22\textwidth]{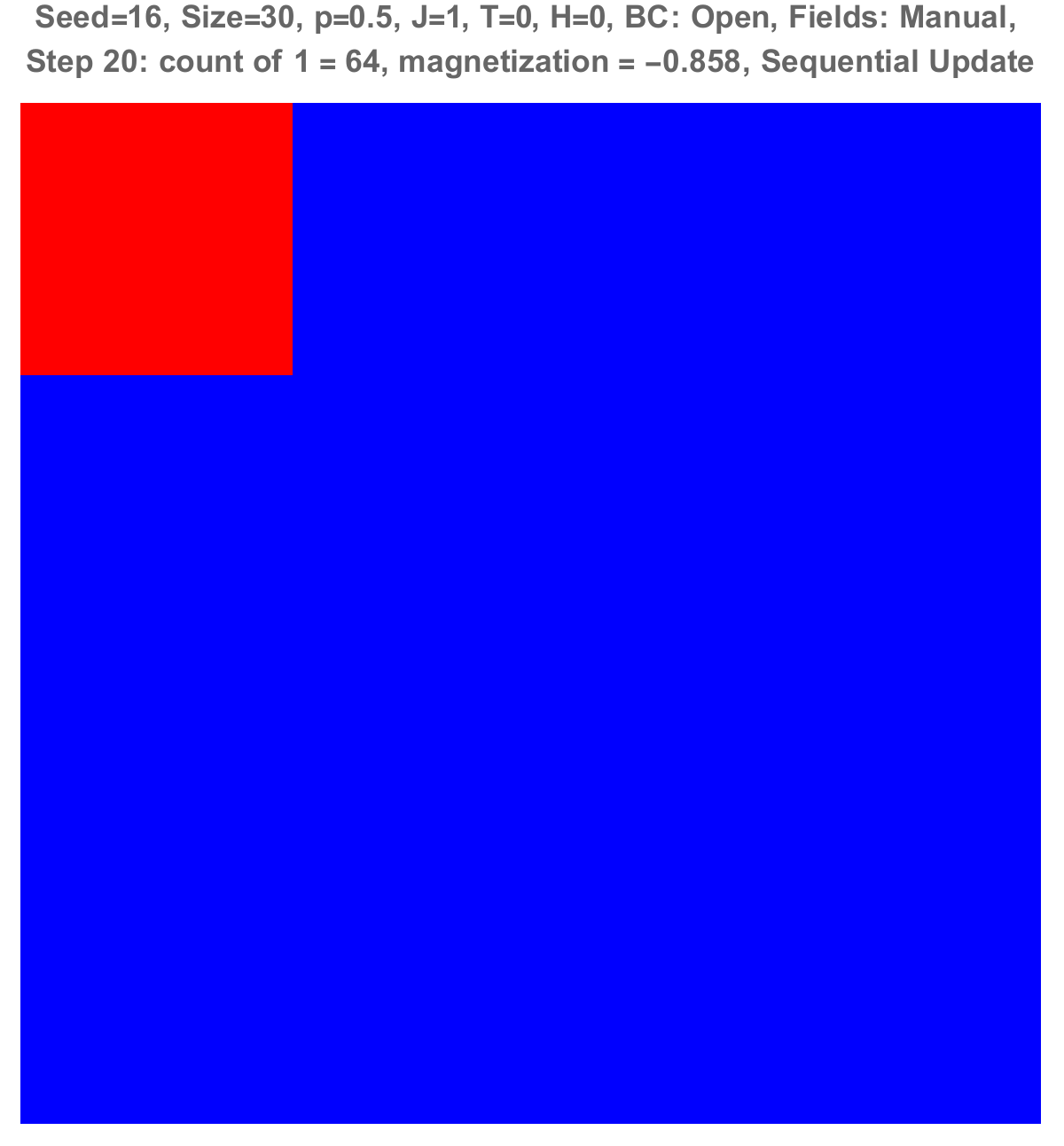}}
\\
\subfigure[]{\includegraphics[width=0.3\textwidth]{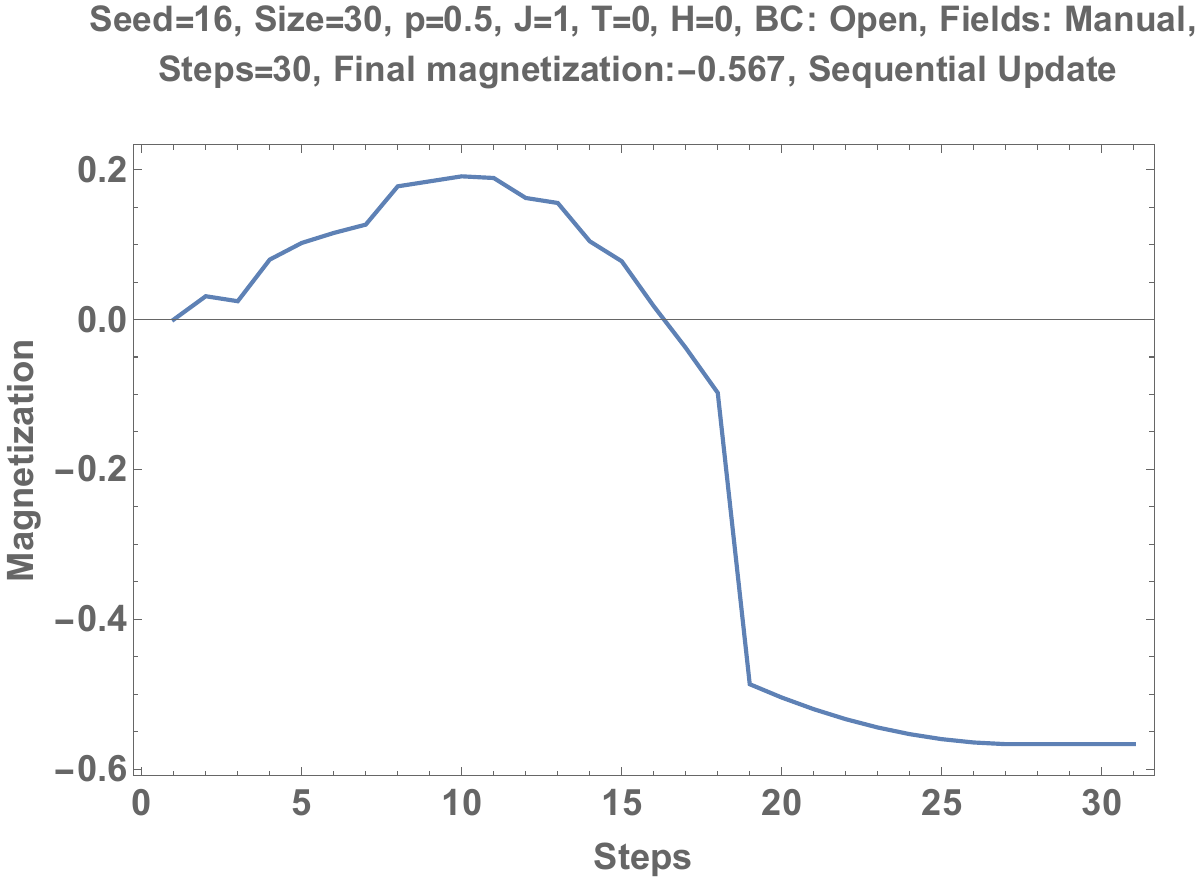}}
\subfigure[]{\includegraphics[width=0.22\textwidth]{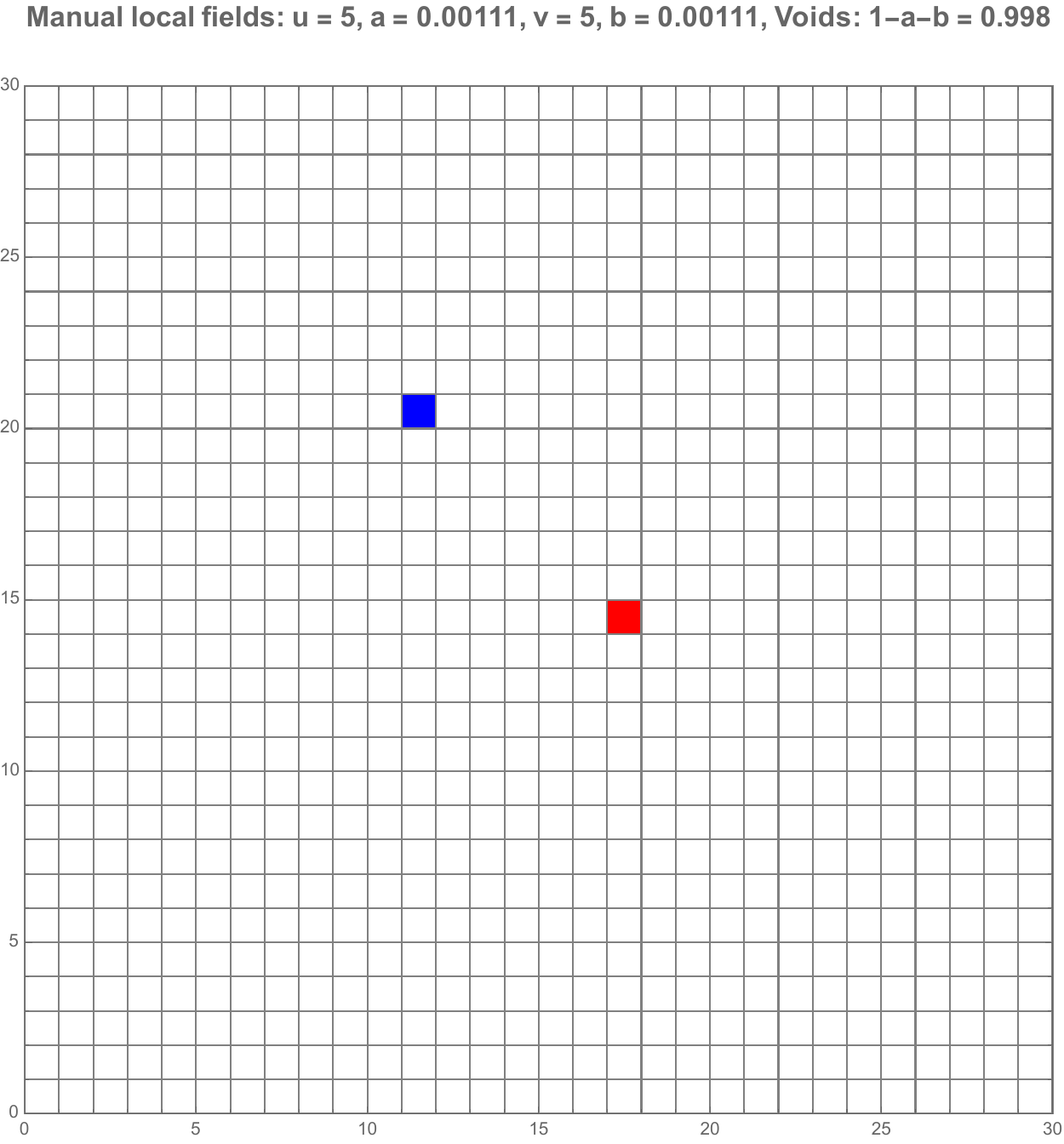}}
\subfigure[]{\includegraphics[width=0.22\textwidth]{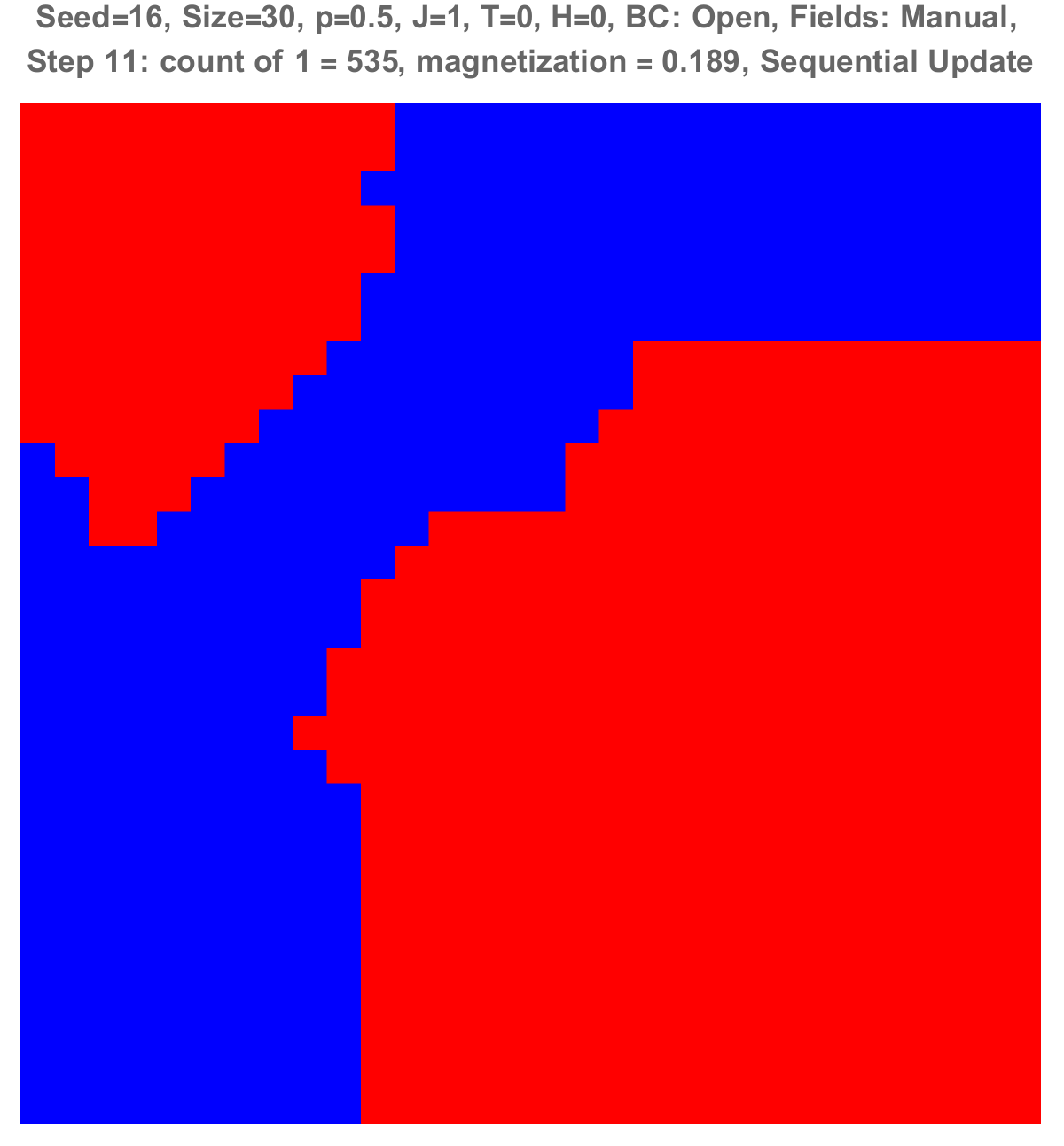}} 
\subfigure[]{\includegraphics[width=0.22\textwidth]{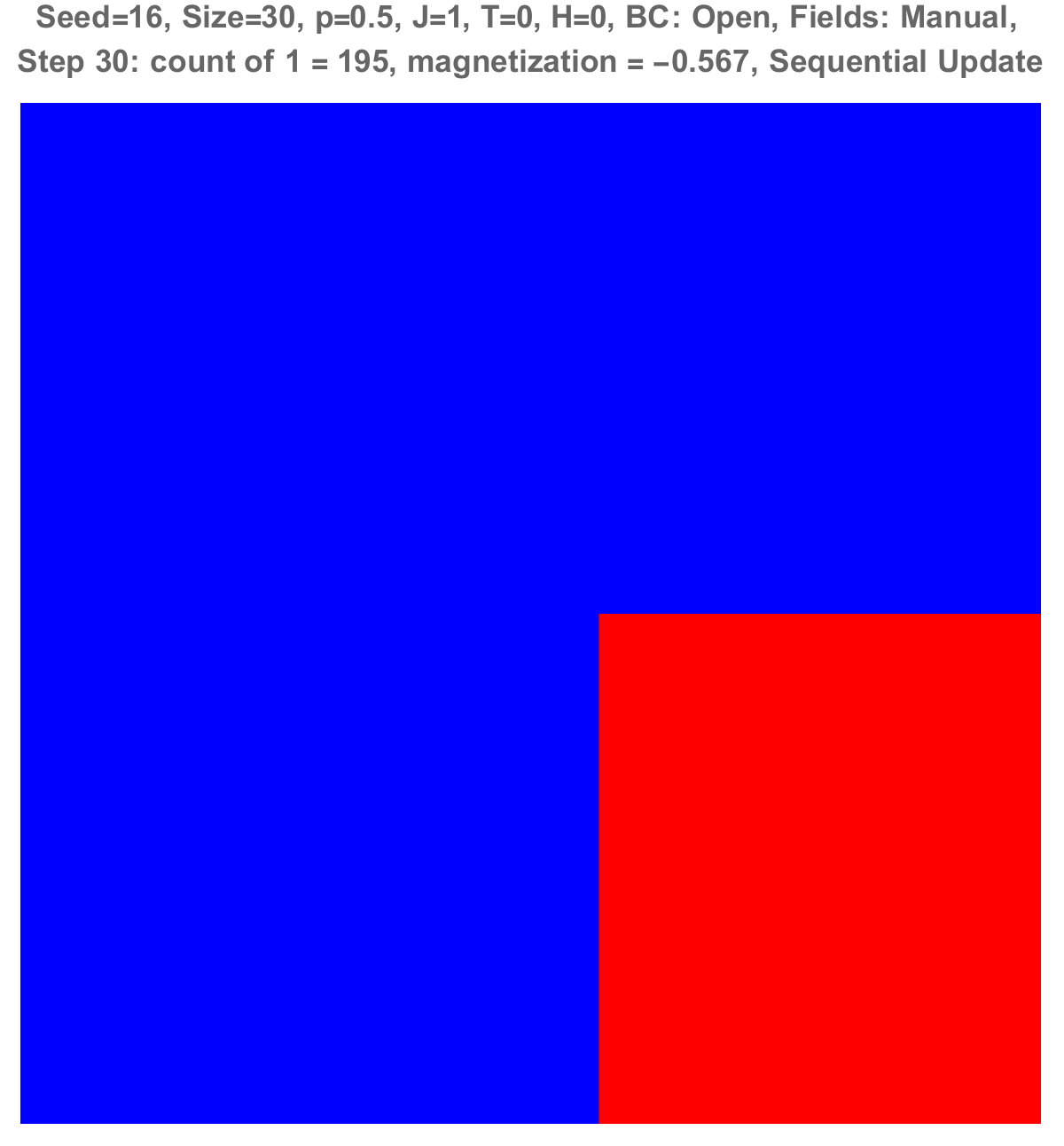}}
\\
\subfigure[]{\includegraphics[width=0.3\textwidth]{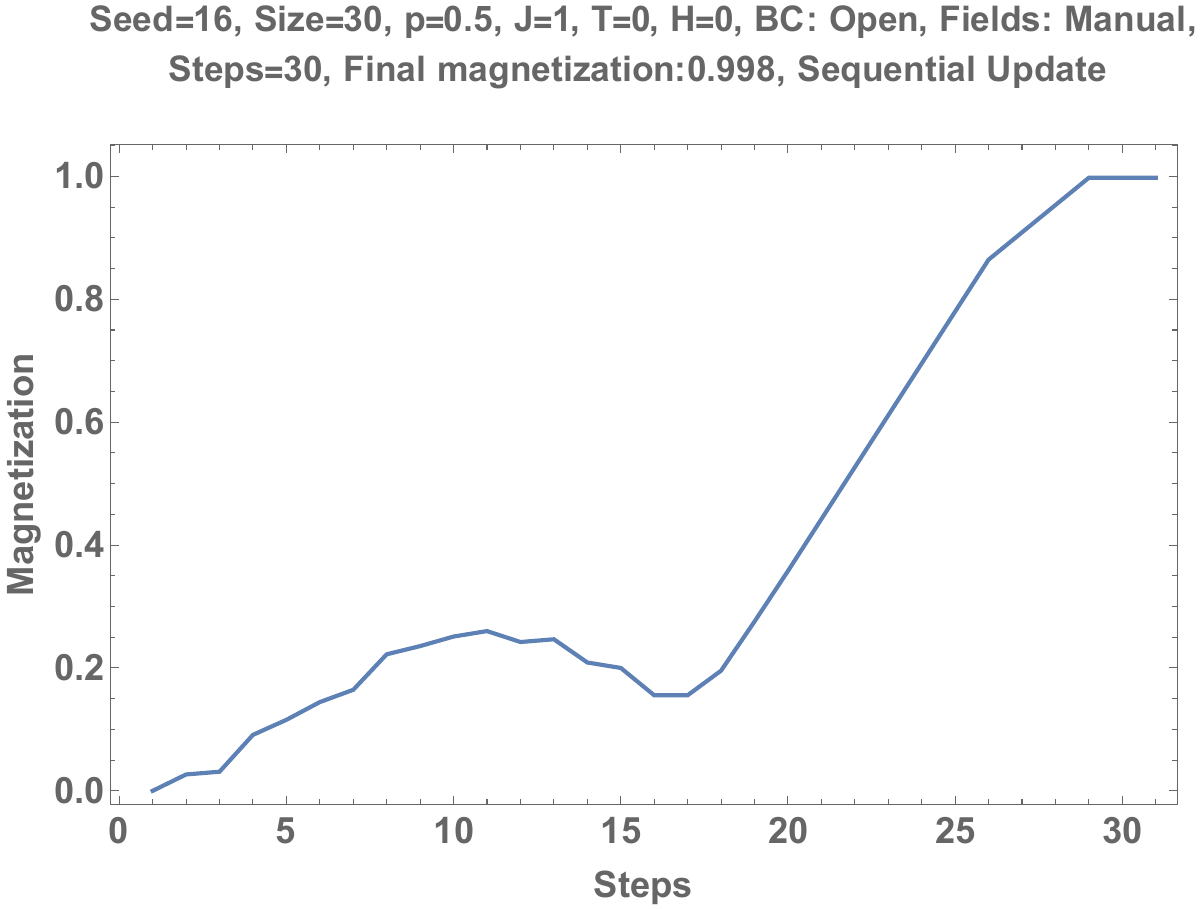}}
\subfigure[]{\includegraphics[width=0.22\textwidth]{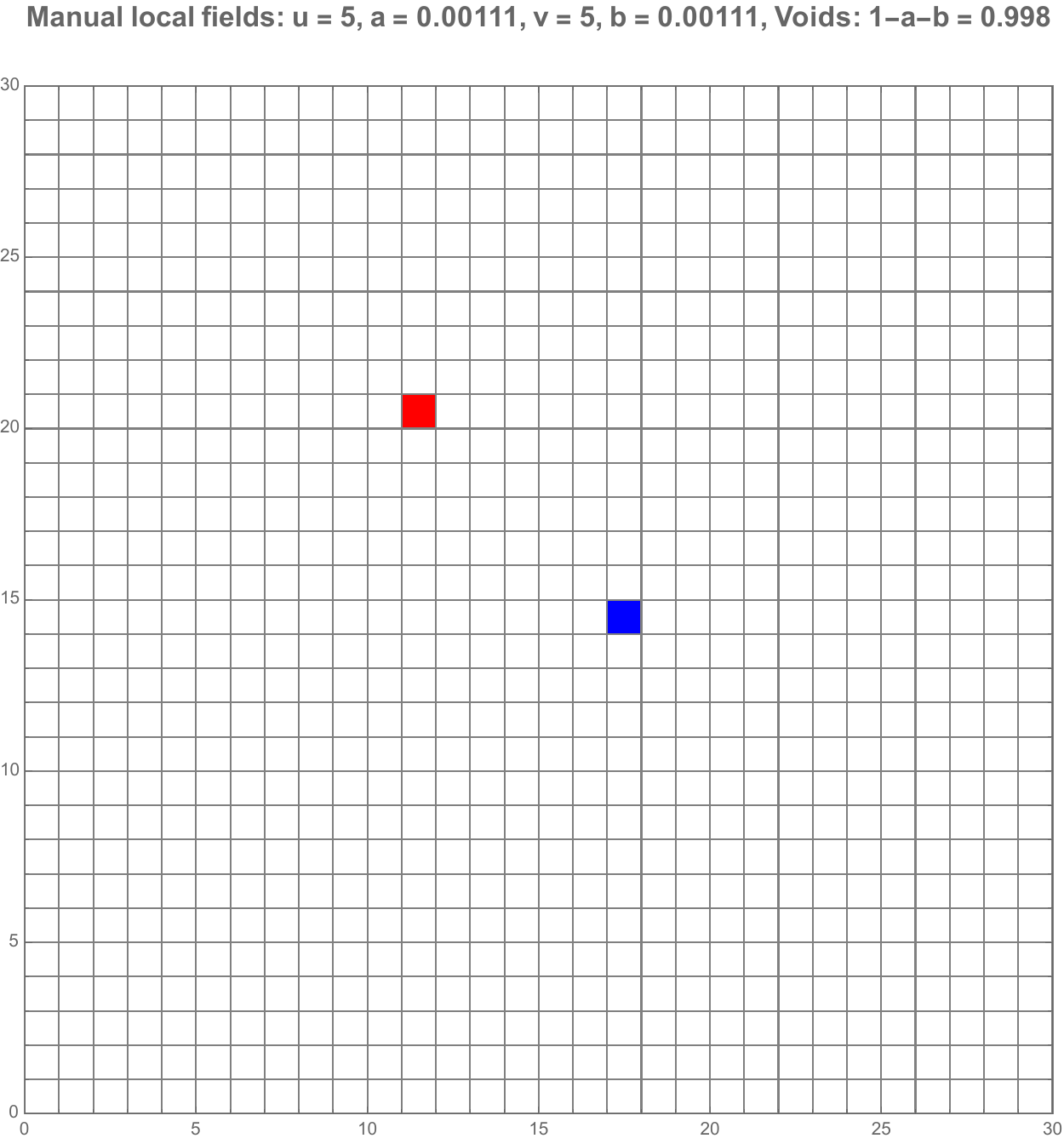}}
\subfigure[]{\includegraphics[width=0.22\textwidth]{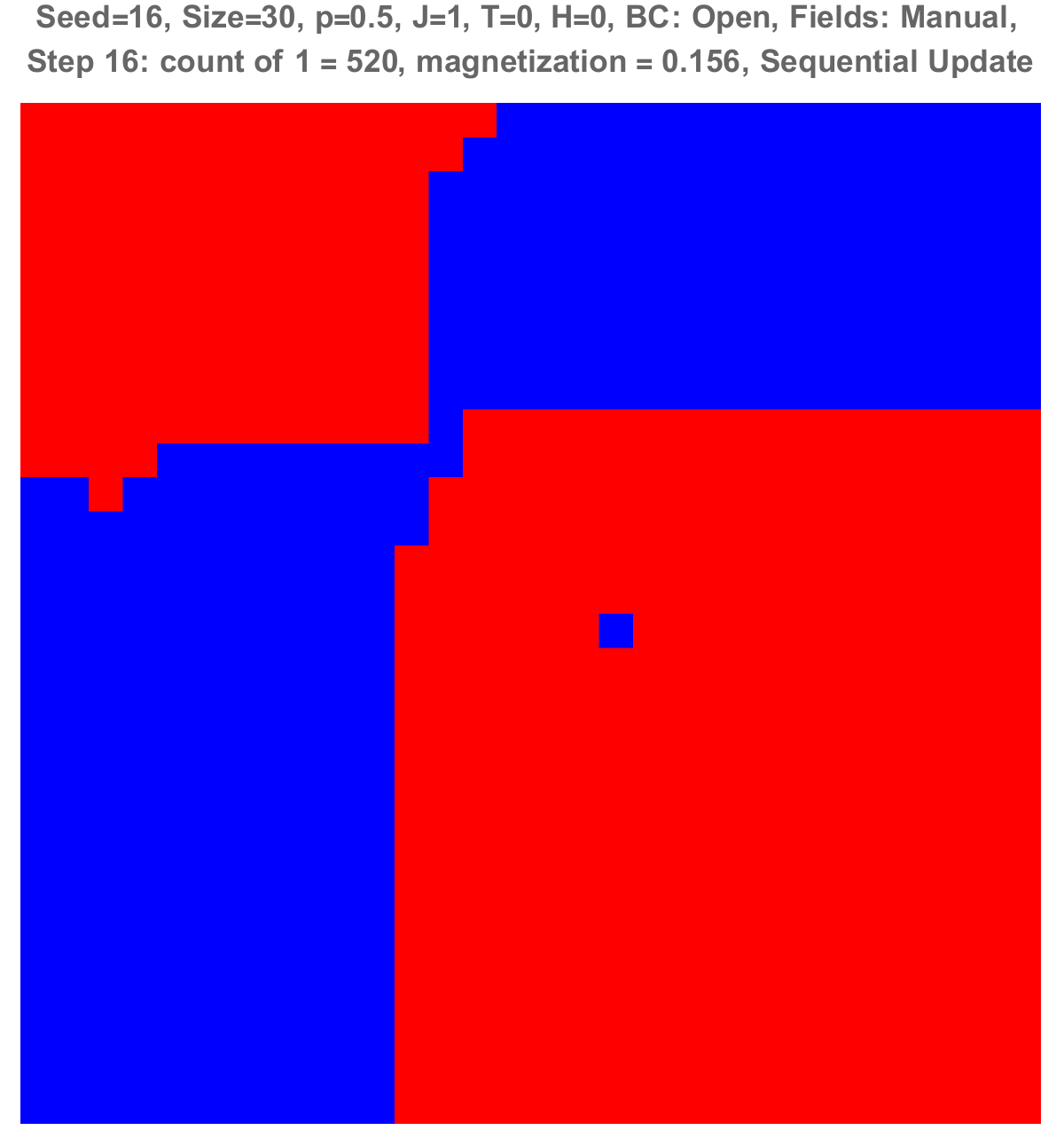}}
\subfigure[]{\includegraphics[width=0.22\textwidth]{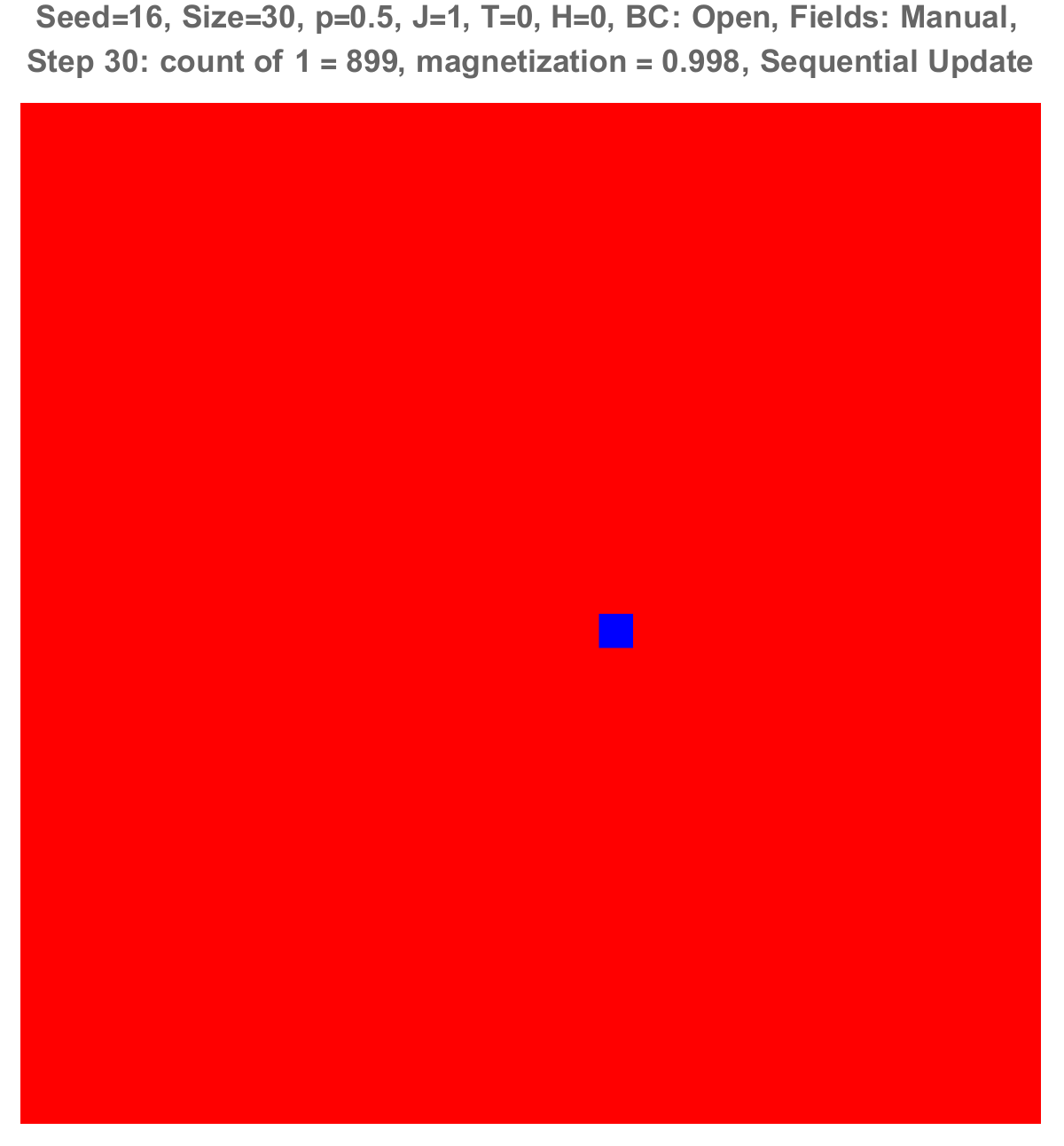}}
\end{figure}

\newpage 

\noindent\captionof{figure}{A simulation with an initial configuration with half +1 and half -1 ($p=0.50$, subpart b)) without local fields. The dynamics is shown as a function of MC steps in subpart (a). The +1 choice gets an increased support till about 10 MC steps and then starts loosing support before shrinking to zero within only a few MC steps. Subpart (c) shows the existence of two domains with A being majority (485 +1 versus 415 -1) after a completion of 15 MC steps. Few more MC steps leads to an unexpected total victory (spontaneous symmetry breaking) of B as seen in subpart (d). In the second row two local opposite fields have been added in subpart (f). More MC steps keeps A being majority (subpart (g)). The final outcome has a small A domain (subpart (h)). In the third row the locations of the red and blue local fields have been shifted (subpart (j)).The results are similar to the above ones with now two separate red domains separated by a blue domain (subpart (k)). The final small A domain is in the lower right part of the grid (subpart (l)). In the fourth row the field colors have been swapped (subpart (n)) During the first 15 MC steps, the dynamics (subparts (m, o))  is similar to subparts (i, k). However, after the beginning of the decrease, instead of a red shrinking, the red is boosted at once to invade the full sample (subpart (p)).}
\label{aa}

\subsubsection{Figure (\ref{bb})}

Using the same initial distribution of red and blue sites from subpart (b) of Figure (\ref{aa}) I add two red and two blue local fields located in the upper left part of the grid. Four very similar field locations are used as shown in subparts (b, e, h, k). Respective dynamics are exhibited in  subparts (a, d, g, j). Subparts (c, f, i, l) show the related final outcomes with drastic contrasts in the winning color. The results illustrate the fragility of the final state against the application of two local fields at strategic locations. This fact reveals the vulnerability of a completely broken symmetry in the face of just two well-positioned local fields.

\begin{figure}
\centering
\vspace{-3cm}
\subfigure[]{\includegraphics[width=0.32\textwidth]{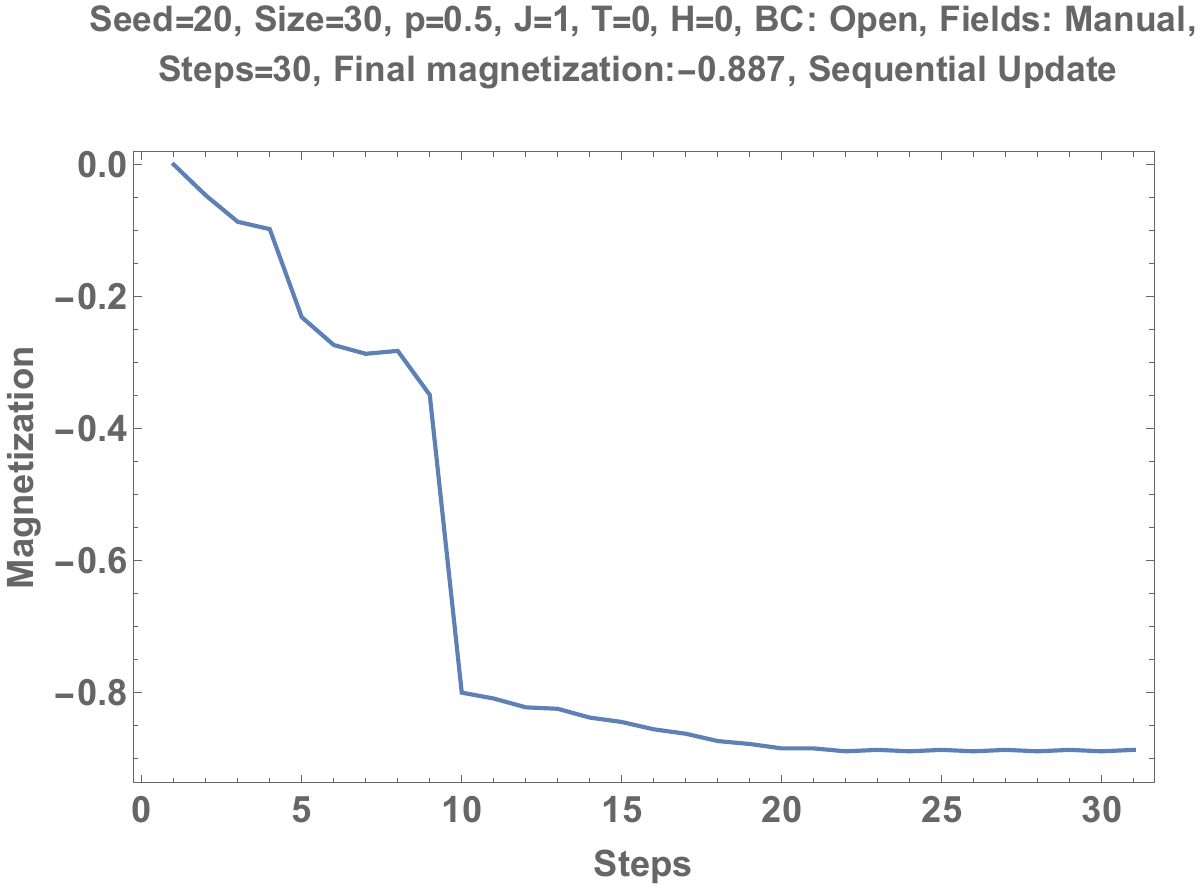}}
\subfigure[]{\includegraphics[width=0.32\textwidth]{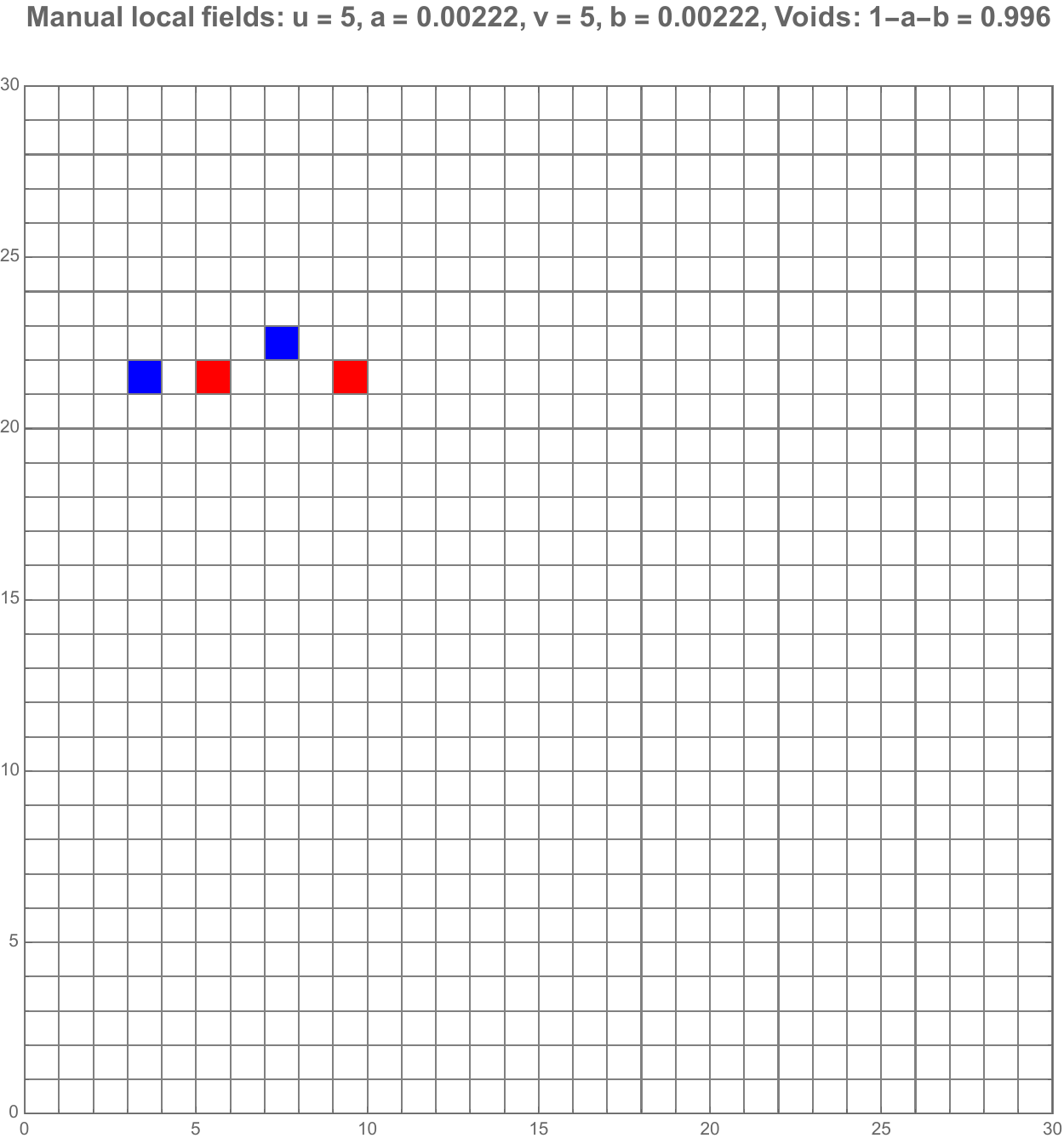}}
\subfigure[]{\includegraphics[width=0.32\textwidth]{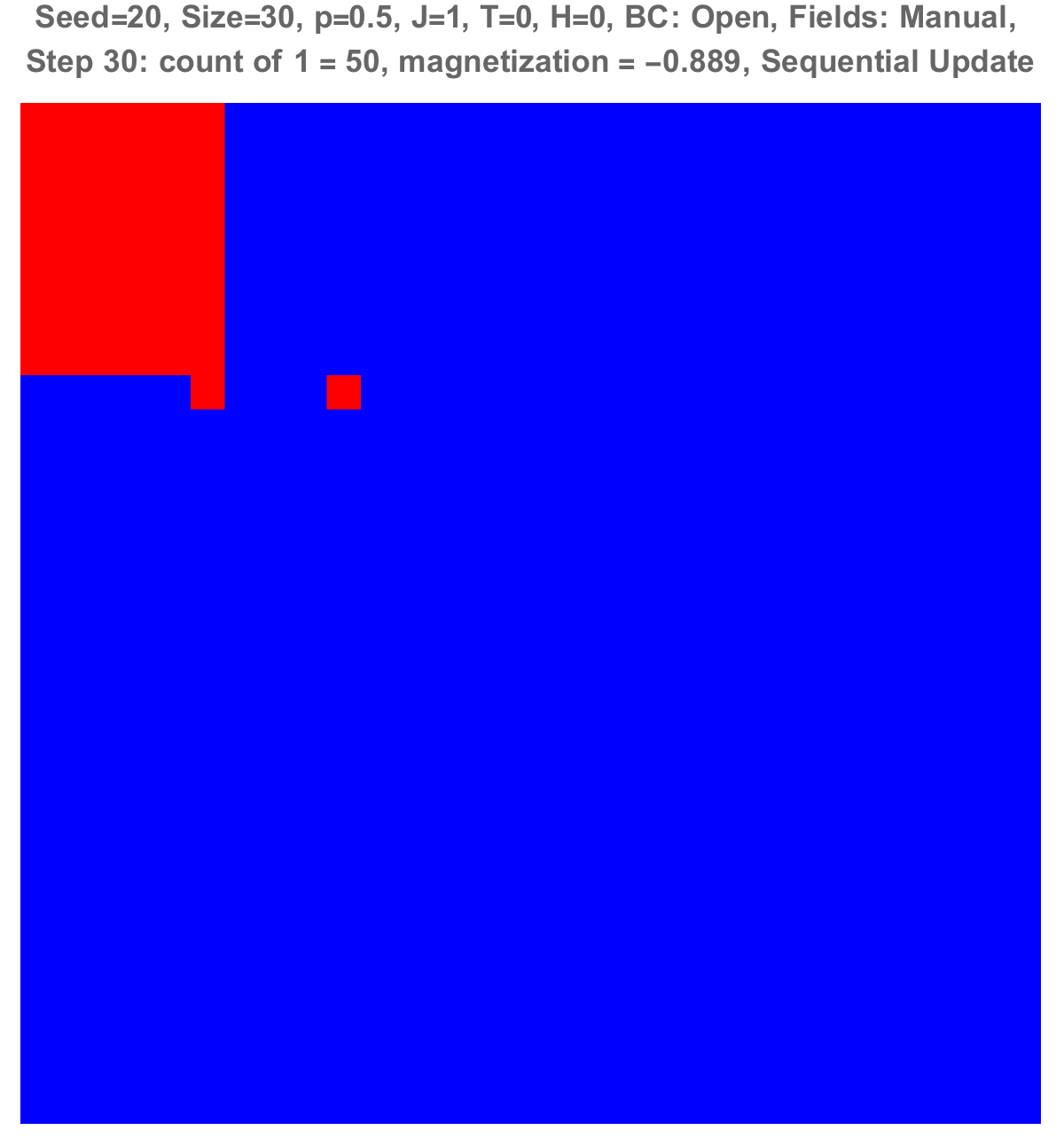}}
\\
\subfigure[]{\includegraphics[width=0.32\textwidth]{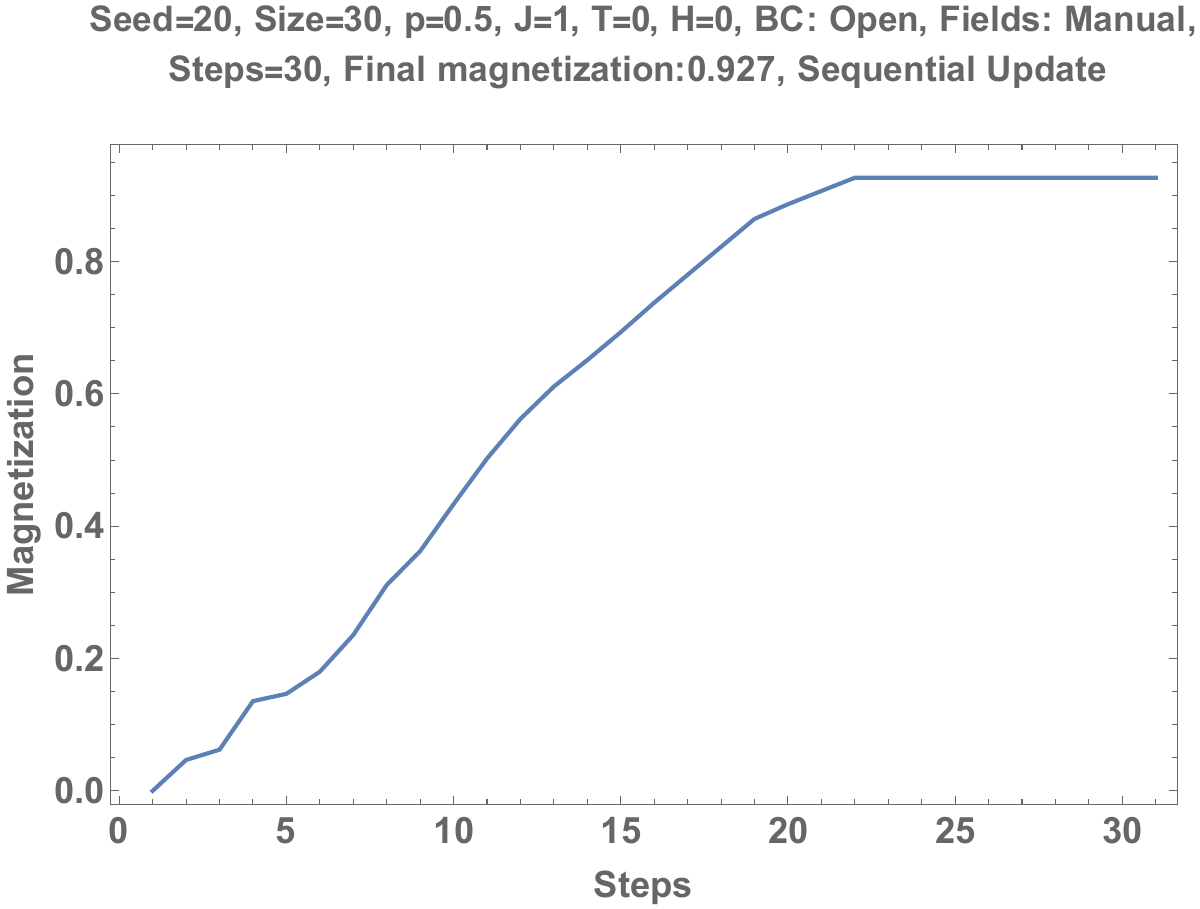}}
\subfigure[]{\includegraphics[width=0.32\textwidth]{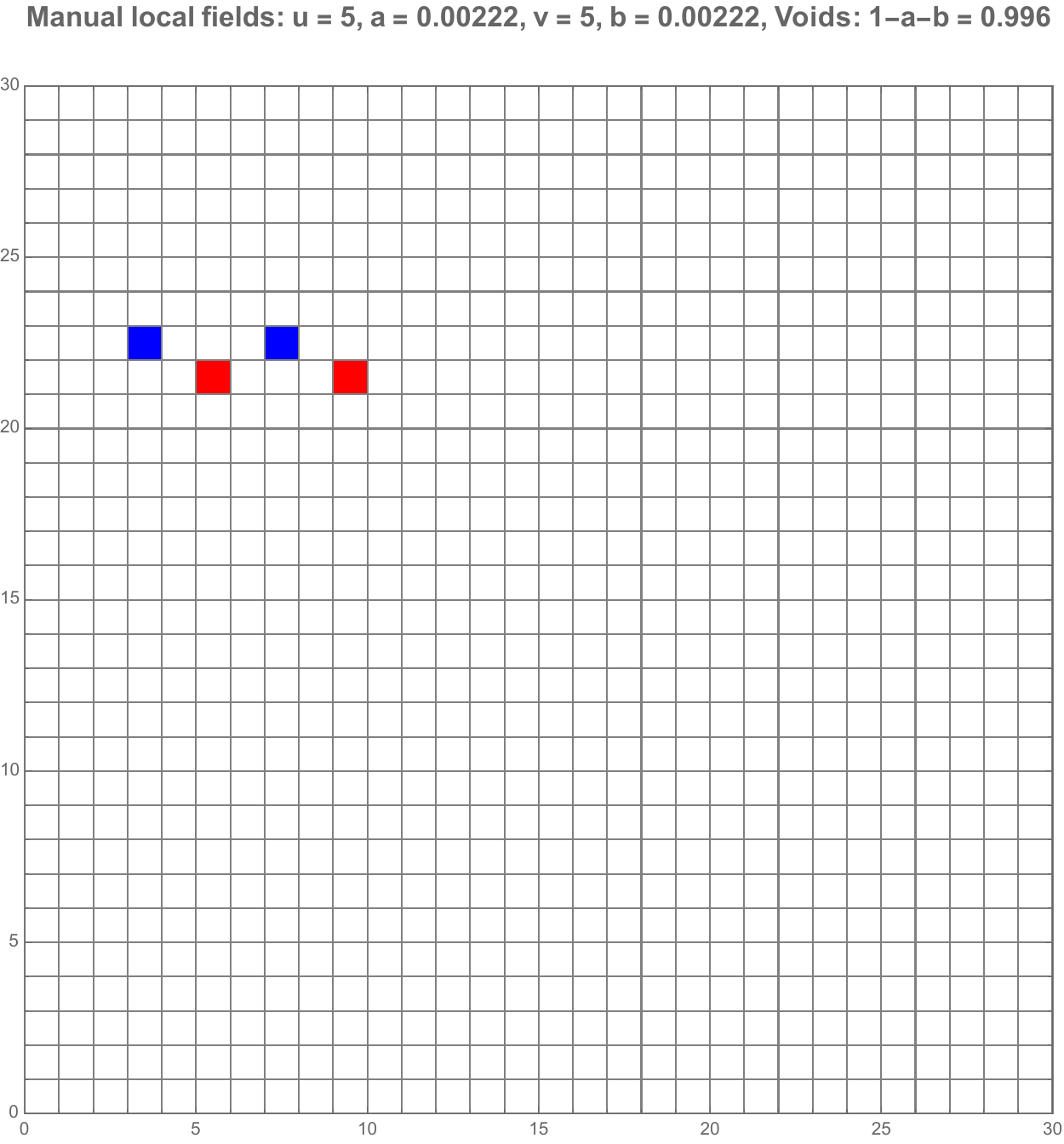}}
\subfigure[]{\includegraphics[width=0.32\textwidth]{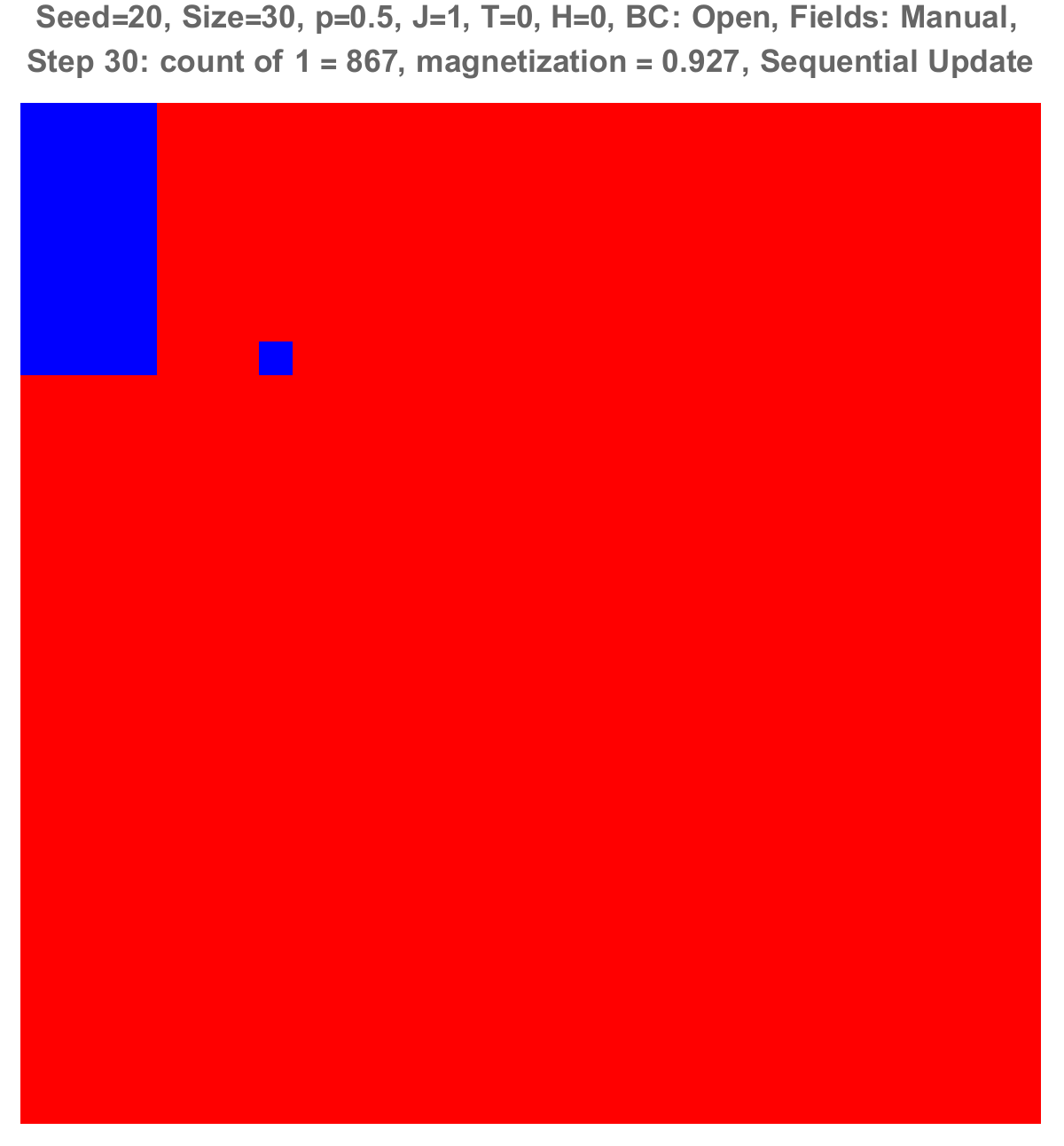}}
\\
\subfigure[]{\includegraphics[width=0.32\textwidth]{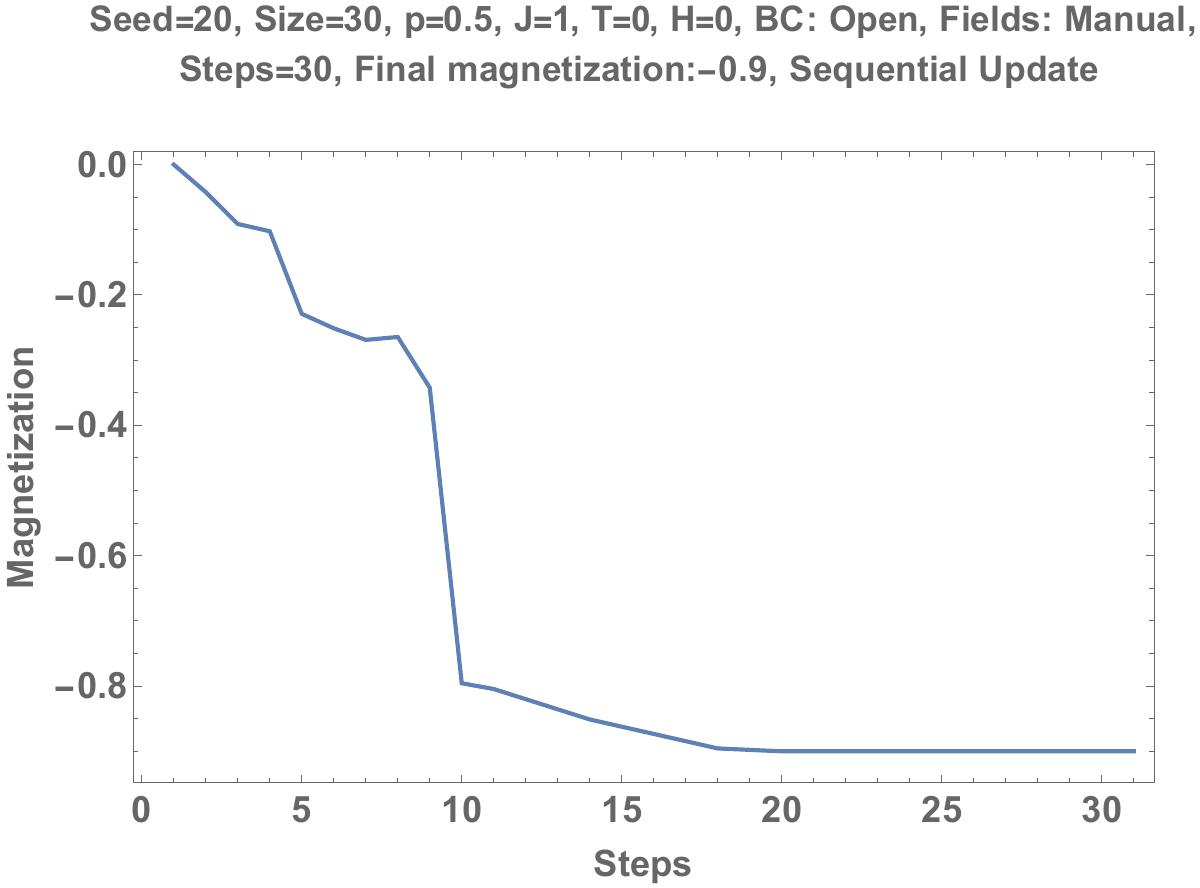}}
\subfigure[]{\includegraphics[width=0.32\textwidth]{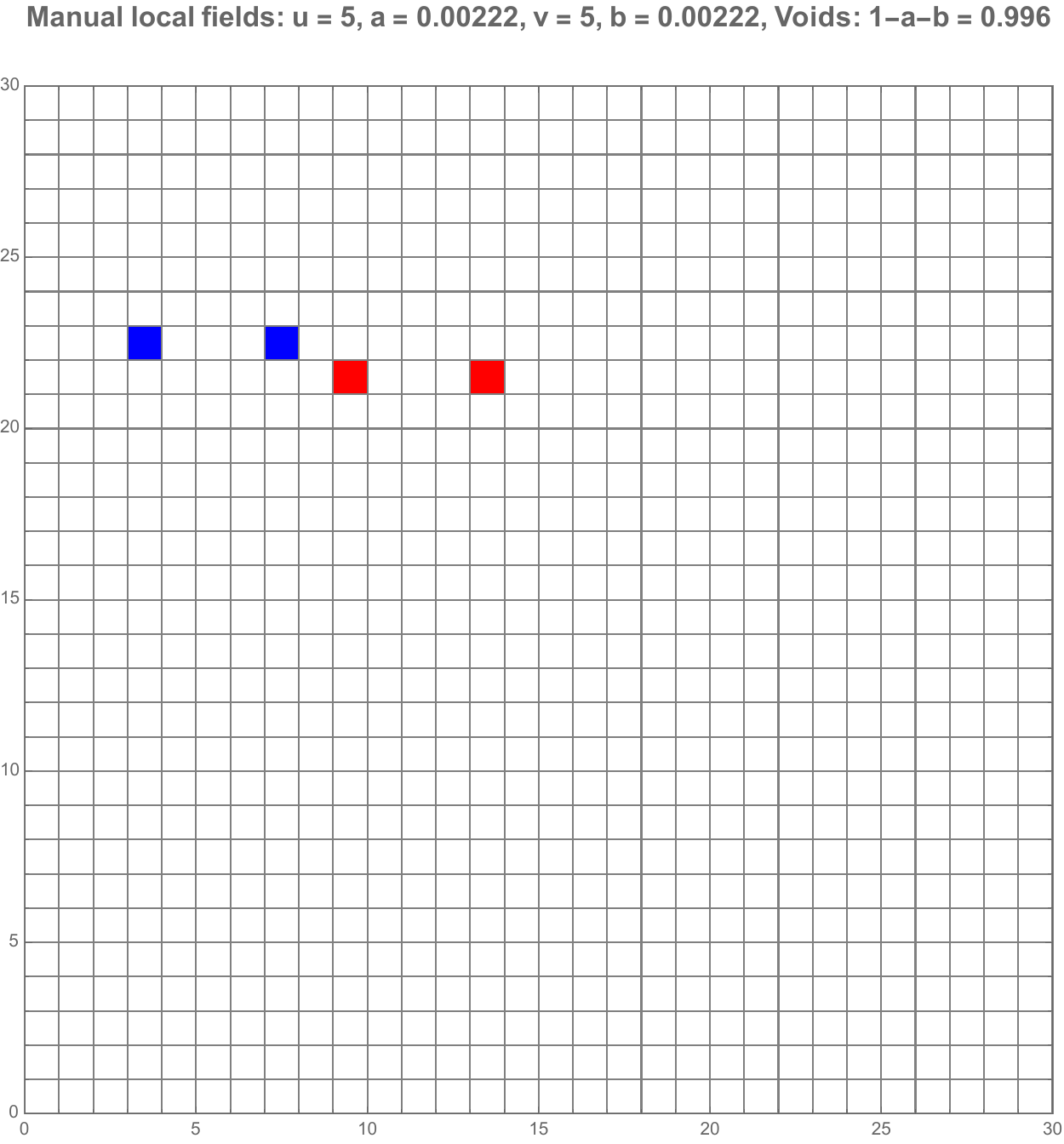}}
\subfigure[]{\includegraphics[width=0.32\textwidth]{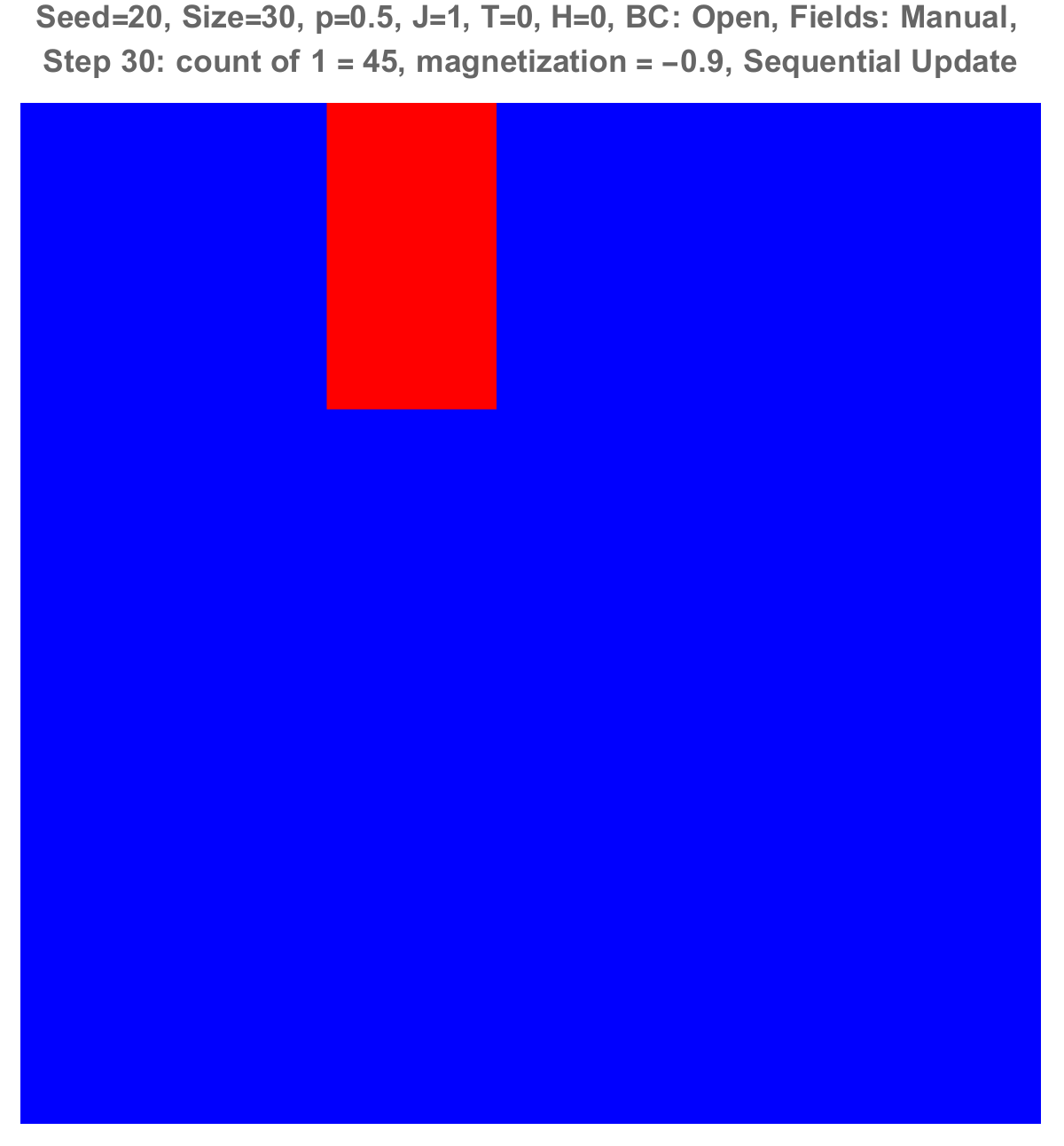}}
\\
\subfigure[]{\includegraphics[width=0.32\textwidth]{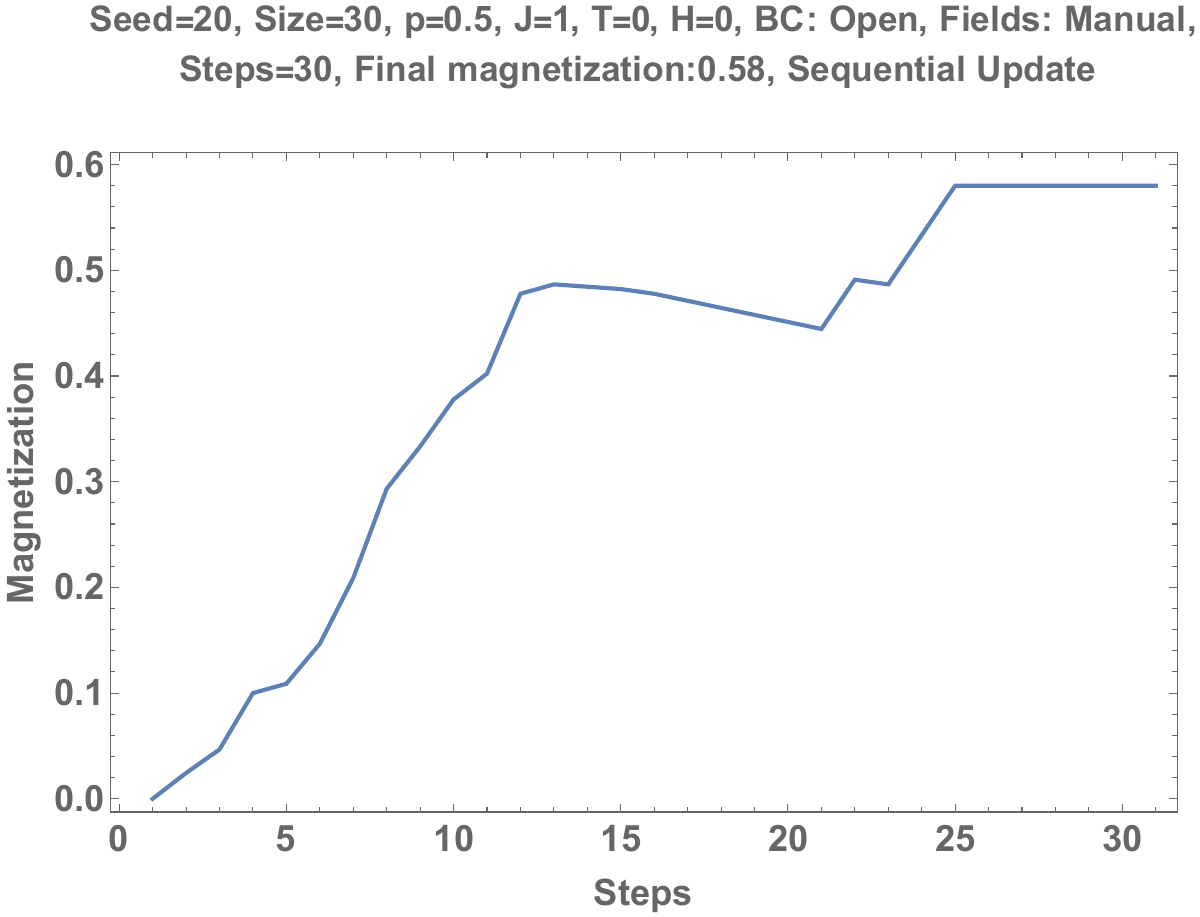}}
\subfigure[]{\includegraphics[width=0.32\textwidth]{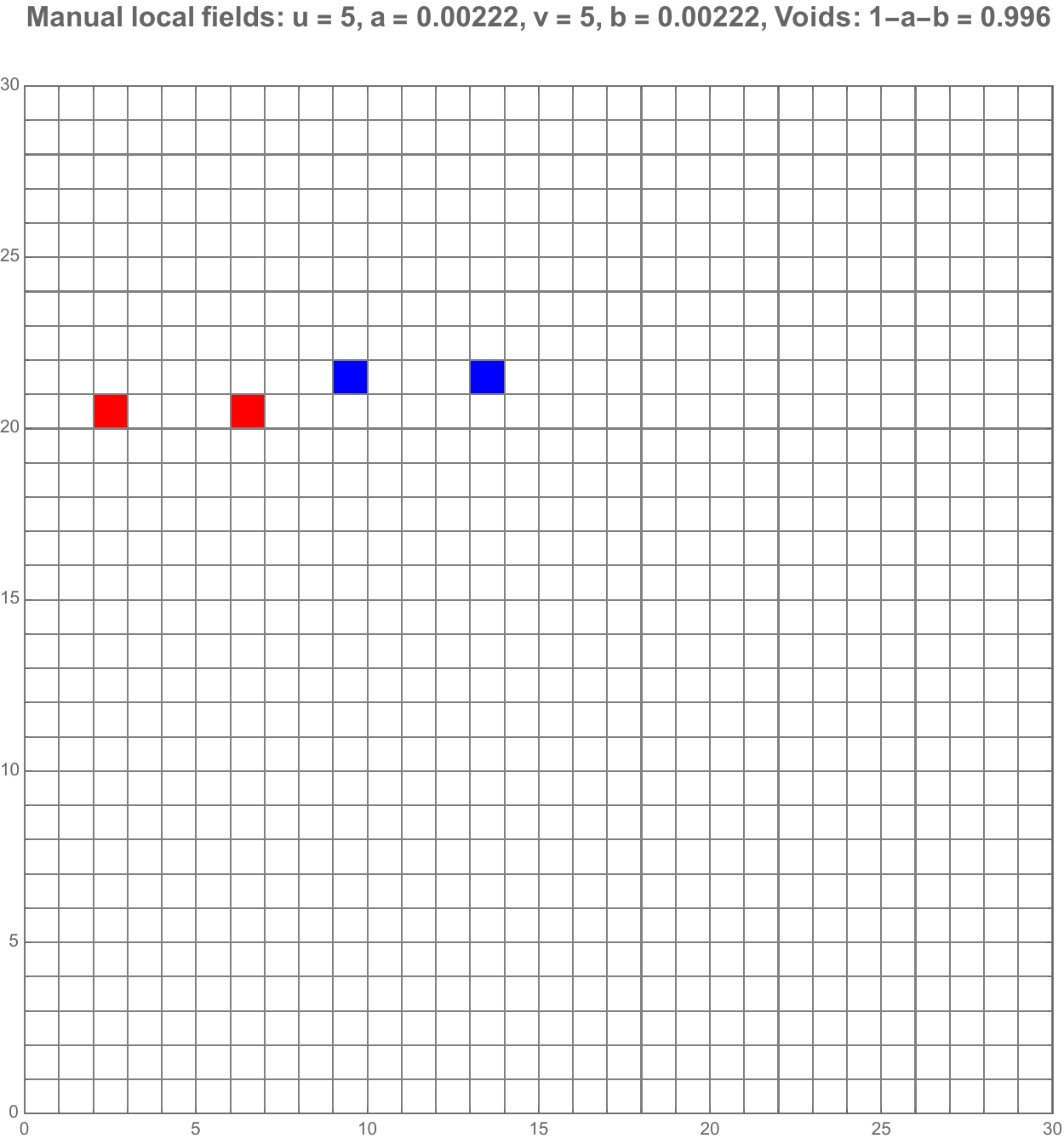}}
\subfigure[]{\includegraphics[width=0.32\textwidth]{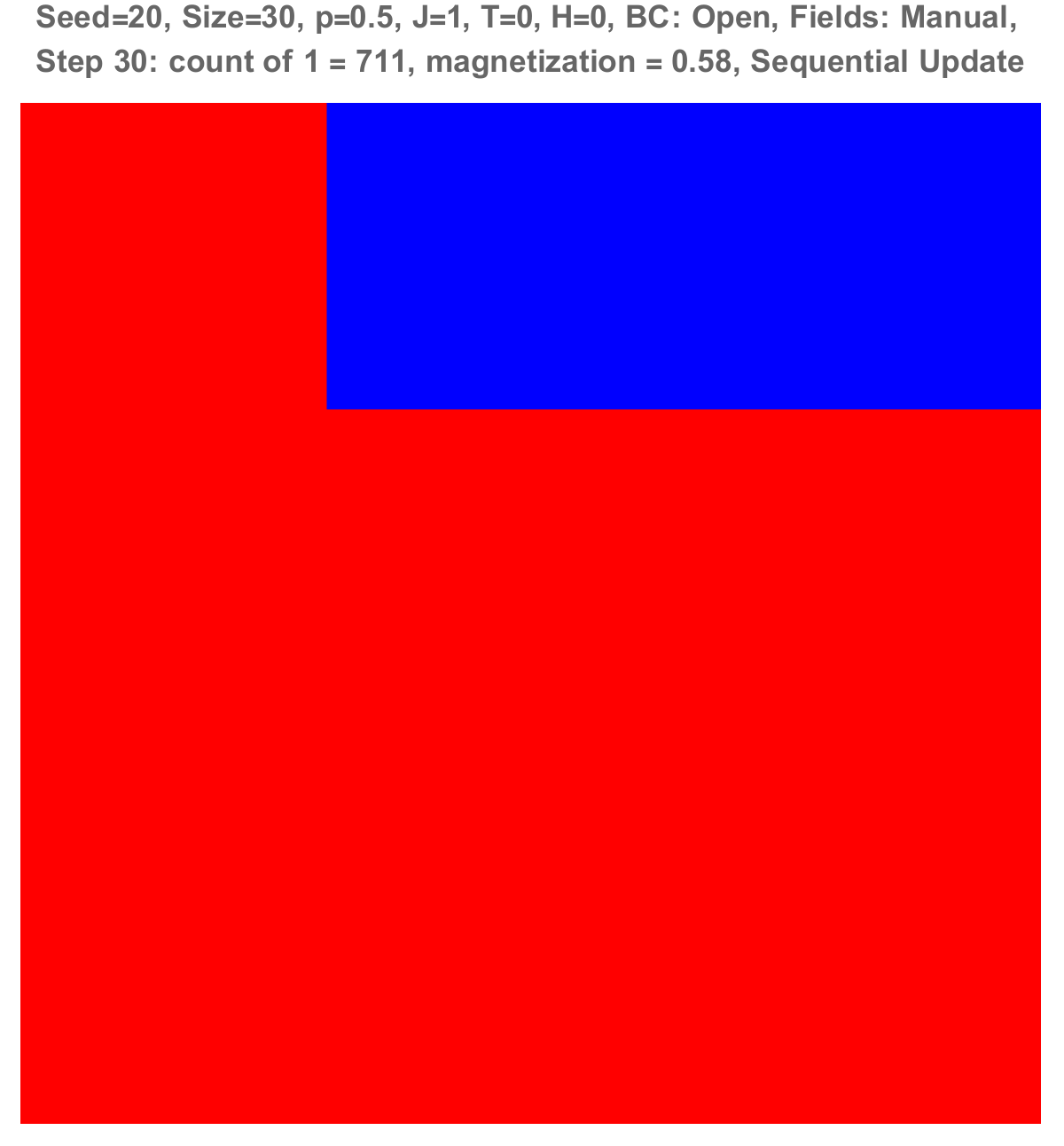}}
\end{figure}

\newpage 

\noindent\captionof{figure}{Same initial distribution of red and blue sites as in subpart (b) of Figure (\ref{aa}). Two red and two blue local fields are applied. Subparts (b, e, h, k) show four different field locations. Respective dynamics are exhibited in  subparts (a, d, g, j). Subparts (c, f, i, l) show the related final outcomes.This fact reveals the vulnerability of a completely broken symmetry in the face of just two well-positioned local fields.}
\label{bb}

\begin{figure}
\centering
\vspace{-4cm}
\subfigure[]{\includegraphics[width=0.32\textwidth]{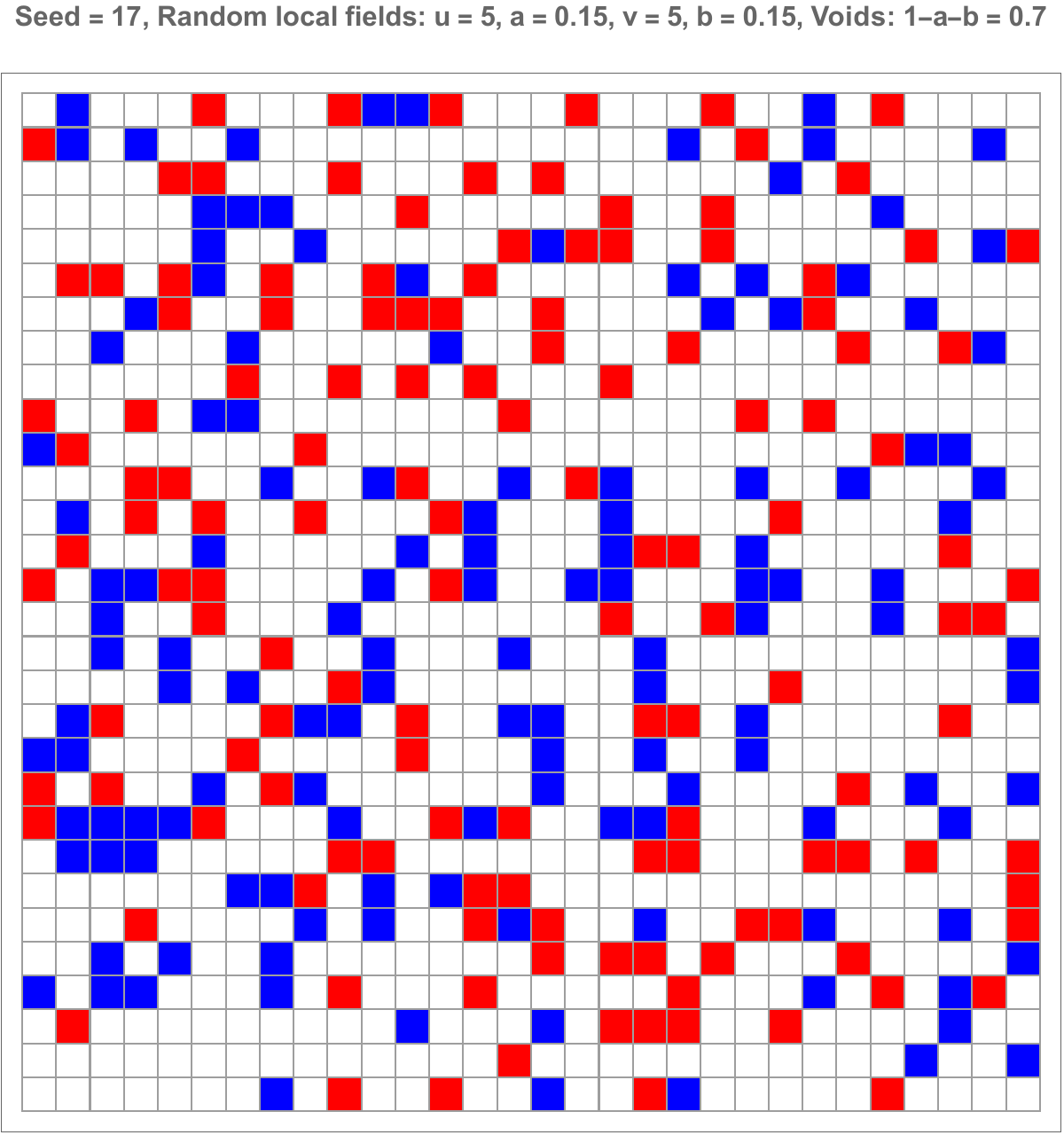}}
\subfigure[]{\includegraphics[width=0.32\textwidth]{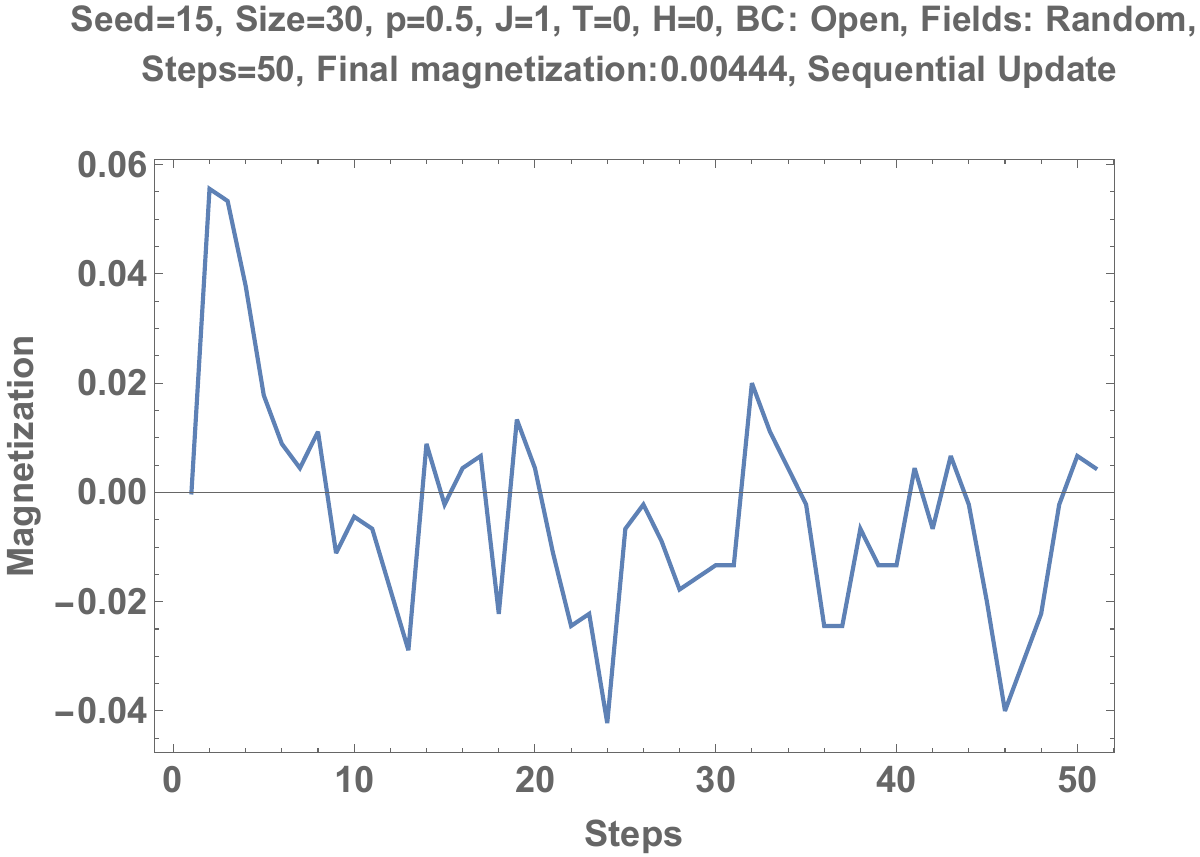}}
\subfigure[]{\includegraphics[width=0.32\textwidth]{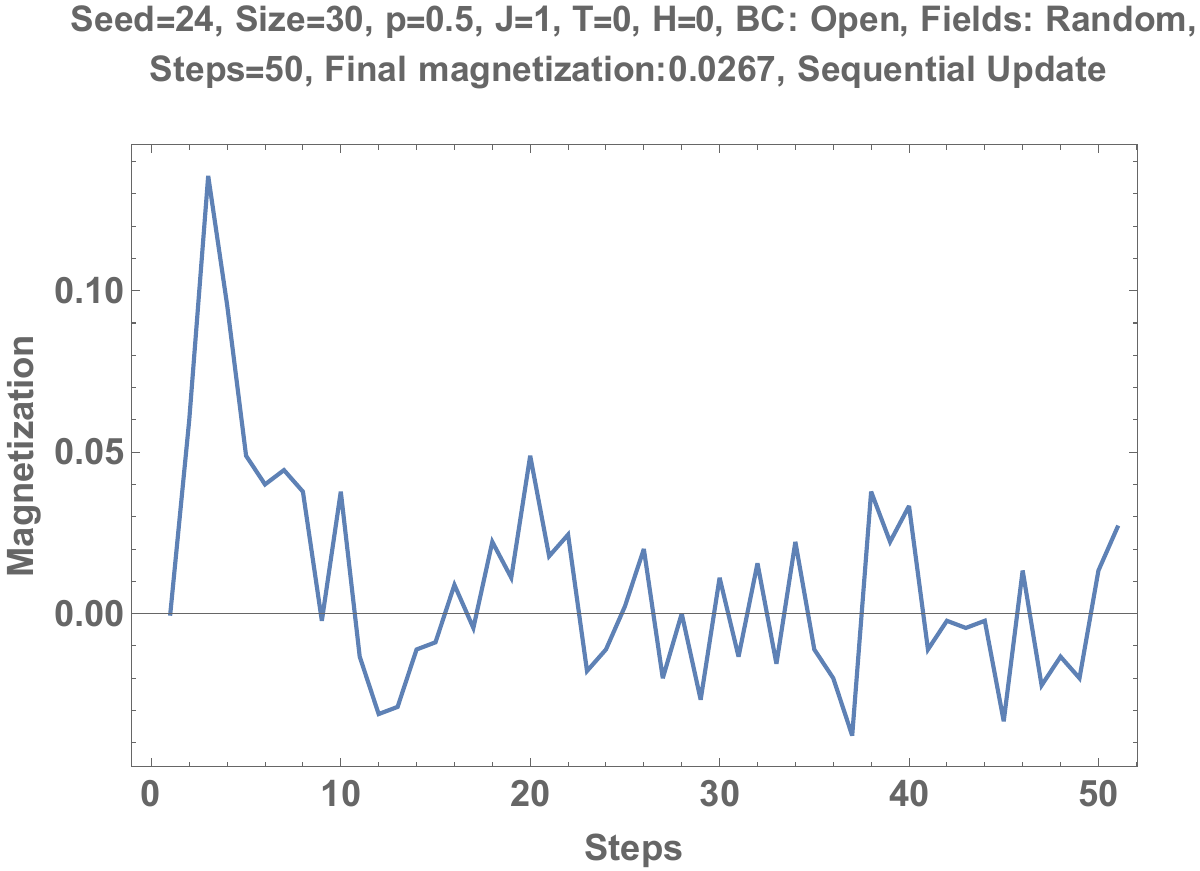}}
\\
\subfigure[]{\includegraphics[width=0.32\textwidth]{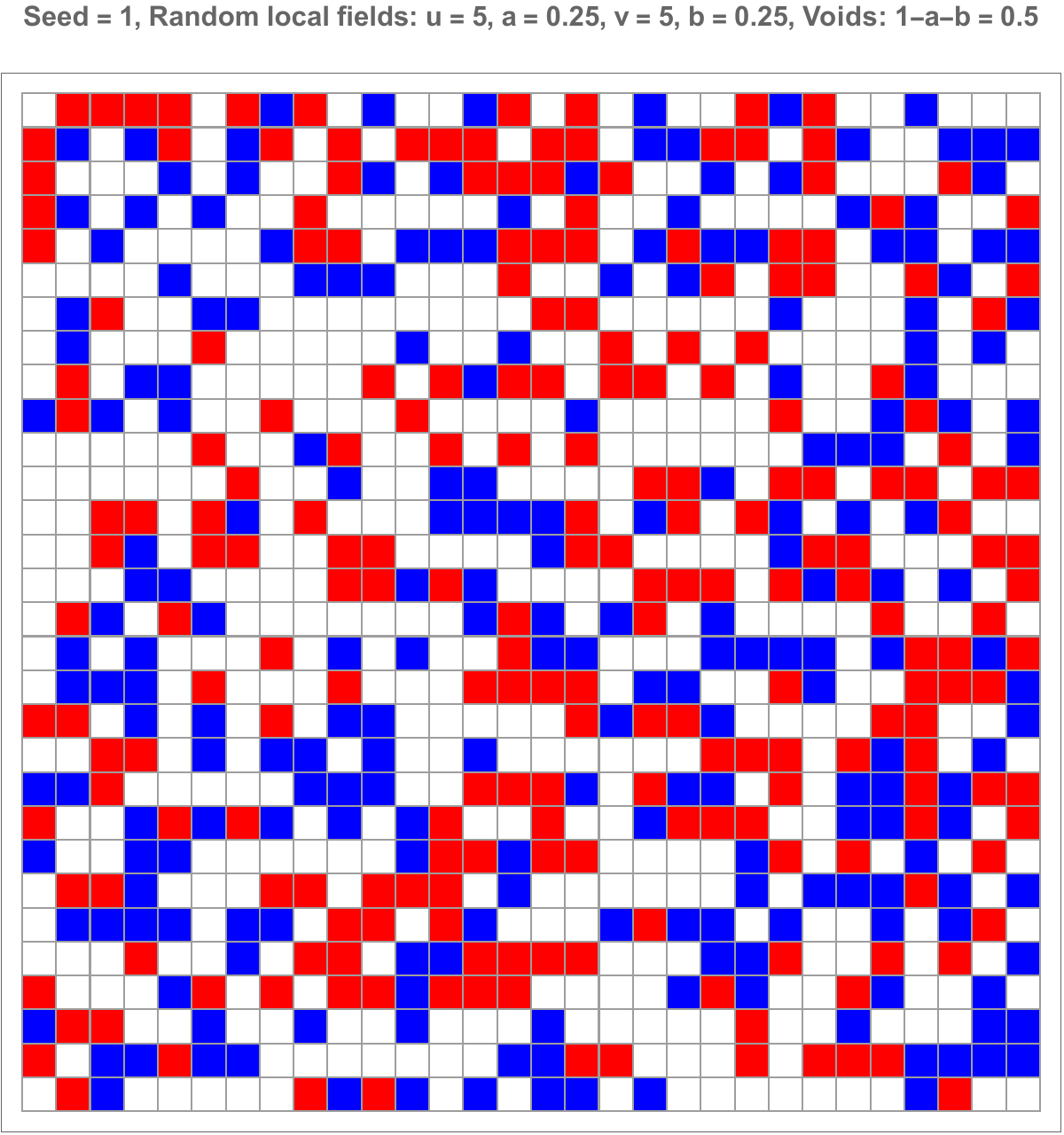}}
\subfigure[]{\includegraphics[width=0.32\textwidth]{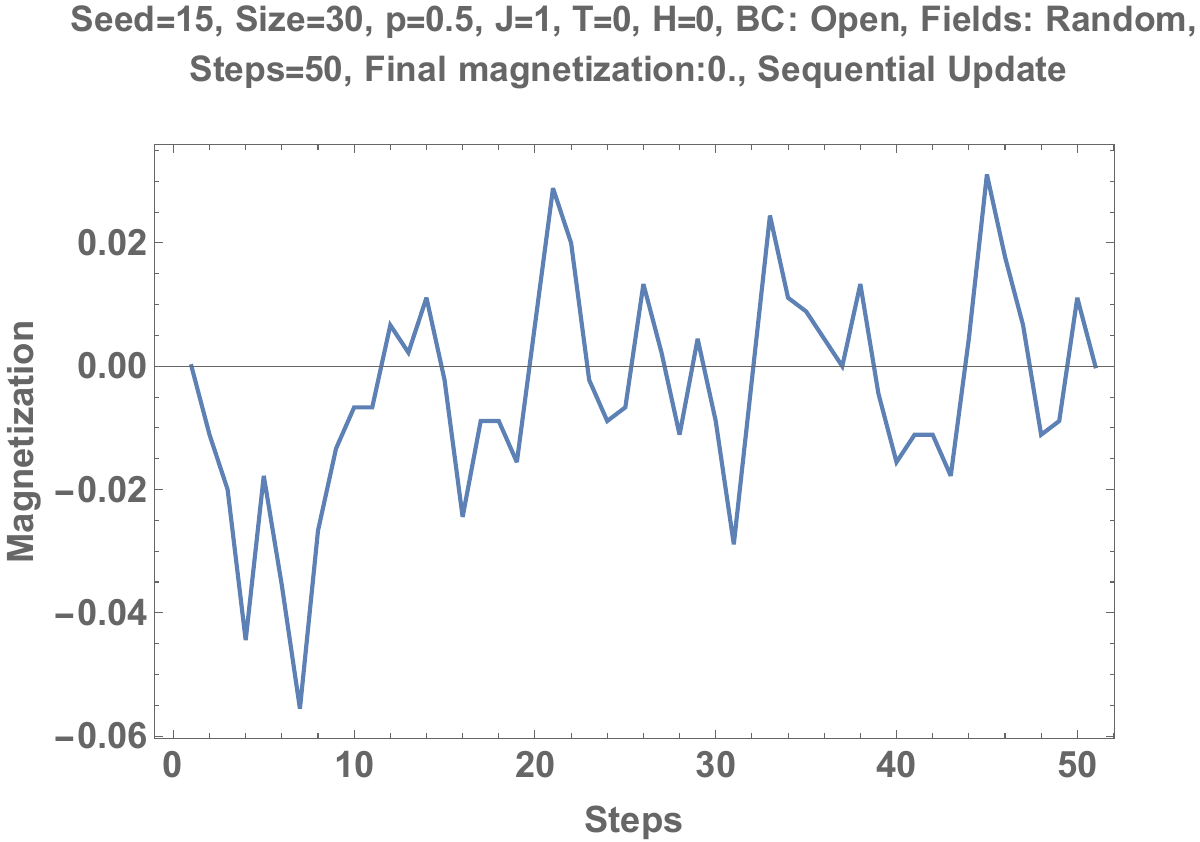}}
\subfigure[]{\includegraphics[width=0.32\textwidth]{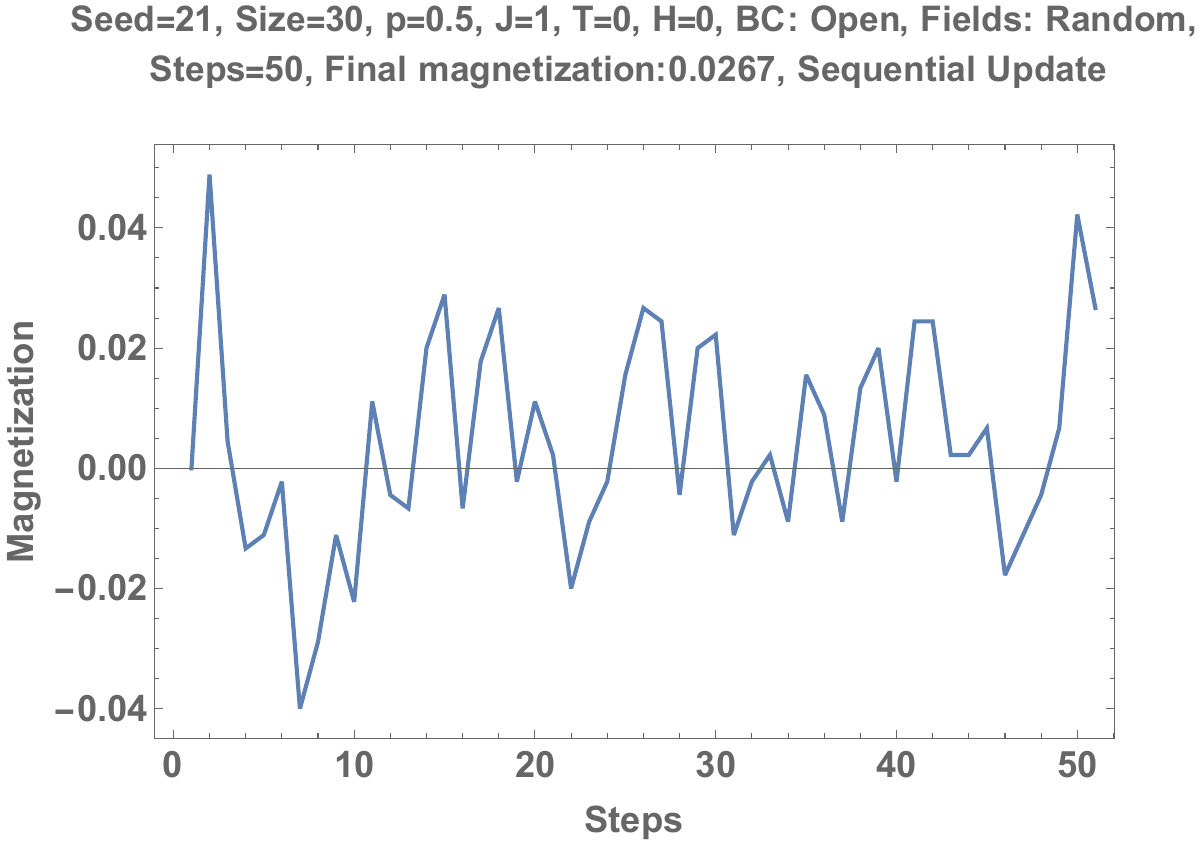}}
\caption{Simulations with proportions $p=0.15, 0.25$ of red and blue local fields shown in subparts (a, d). Subparts (b, c, e, f) exhibit the related dynamics for two different initial distributions of initial choices with $p=0.50$. Yet, the outcome is still random with strong fluctuations with respect to the winning majority as seen in the subparts.}
\label{cc1}
\end{figure}

\subsubsection{Figure (\ref{cc1})}

Above results have revealed a fragility in the dynamics of spontaneous symmetry breaking associated with just two local fields located at some tipping sites. The effect was found to be of a hazardous nature due to the difficulty in identifying the location of those tipping sites. 

To overpass this flaw, I run two simulations with large and equal proportions of red and blue local fields, respectively 0.15 and 0.25 \% as shown in subparts (a, d) of Figure (\ref{cc1}). Subparts (b, c, e, f) exhibit the related dynamics for two different initial distributions of initial choices with $p=0.50$. Yet, the outcome is still random with strong fluctuations with respect to the winning majority as seen in the subparts.

\subsection{Asymmetric initial configuration and local fields}

\begin{figure}
\centering
\subfigure[]{\includegraphics[width=0.32\textwidth]{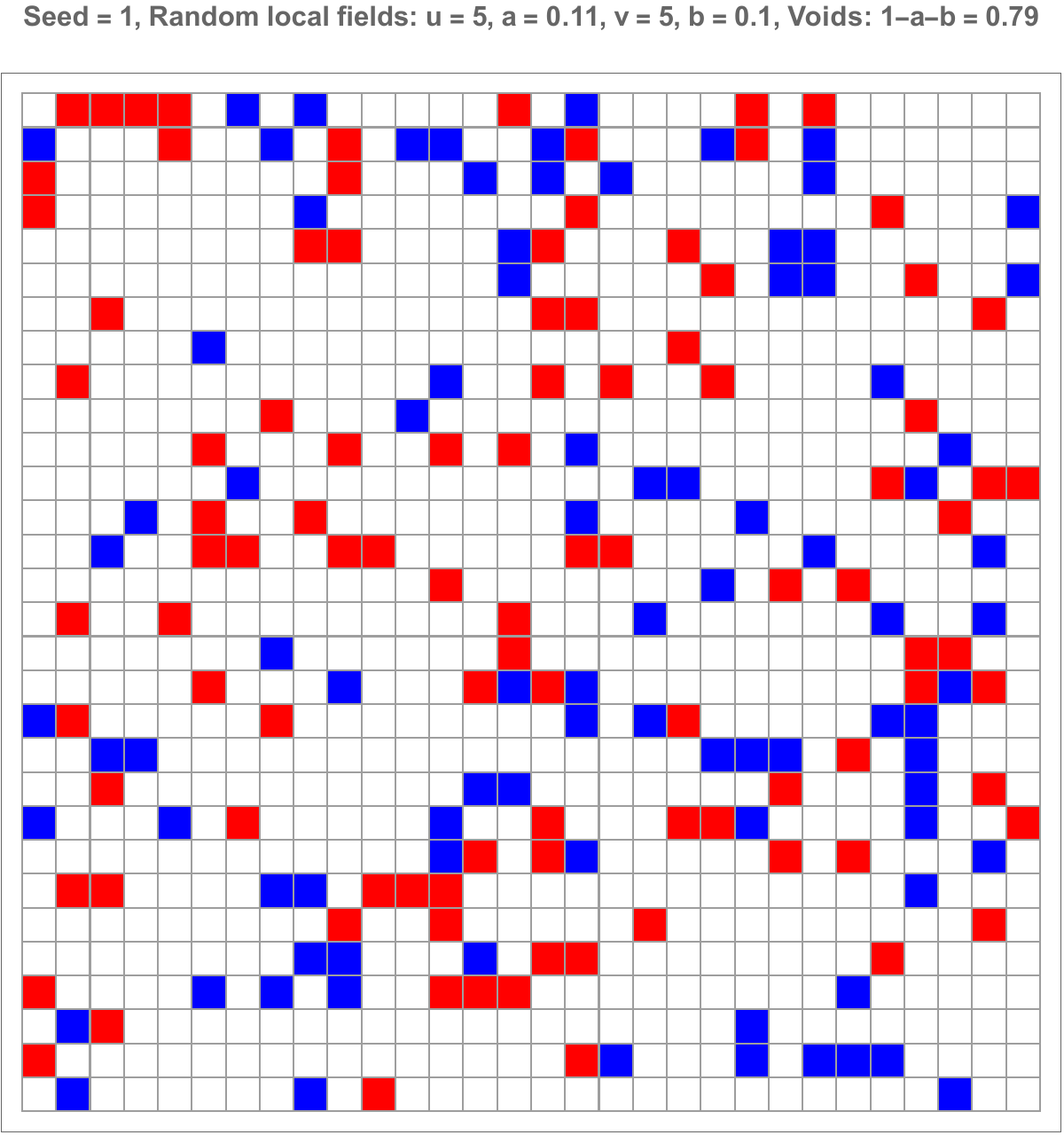}}
\subfigure[]{\includegraphics[width=0.32\textwidth]{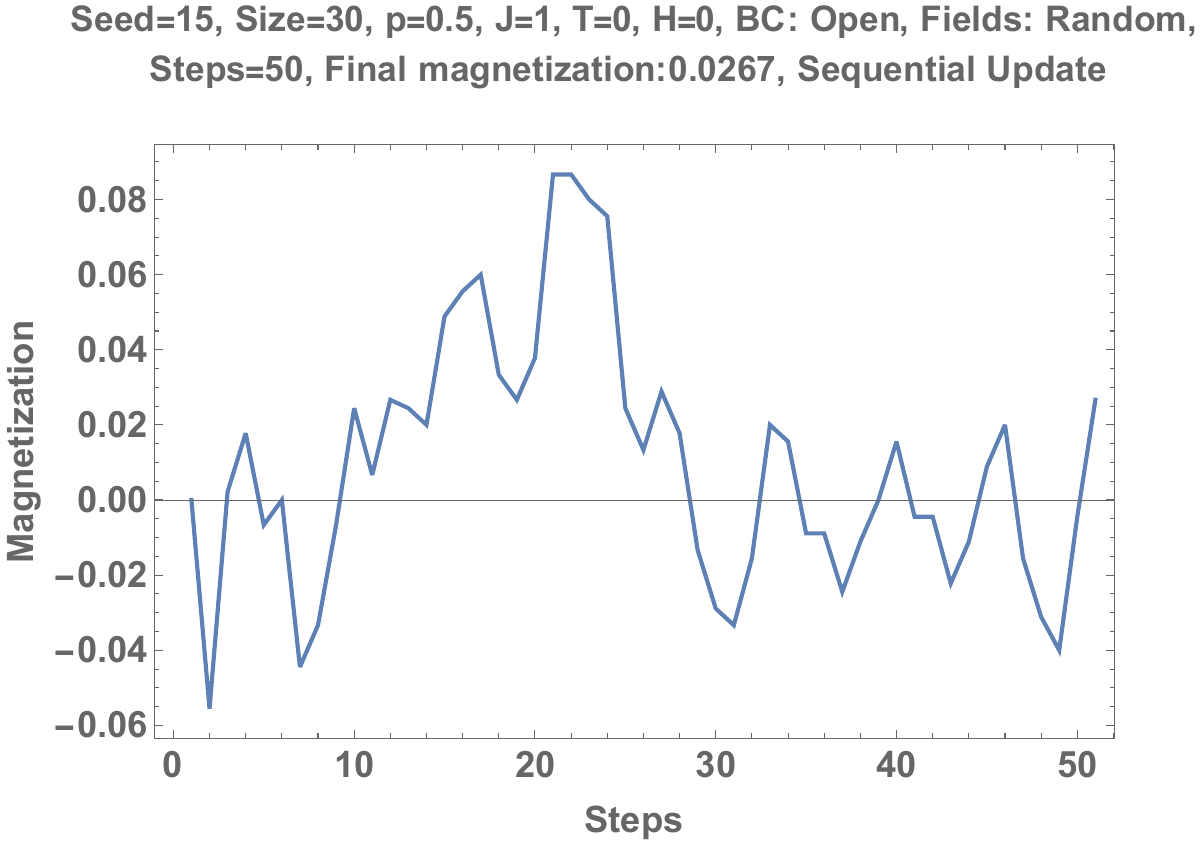}}
\subfigure[]{\includegraphics[width=0.32\textwidth]{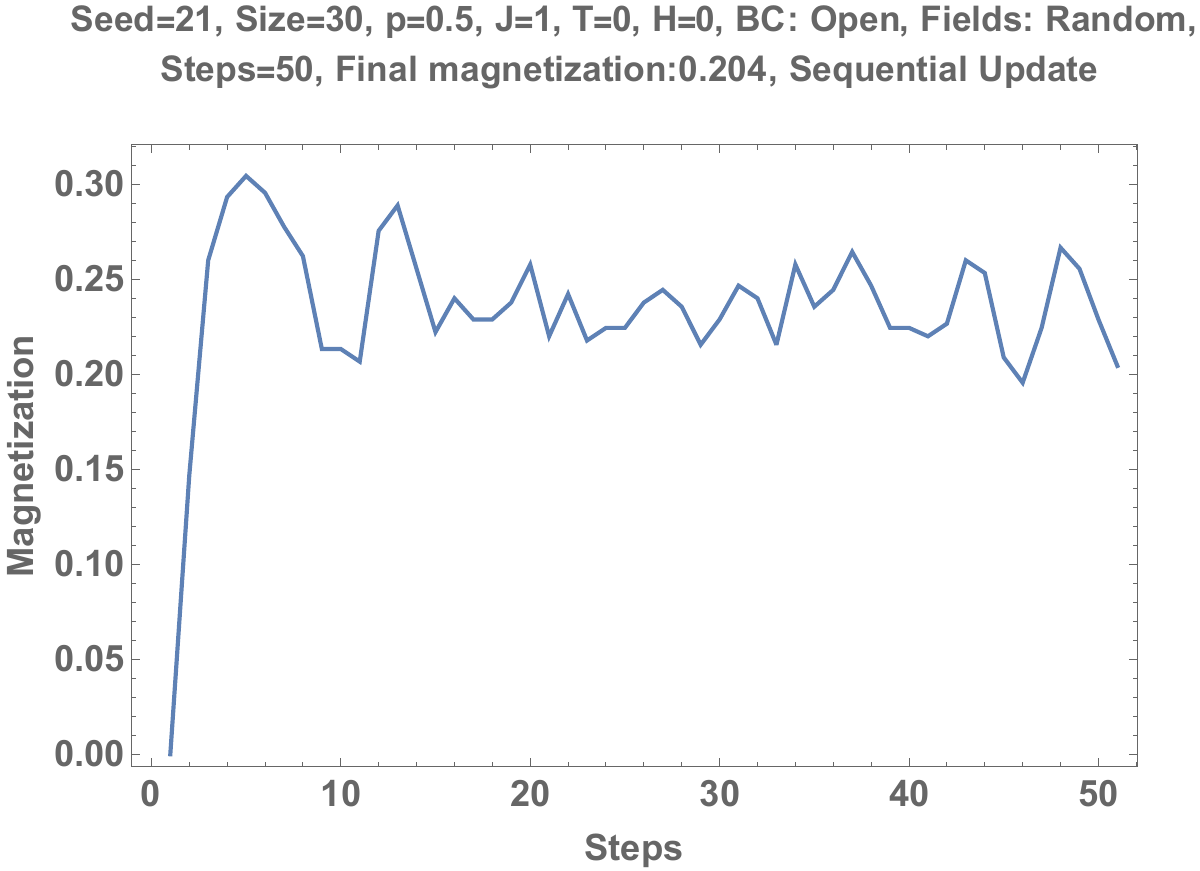}}
\\
\subfigure[]{\includegraphics[width=0.32\textwidth]{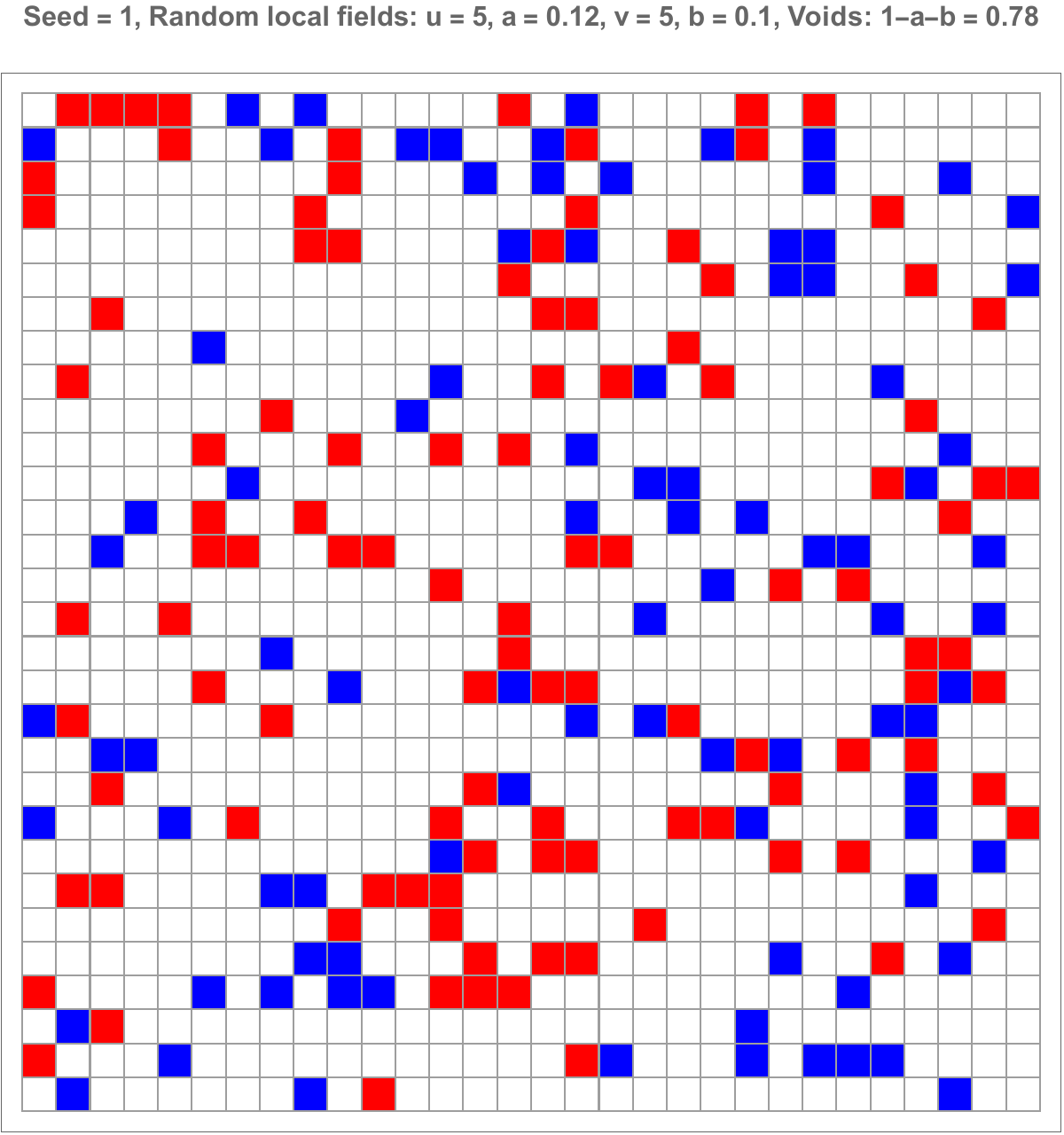}}
\subfigure[]{\includegraphics[width=0.32\textwidth]{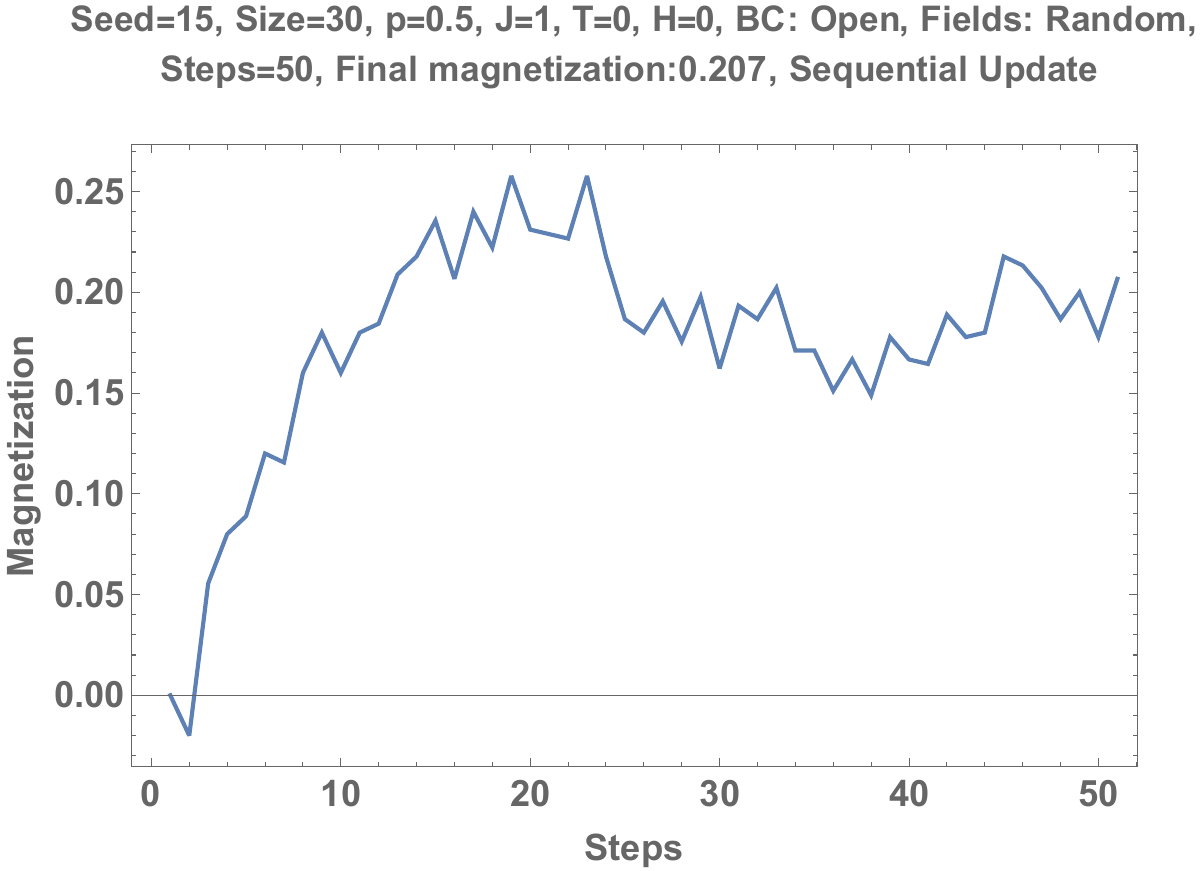}}
\subfigure[]{\includegraphics[width=0.32\textwidth]{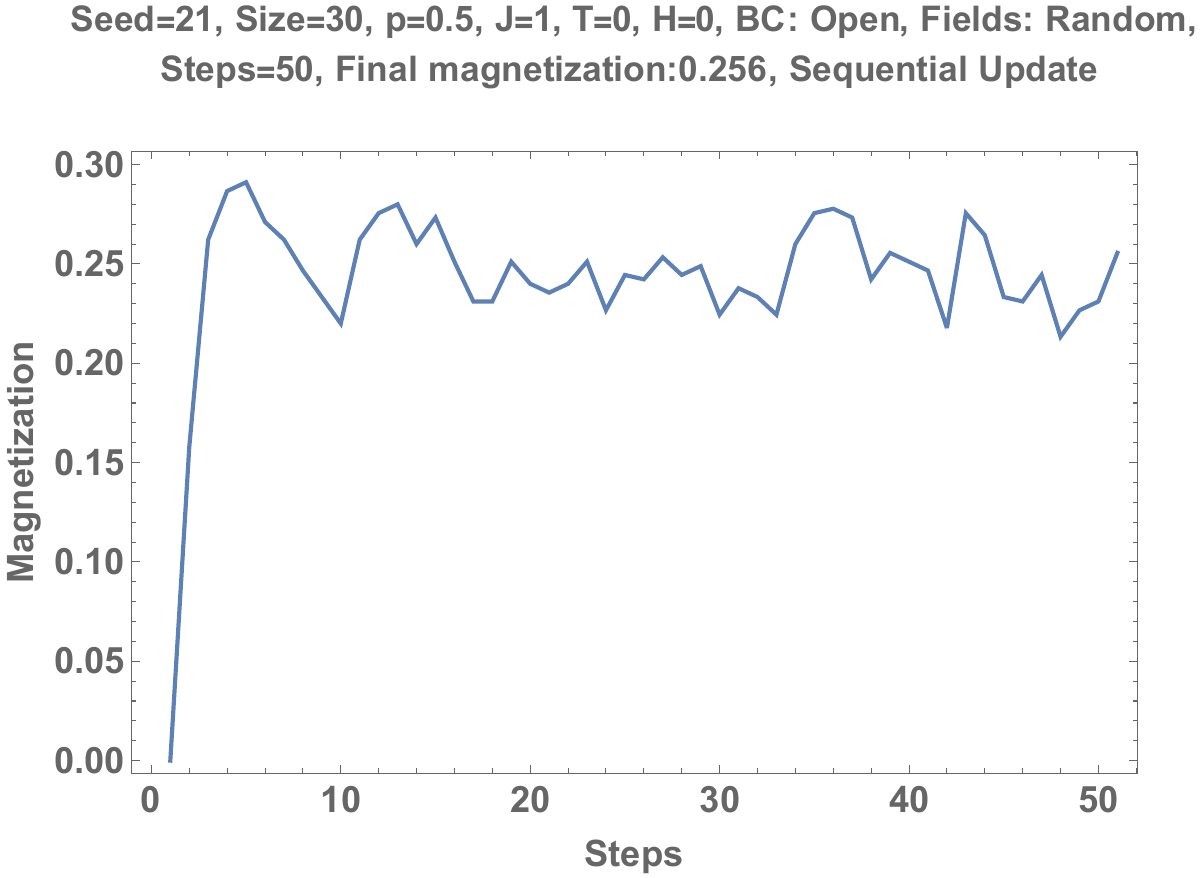}}
\caption{Subpart (a) show a distribution $a=0.11$ and $b=0.10$ of red and blue fields with two different distributions of initial choices. The related dynamics are shown in subparts (b, c). Subpart (d) has $a=0.12$ and $b=0.10$ with the associated dynamics shown in subparts (e, f). While an extra 1\% of red fields ensure a red majority for one initial distribution (subpart (c)), it does not in the other (subpart (b)). In contrast, subparts (d, e, f) show that an extra 2\% ensures a red winning in both cases.}
\label{cc2}
\end{figure}

\subsubsection{Figure (\ref{cc2})}

I now investigate asymmetric cases, first in the proportions of red and blue local fields with still an equal proportions of red and blue initial choices. 

Subpart (a) show a distribution $a=0.11$ and $b=0.10$ of red and blue fields with two different distributions of initial choices. The related dynamics are shown in subparts (b, c). Subpart (d) has $a=0.12$ and $b=0.10$ with the dynamics shown in subparts (e, f). 

While an extra 1\% of red fields ensure a red majority for one initial distribution (subpart (c)), it does not in the other (subpart (b)).

In contrast, subparts (d, e, f) show that an extra 2\% ensures a red winning in both cases. 

\subsubsection{Figure (\ref{dd})}

All above cases start from a balanced initial distribution ($p=0.50$). Some differ in terms of red and blue site locations. Each specific distribution is identified by the Seed parameter to ensure tracking. The goal was to investigate the distortion of the expected spontaneous symmetry breaking by applying some local breaking fields. 

In the following cases I start with a totally broken initial distribution where all sites are blue ($p=0$) to identity the possibility to revere the collective blue choice. To this end, I am applying only red local fields.

Figure (\ref{dd}) shows three cases with respectively $a=0.15, 0.09, 0.08$ for the density of red local fields. The related locations are seen in subparts (b, e, h). Locations are random in first three and selected in last one. Associated dynamics are exhibited in subparts (a, d, g). Subpart (c) shows the grid after six MC steps before it turns all red after about 15 MC steps. Subparts (f) with $a=0.09$ shows the same final grid as for $a=0.015$. However, with one percent less in the proportion of local red field ($a=0.08$) the reversal from complete blue to complete red is lost as illustrated by subpart (i).

\begin{figure}
\centering
\vspace{-3cm}
\subfigure[]{\includegraphics[width=0.32\textwidth]{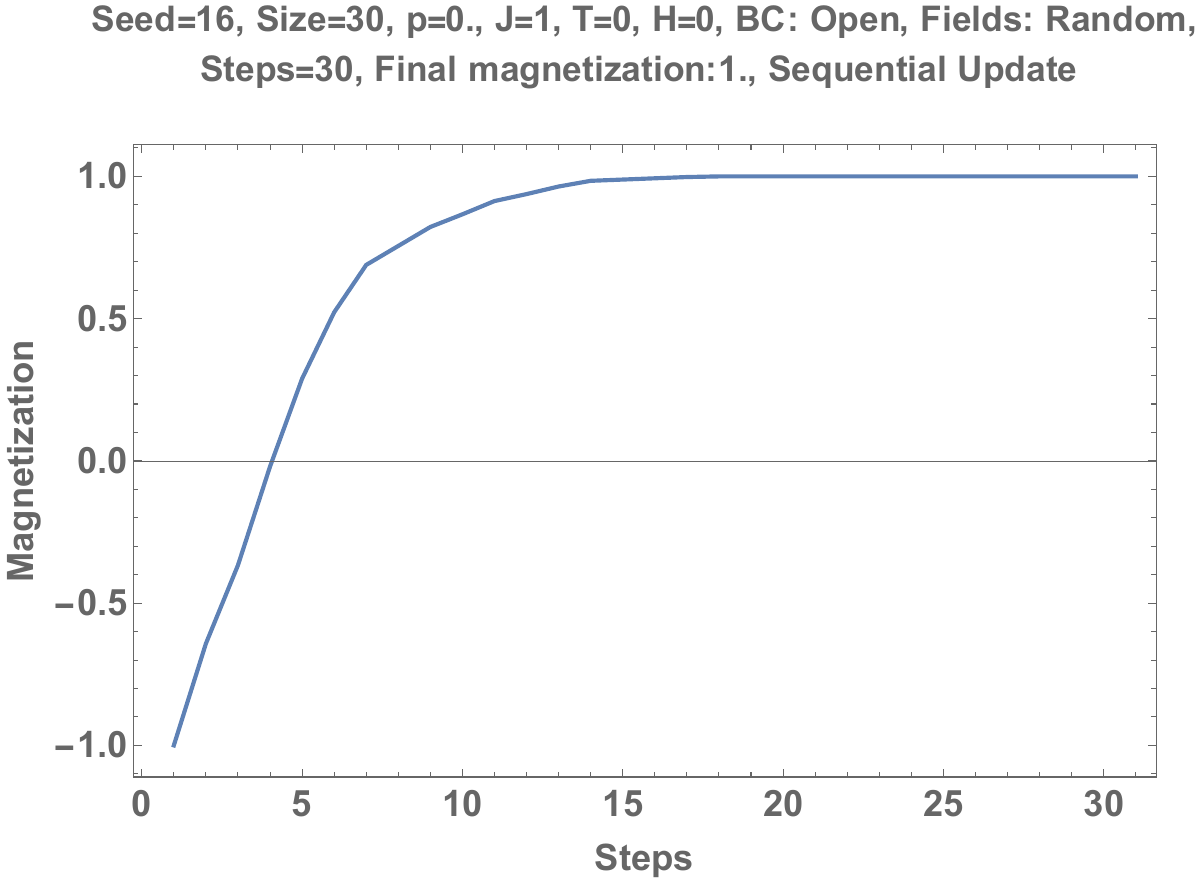}}
\subfigure[]{\includegraphics[width=0.32\textwidth]{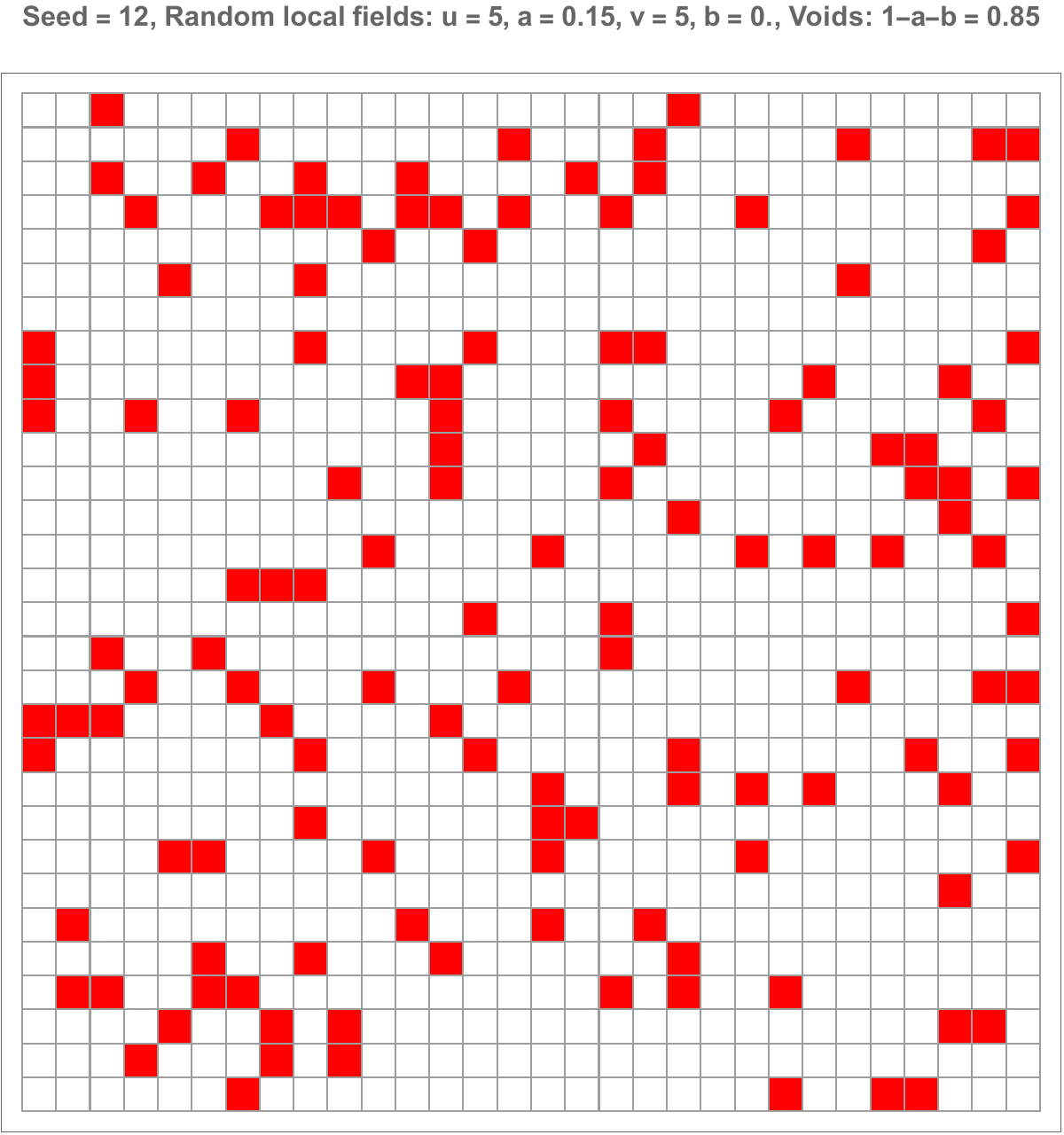}}
\subfigure[]{\includegraphics[width=0.32\textwidth]{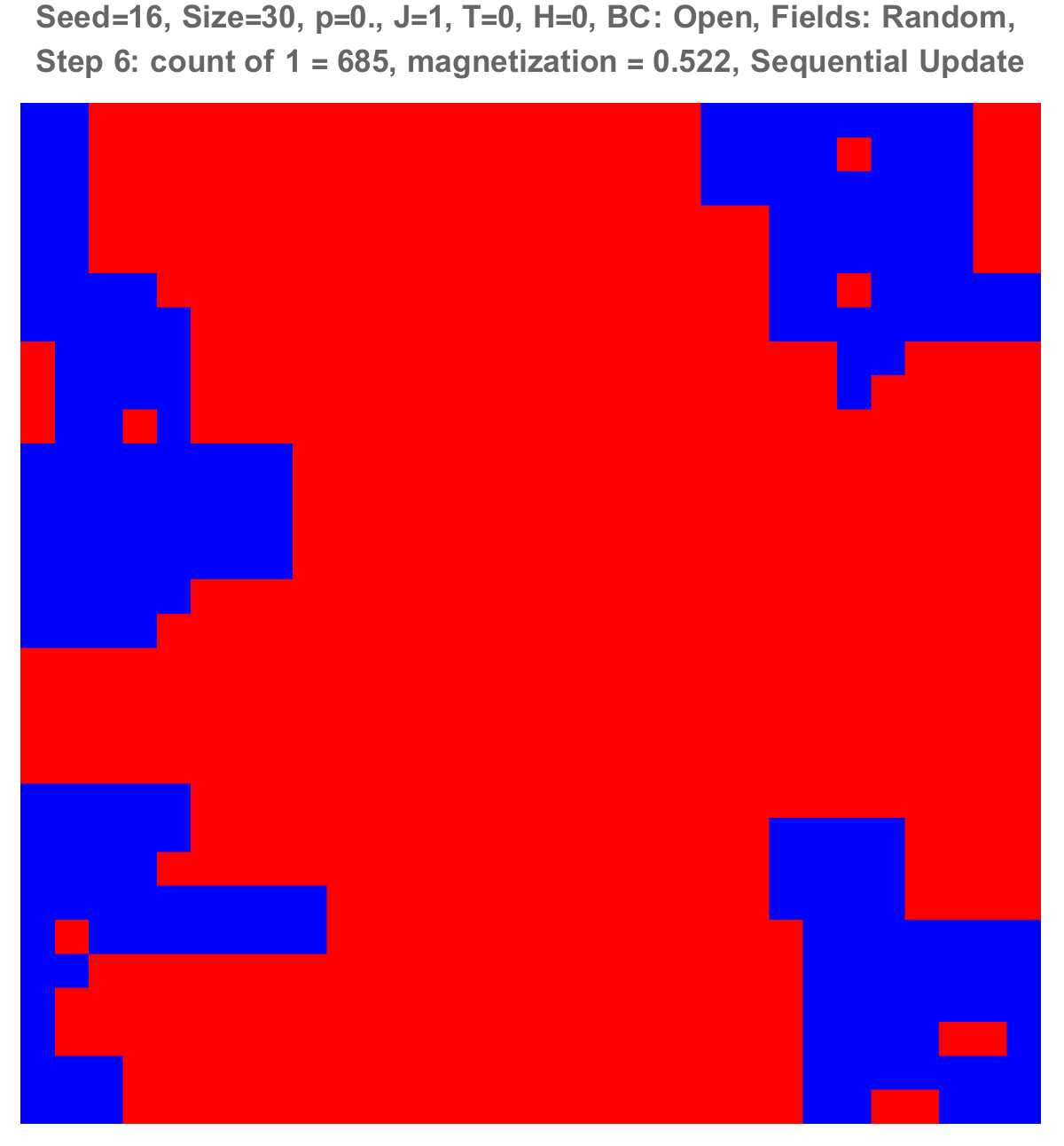}}
\\
\subfigure[]{\includegraphics[width=0.32\textwidth]{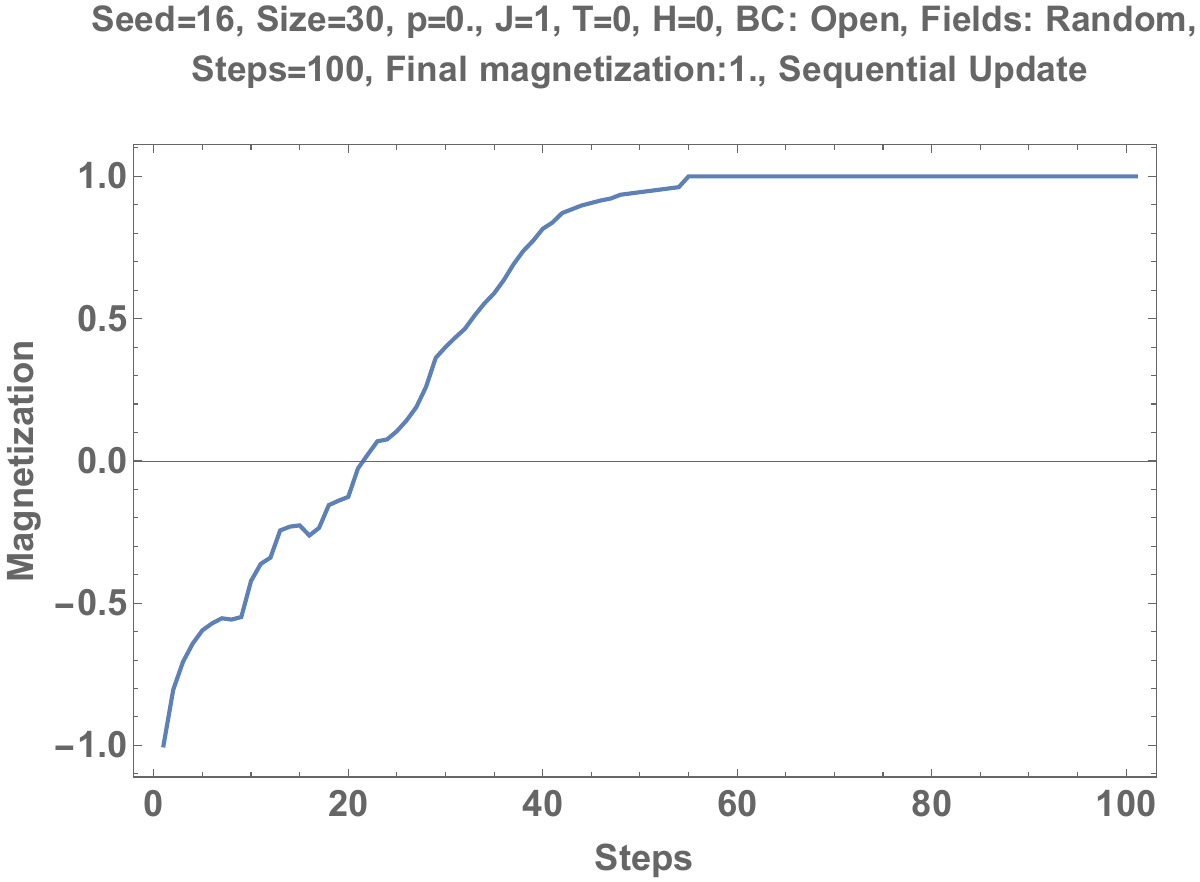}}
\subfigure[]{\includegraphics[width=0.32\textwidth]{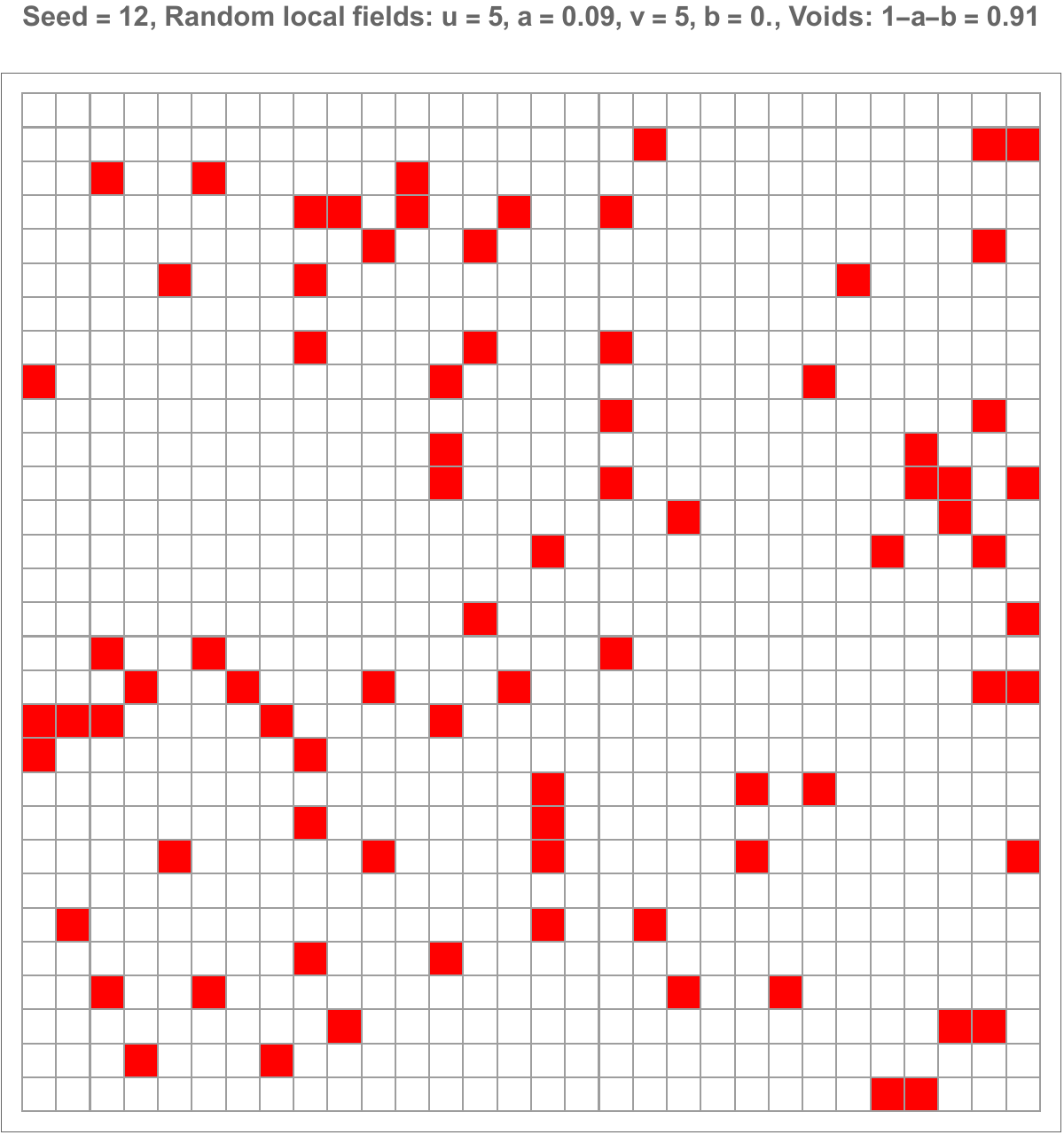}}
\subfigure[]{\includegraphics[width=0.32\textwidth]{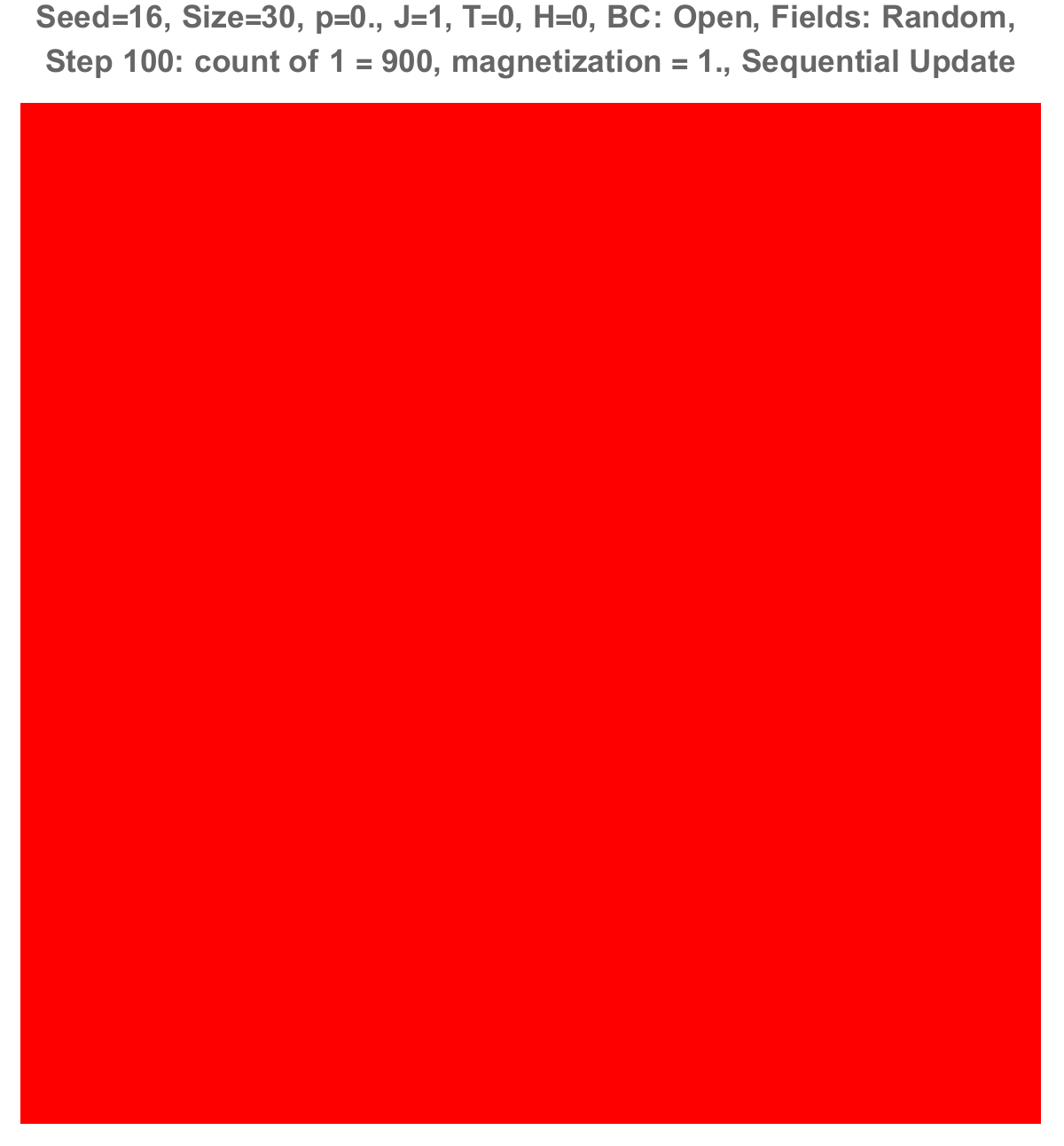}}
\\
\subfigure[]{\includegraphics[width=0.32\textwidth]{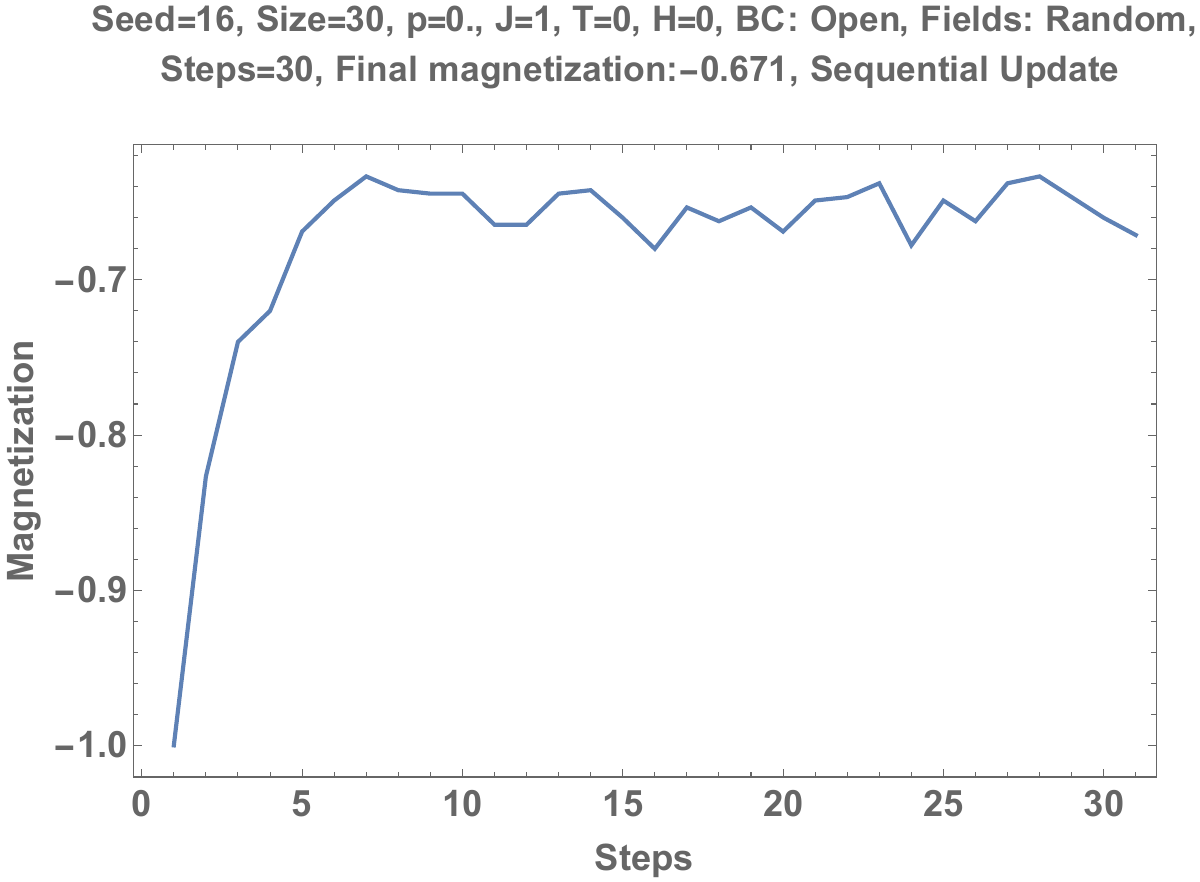}}
\subfigure[]{\includegraphics[width=0.32\textwidth]{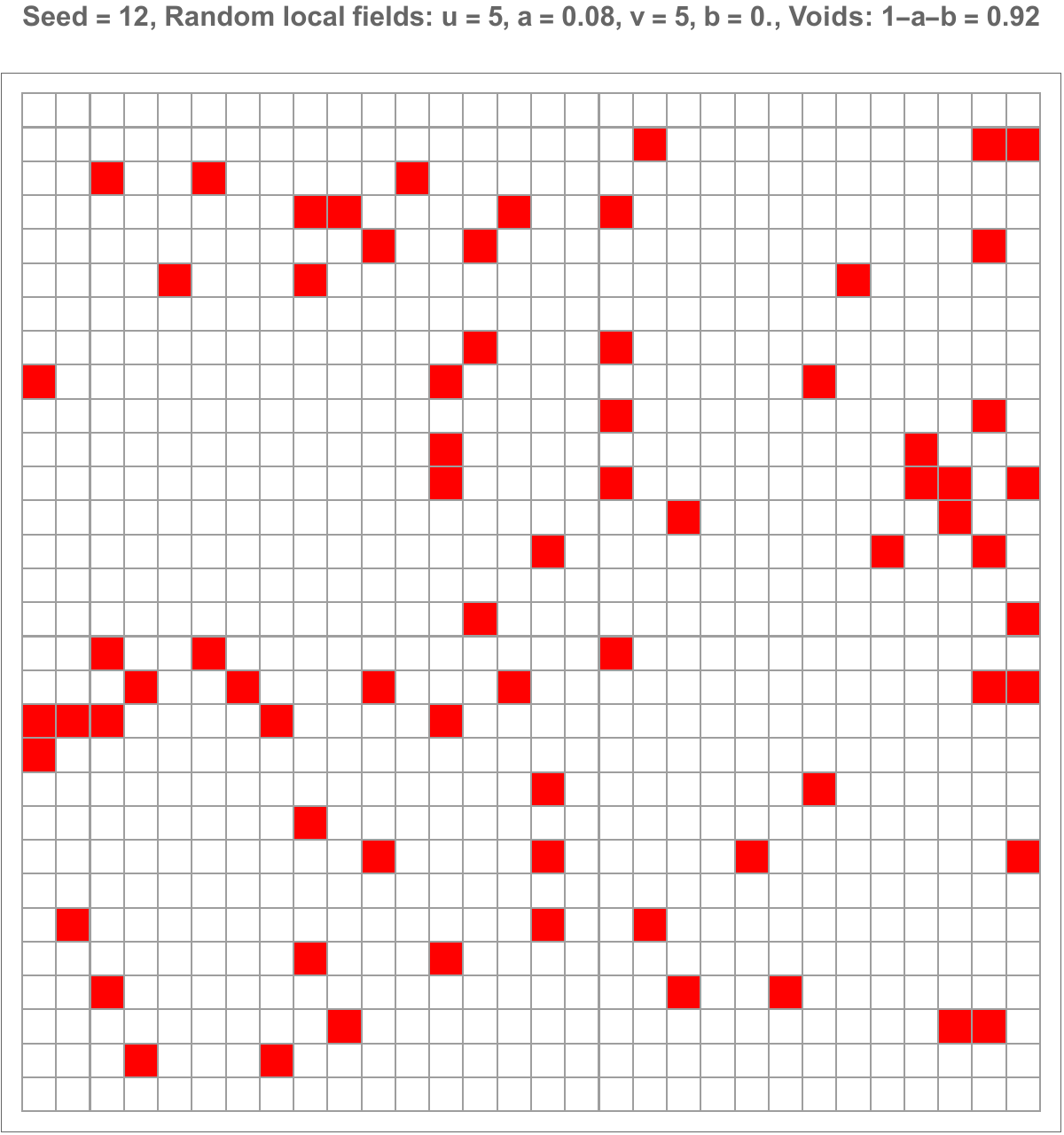}}
\subfigure[]{\includegraphics[width=0.32\textwidth]{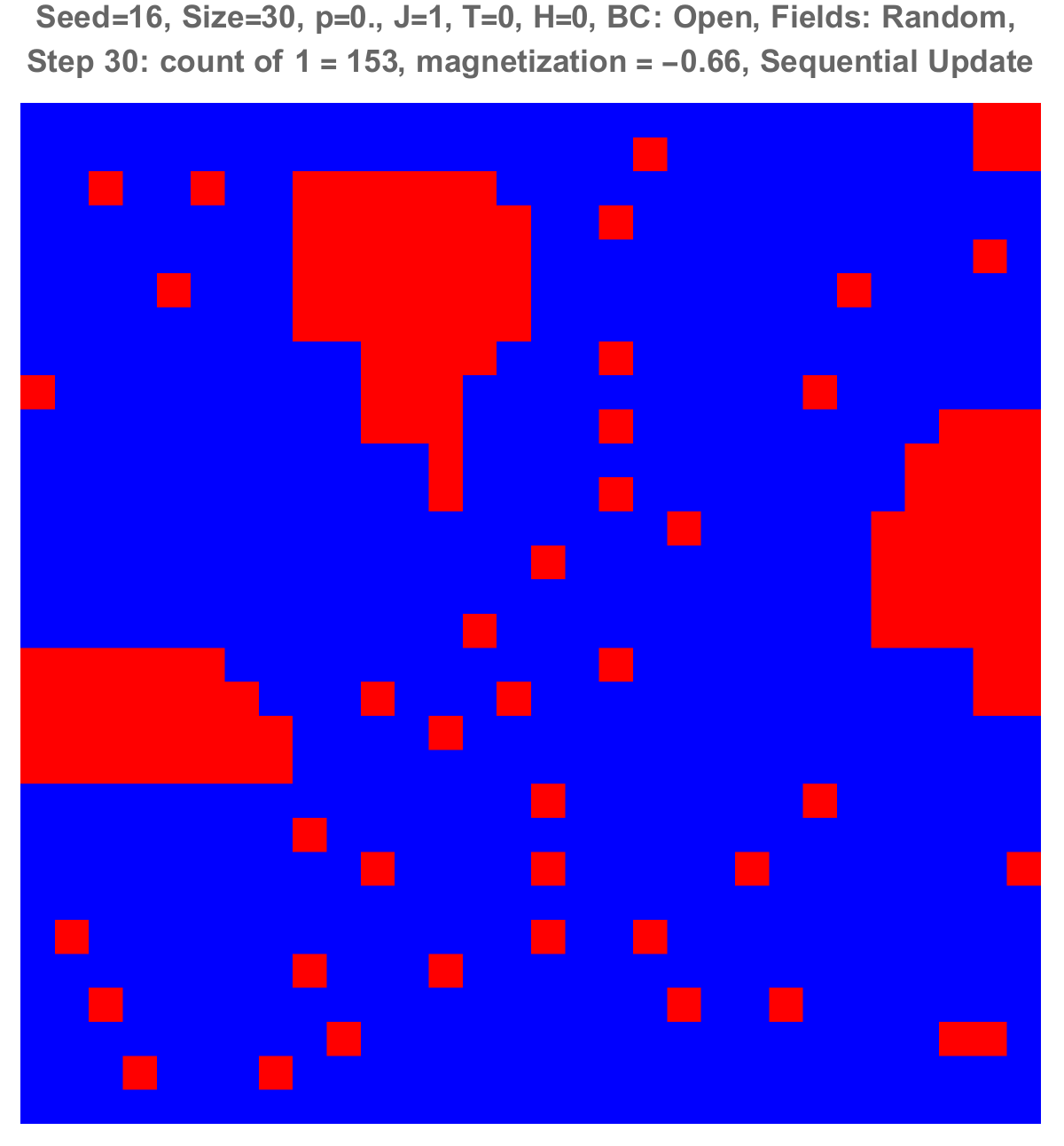}}
\end{figure}

\newpage 

\noindent\captionof{figure}{Three cases with respectively $a=0.15, 0.09, 0.08$ for the density of red local fields. The related locations are seen in subparts (b, e, h). Locations are random in first three and selected in last one. Associated dynamics are exhibited in subparts (a, d, g). Subpart (c) shows the grid after six MC steps before it turns all red after about 15 MC steps. Subparts (f) with $a=0.09$ shows the same final grid as for $a=0.015$. One percent less in the proportion of local red field ($a=0.08$) loses the reversal from complete blue to complete red as illustrated by subpart (i).}
\label{dd}

\subsubsection{Figure (\ref{ee})}

Figure (\ref{ee}) shows red local fields applied at selected locations. Three cases are exhibited with respectively 18 (subpart (b), $a=0.02$), 30 (subpart (e), $a=0.033$), 31 (subpart (h), $a=0.034$) red local fields. Associated dynamics are shown in subparts (a, d, g), Final outcomes are shown in subparts (c, f, i) with respectively 136, 116, 530 red choices among the 900 ones.
Again, one additional single local field located at some tipping site is found to have a substantial impact on the final outcome with the making of a large majority of 530 red choices against a minority of 116 previously.

\begin{figure}
\centering
\vspace{-3cm}
\subfigure[]{\includegraphics[width=0.32\textwidth]{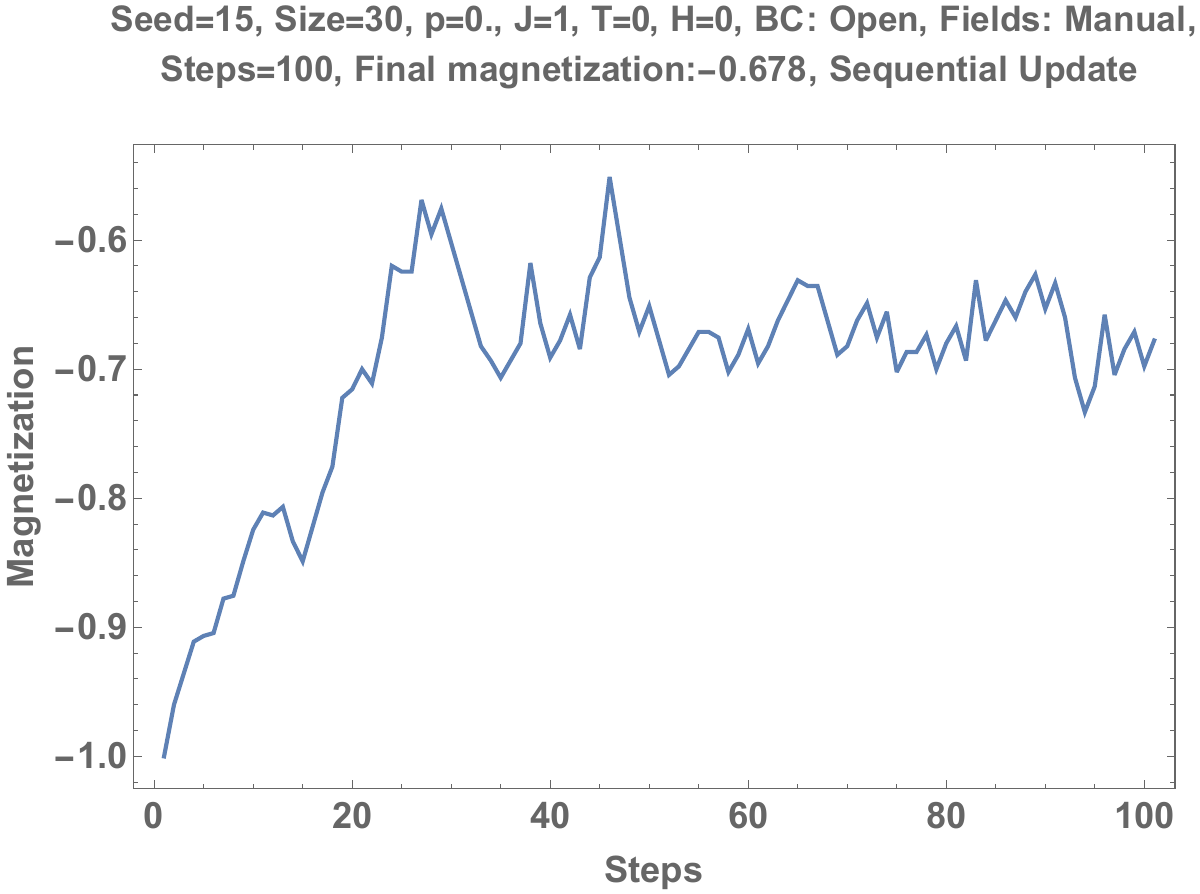}}
\subfigure[]{\includegraphics[width=0.32\textwidth]{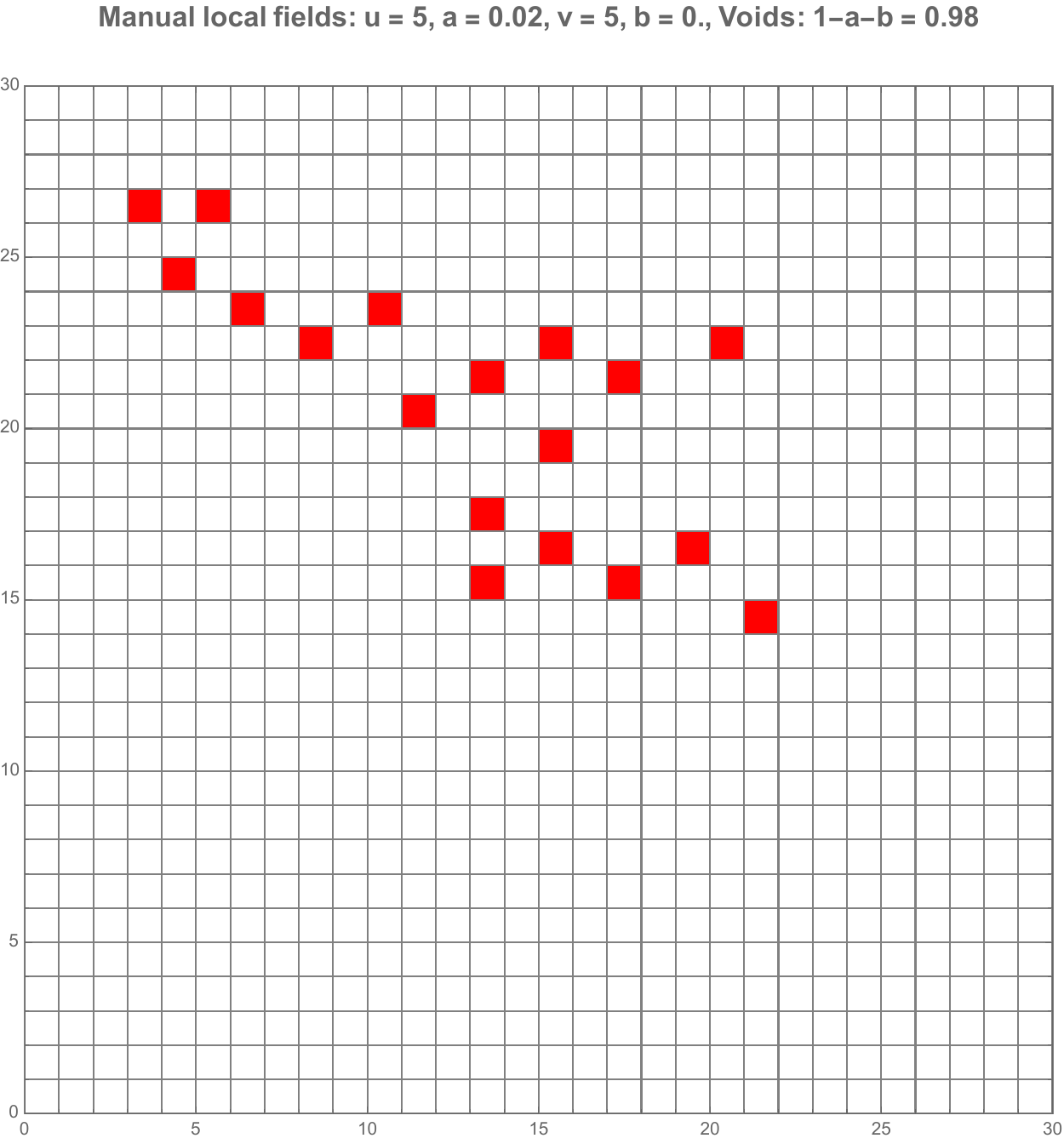}}
\subfigure[]{\includegraphics[width=0.32\textwidth]{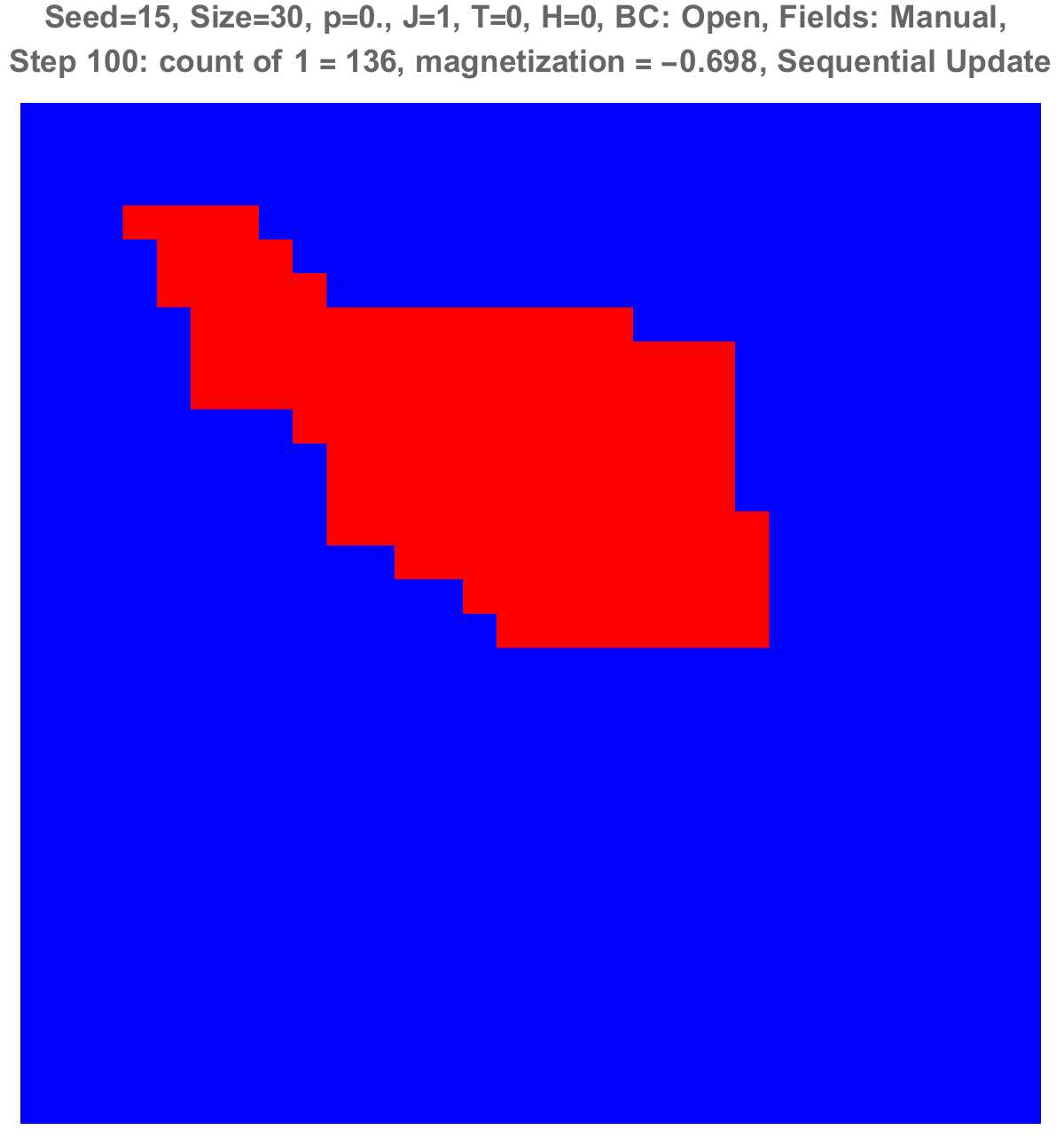}}
\\
\subfigure[]{\includegraphics[width=0.32\textwidth]{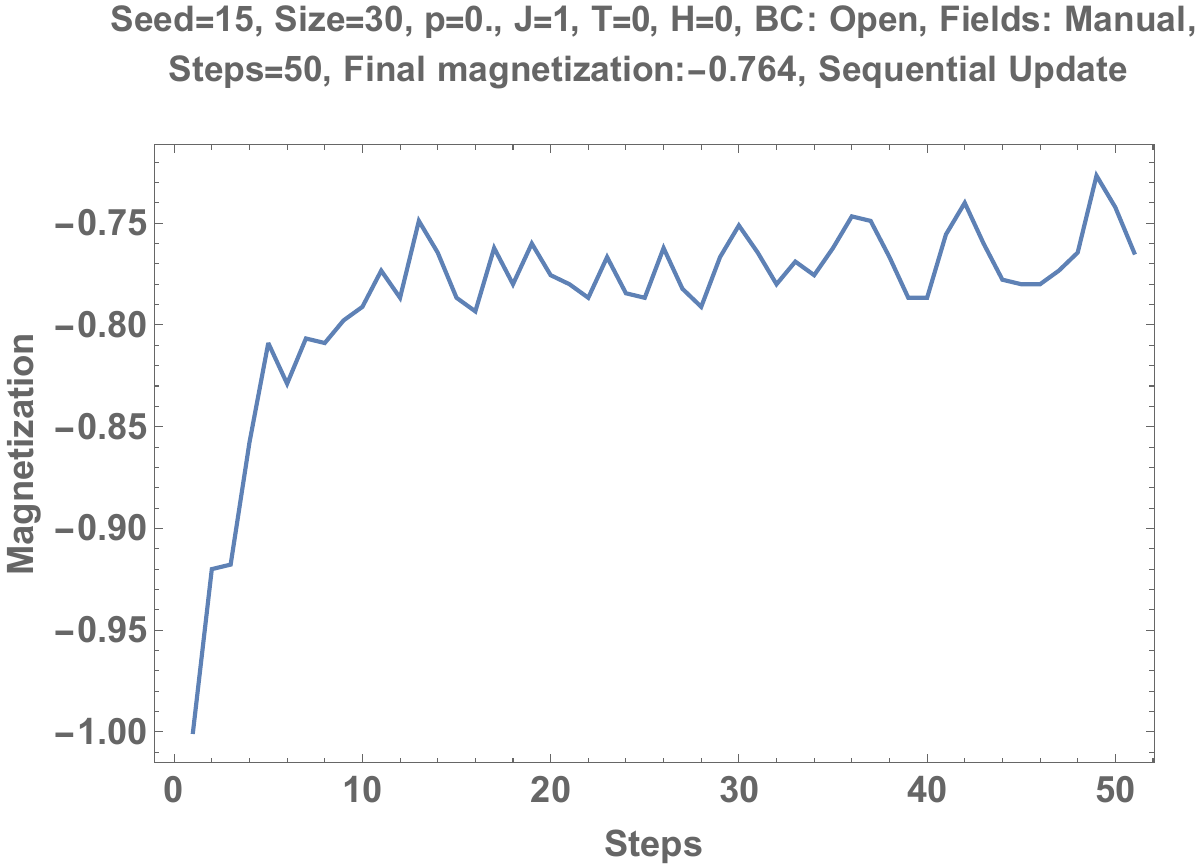}}
\subfigure[]{\includegraphics[width=0.32\textwidth]{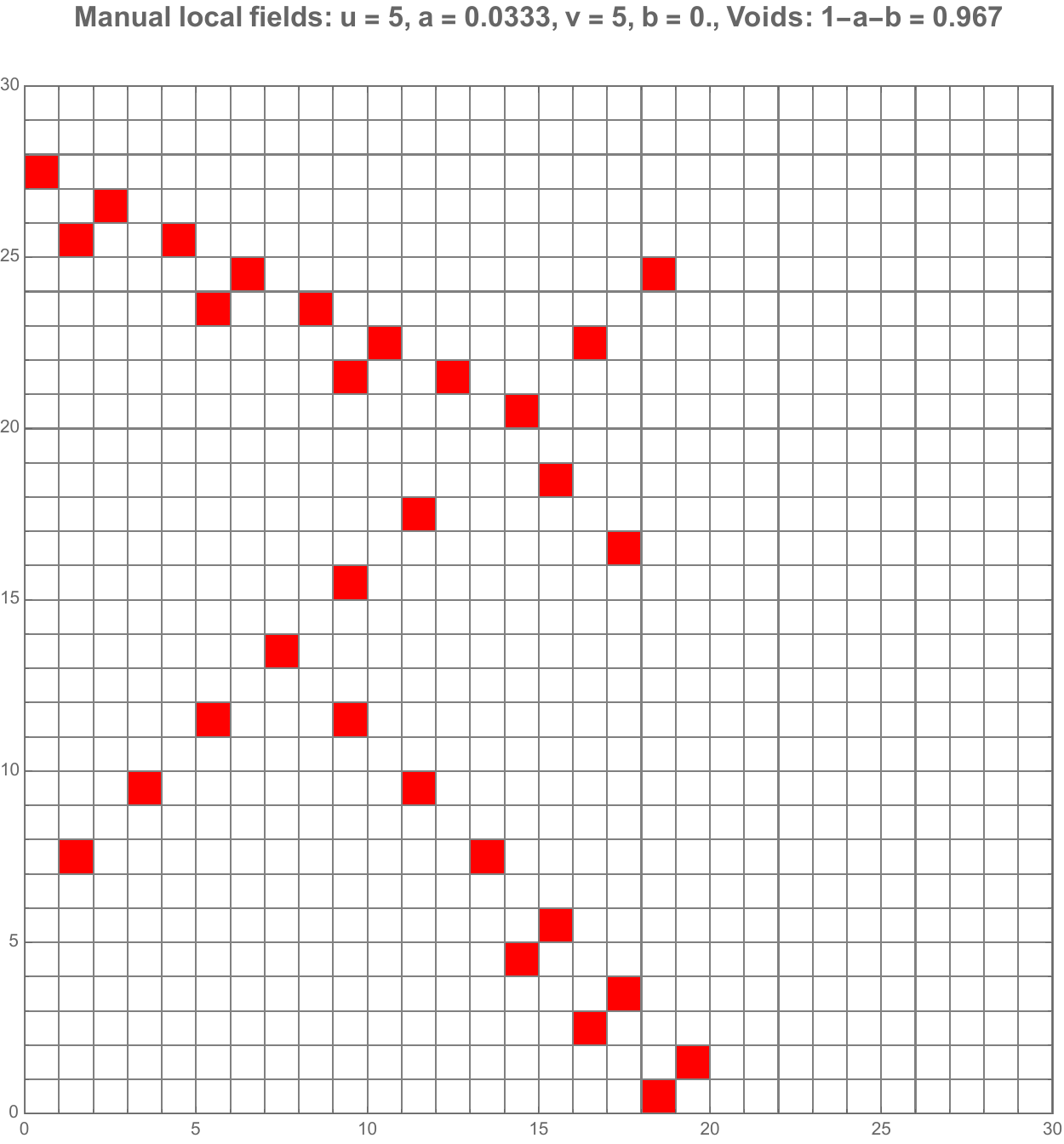}}
\subfigure[]{\includegraphics[width=0.32\textwidth]{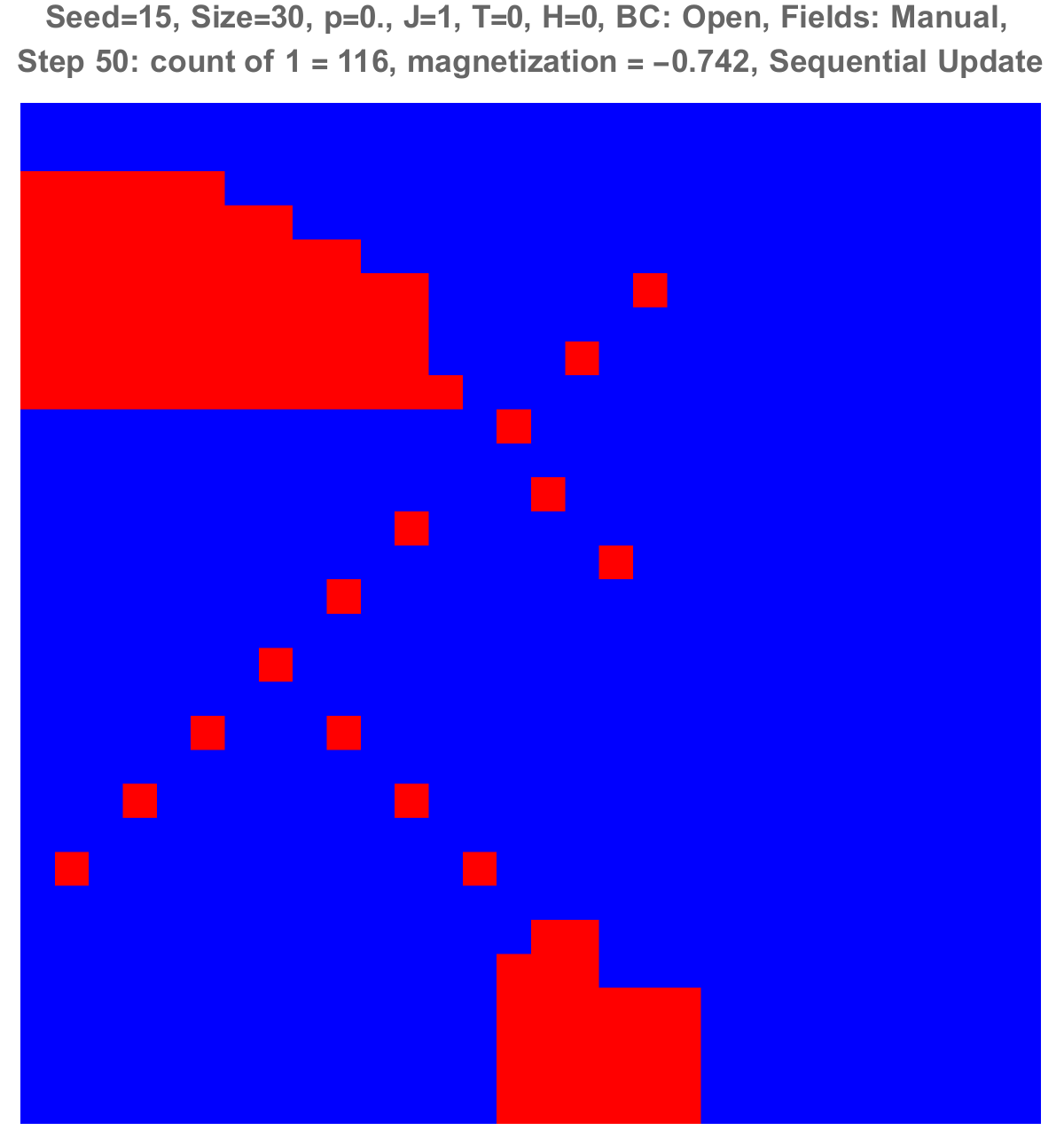}}
\\
\subfigure[]{\includegraphics[width=0.32\textwidth]{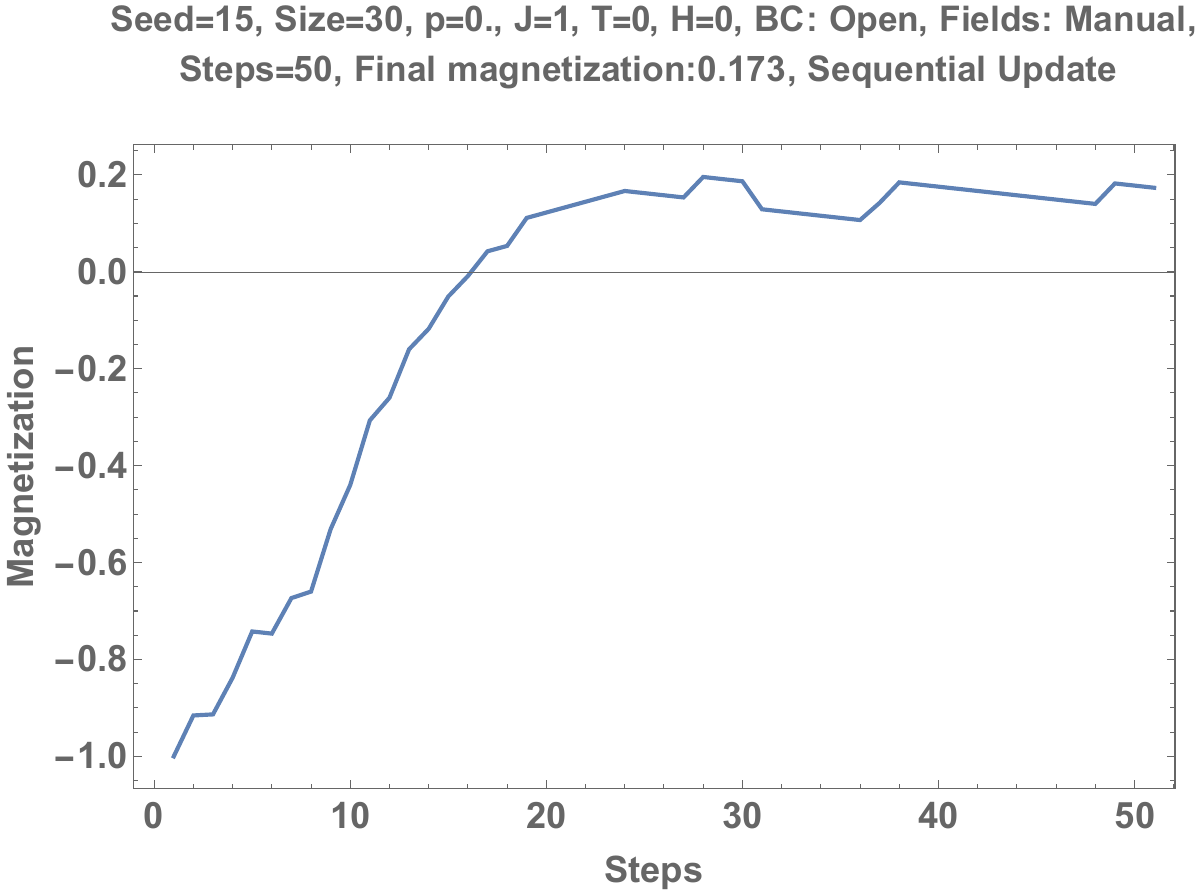}}
\subfigure[]{\includegraphics[width=0.32\textwidth]{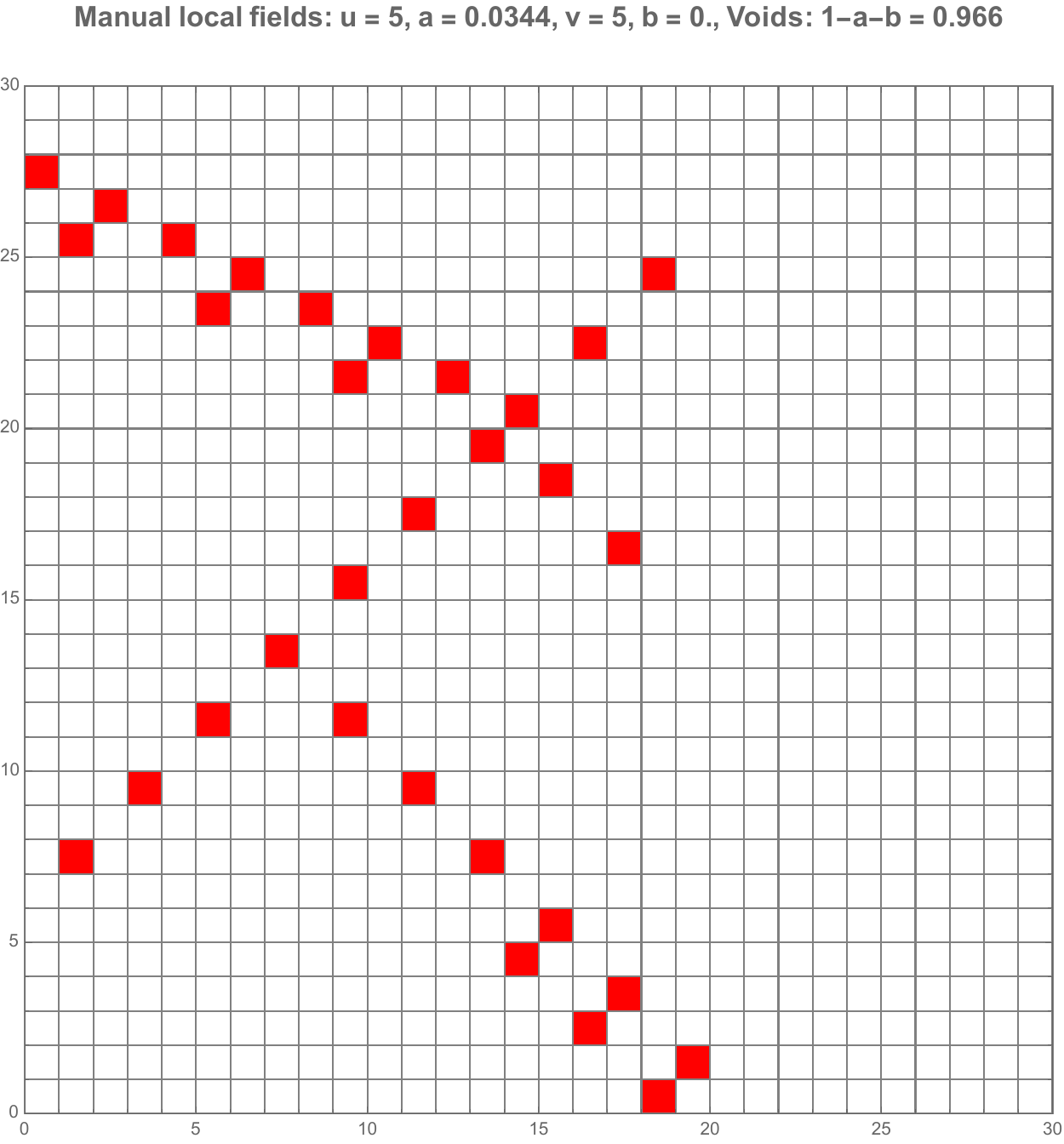}}
\subfigure[]{\includegraphics[width=0.32\textwidth]{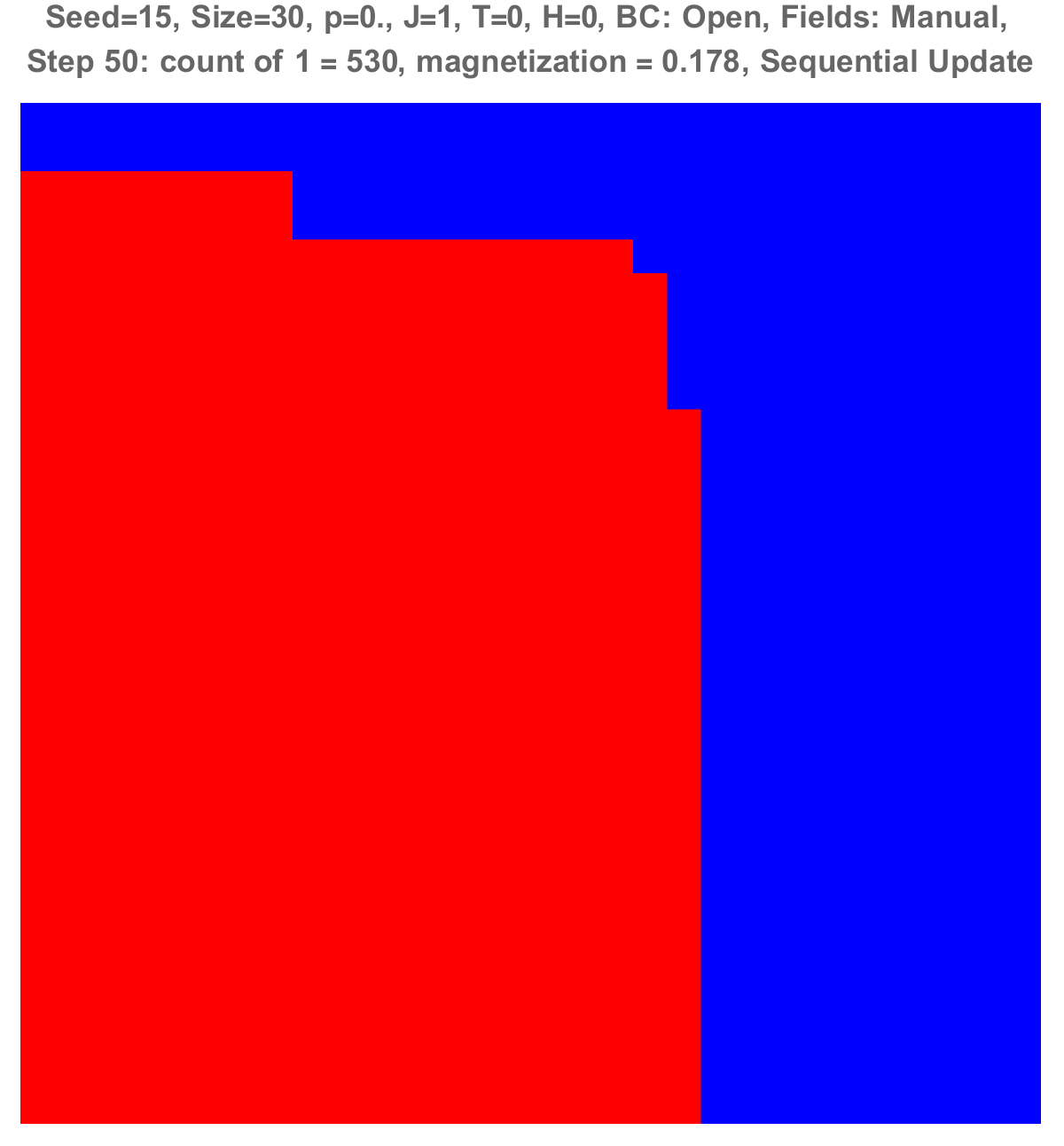}}
\end{figure}

\newpage 

\noindent\captionof{figure}{Red local fields are applied at selected locations: 18 in subpart (b), 30 in subpart (e), 31 in subpart (h), Associated dynamics are shown in subparts (a, d, g), Final outcomes are shown in subparts (c, f, i) with respectively 136, 116, 530 red choices. Last two cases show that a single local field (from 30 to 31) located at some tipping site has a substantial impact on the final outcome with a large majority of 530 red choices against a minority of 116.}
\label{ee}

\subsection{The single local field effect}

\subsubsection{Figure (\ref{ff})}

Subparts (m, n n, o, p) of Figure (\ref{aa}) infer that one single red local field located at a specific site of a distribution of equal proportions of red and blue colors ($p=0.50$) should distort the related spontaneous blue symmetry breaking towards a red total symmetry breaking. The prediction is confirmed by subparts (a, b, c, d) of Figure (\ref{ff})

Subpart (e) shows a different initial distribution of $p=0$ choices (Seed = 17) while subpart (f) exhibits the associated dynamics of symmetry breaking towards blue unanimity. Adding the subpart (c) local red field modifies the dynamics with a stabilization of a minority red domains as seen in subparts (g, h). 

Subpart (i) shows another different initial distribution of $p=0$ choices (Seed = 26) leading to a stabilization of two equal red and blue domains as seen in subpart (j). Here, applying the red local field modifies the dynamics but preserves the final coexistence of two equal domains as shown in subparts (k, l).

Subpart (m) shows another initial distribution $p=0$ leading again to a stabilization of two equal red and blue domains (subpart (n)). However, now the red local field produces a majority of red choices as seen from subparts (o, p).

\begin{figure}
\centering
\vspace{-3cm}
\subfigure[]{\includegraphics[width=0.22\textwidth]{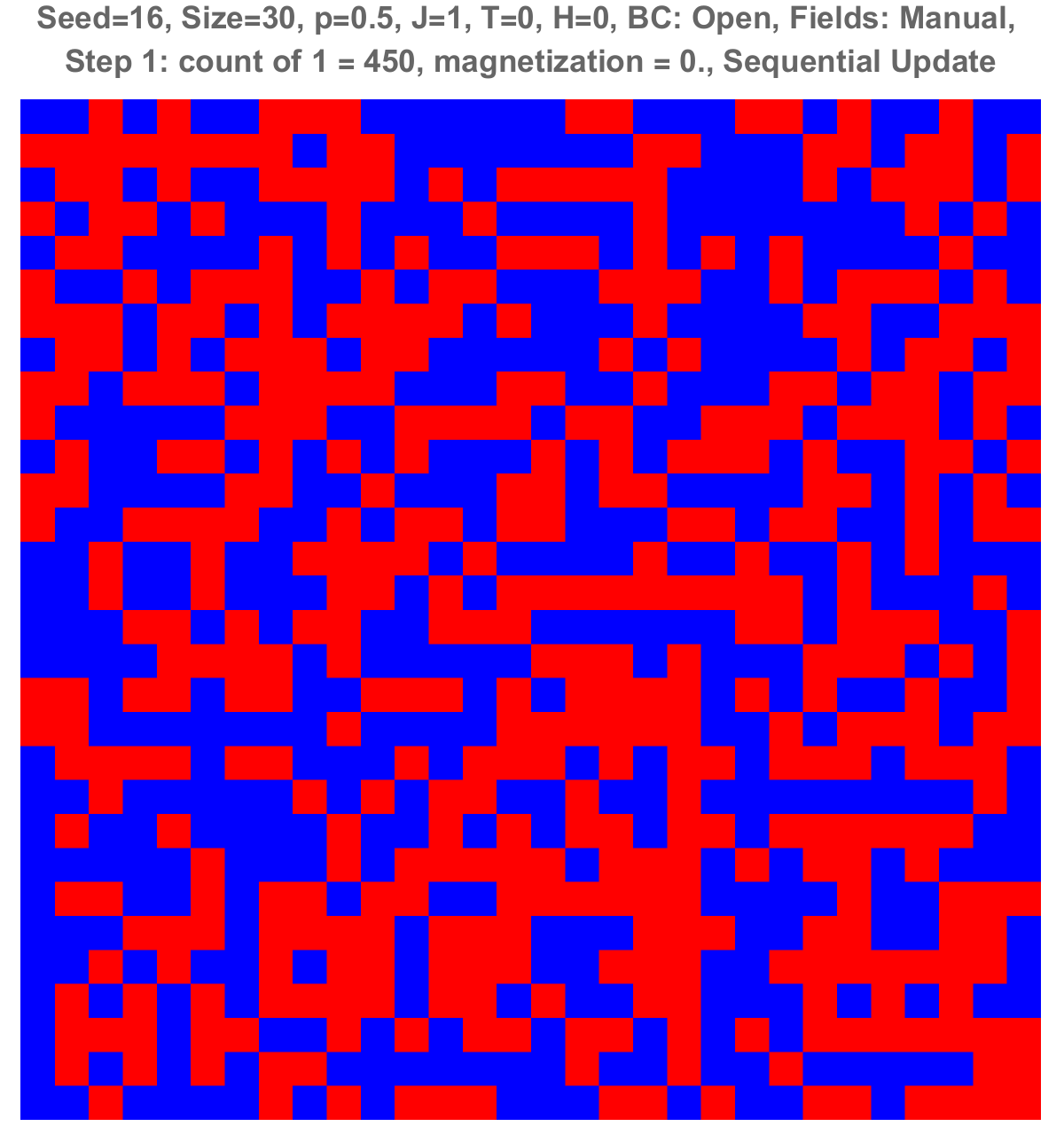}}
\subfigure[]{\includegraphics[width=0.26\textwidth]{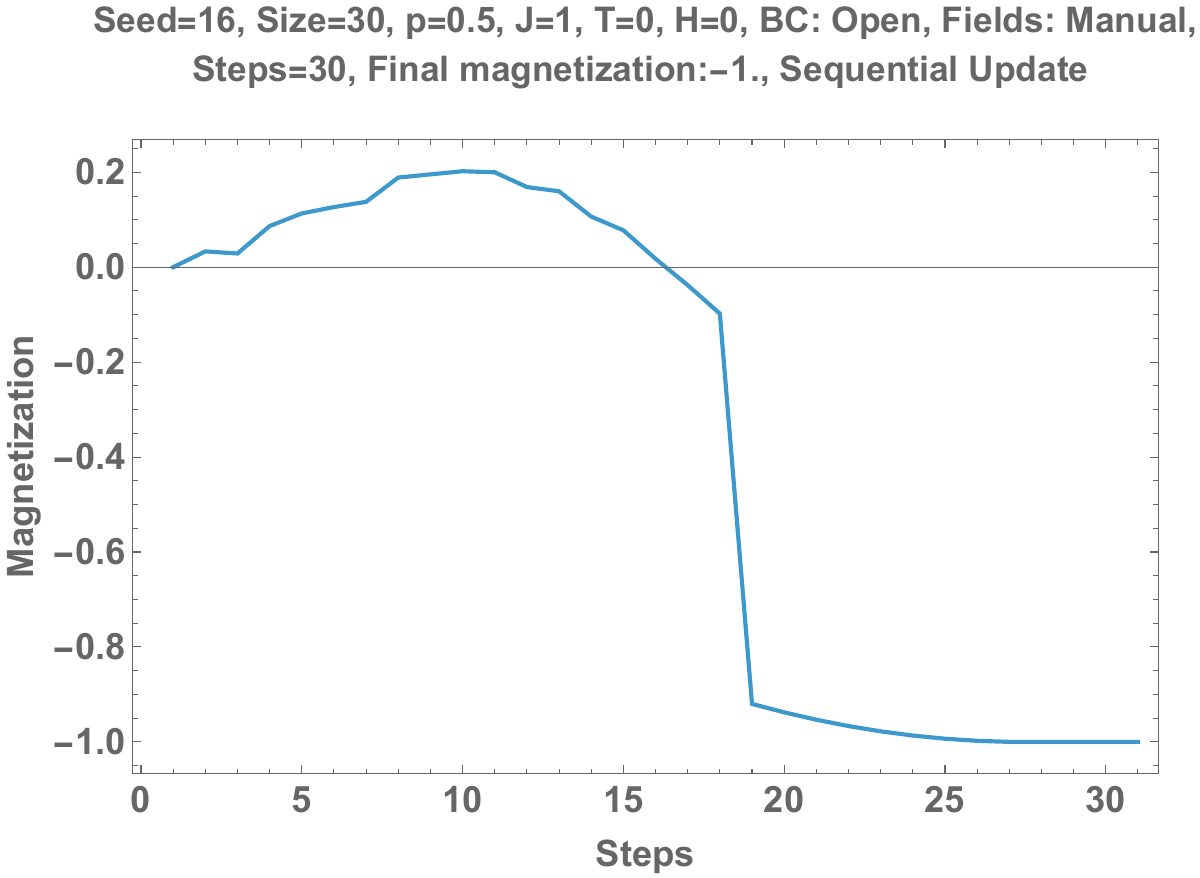}}
\subfigure[]{\includegraphics[width=0.22\textwidth]{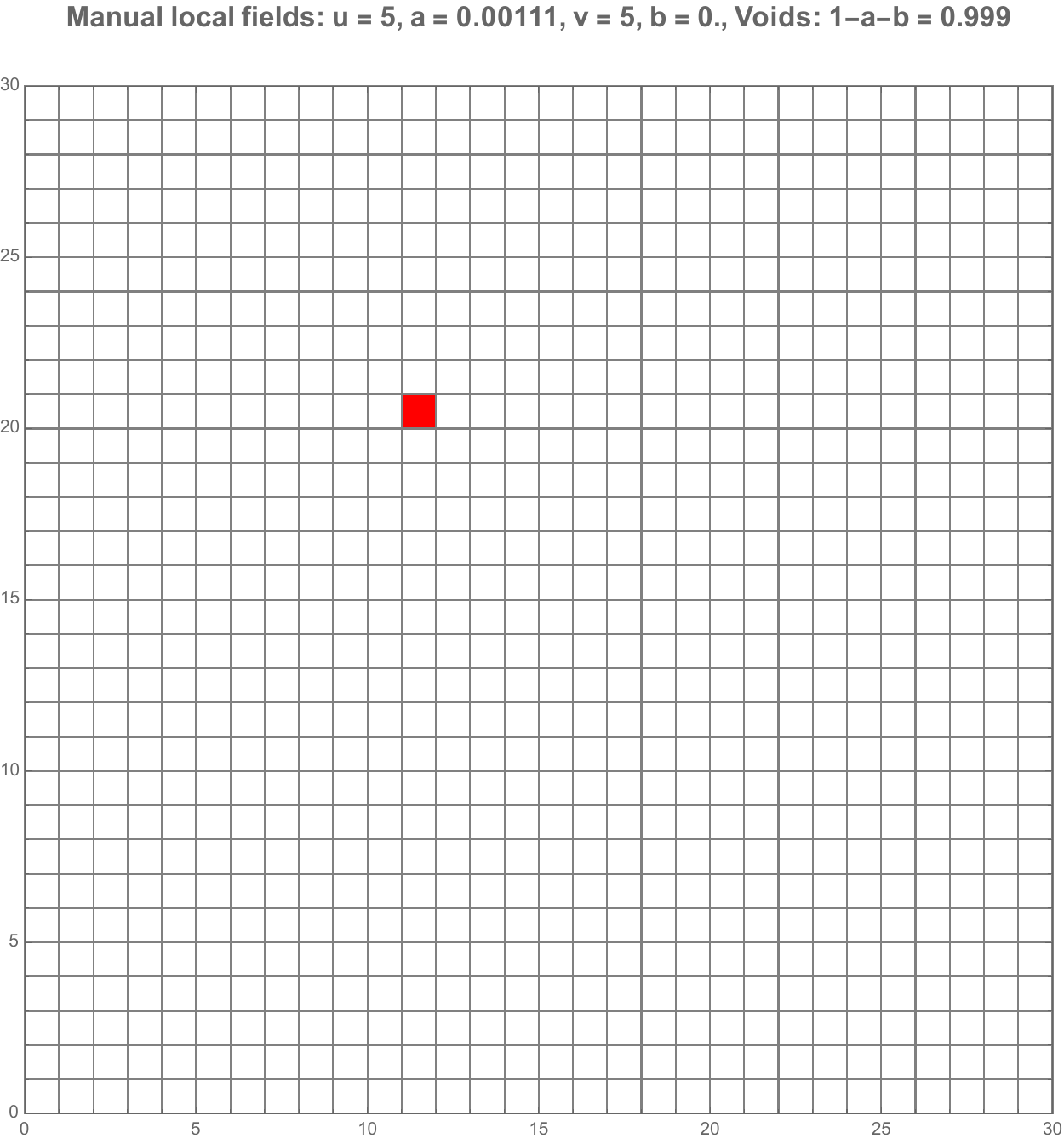}}
\subfigure[]{\includegraphics[width=0.26\textwidth]{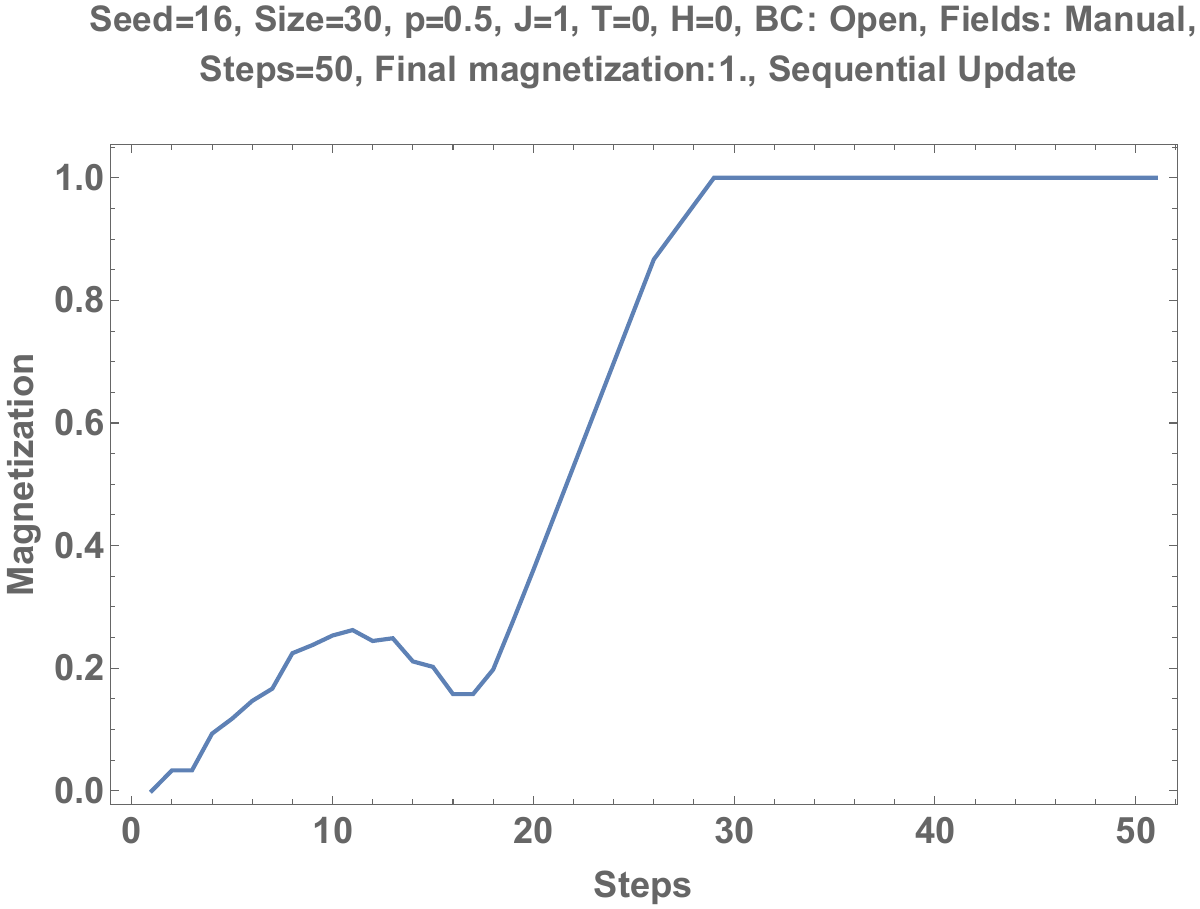}}
\\
\subfigure[]{\includegraphics[width=0.22\textwidth]{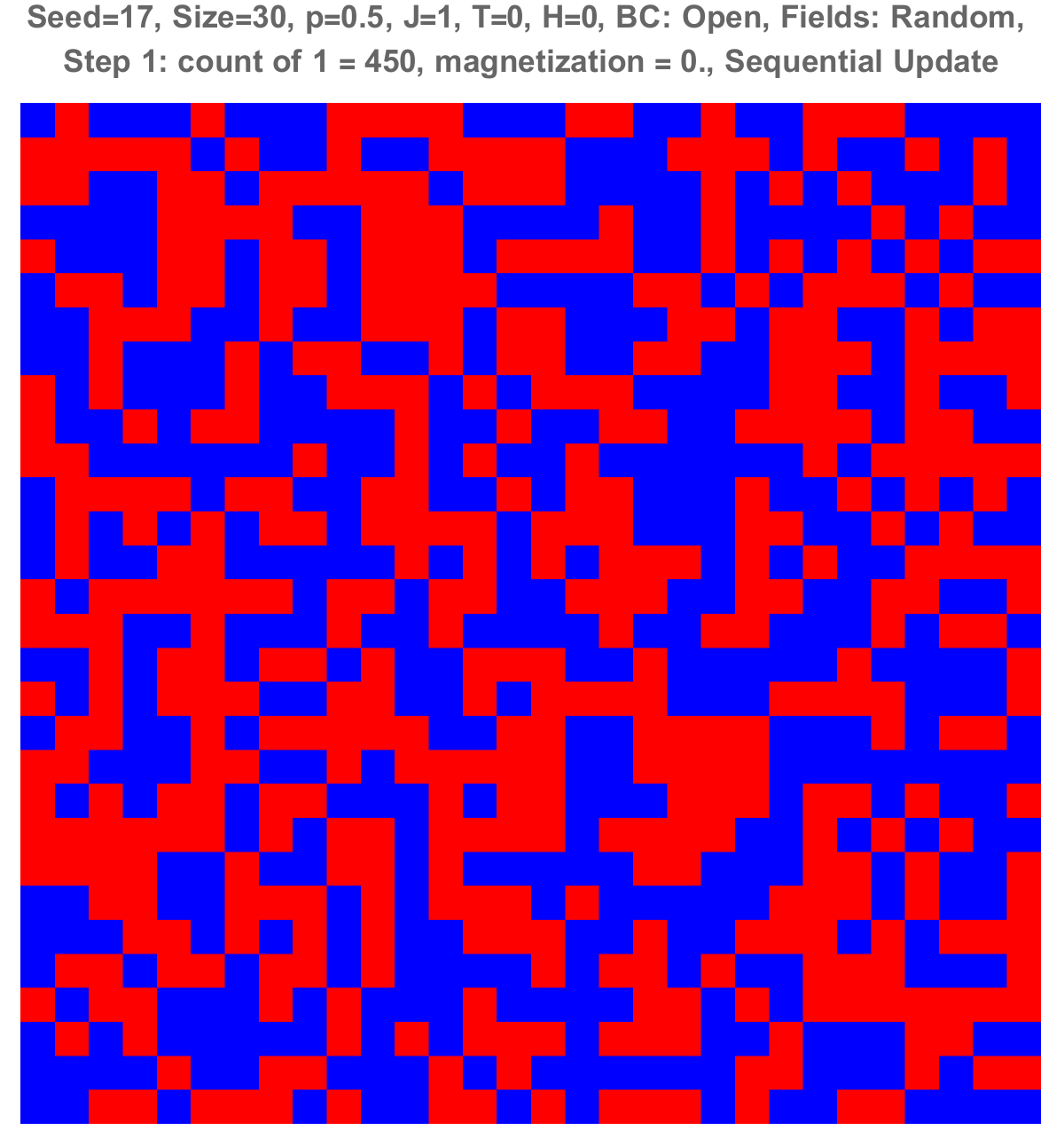}}
\subfigure[]{\includegraphics[width=0.26\textwidth]{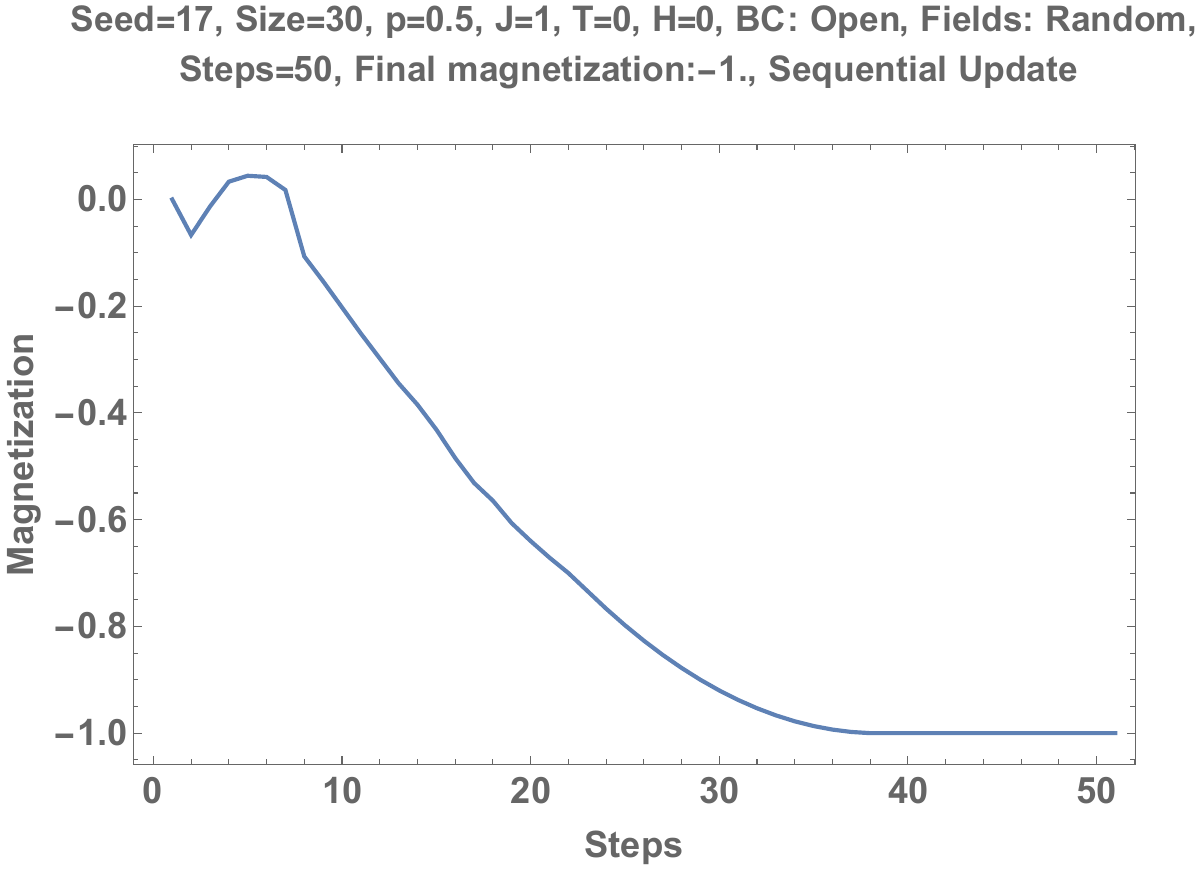}}
\subfigure[]{\includegraphics[width=0.26\textwidth]{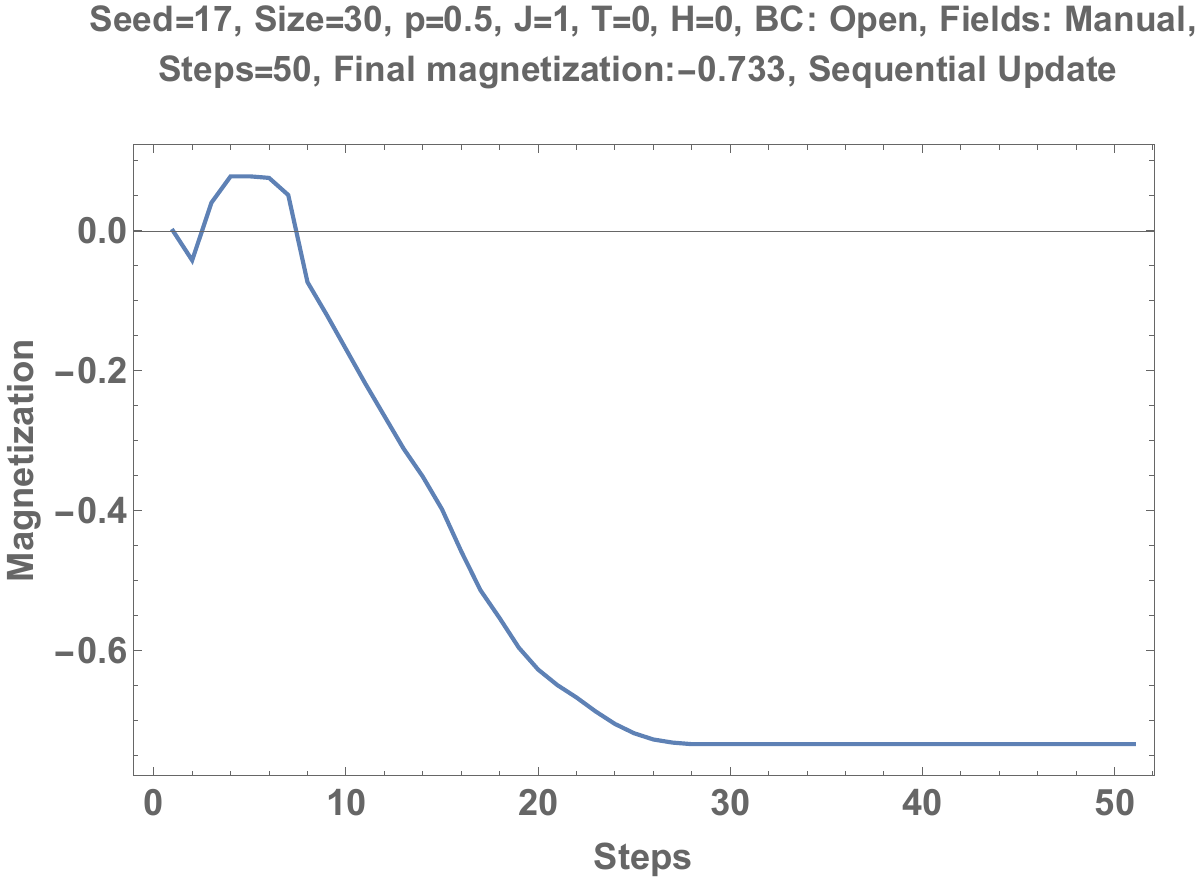}}
\subfigure[]{\includegraphics[width=0.22\textwidth]{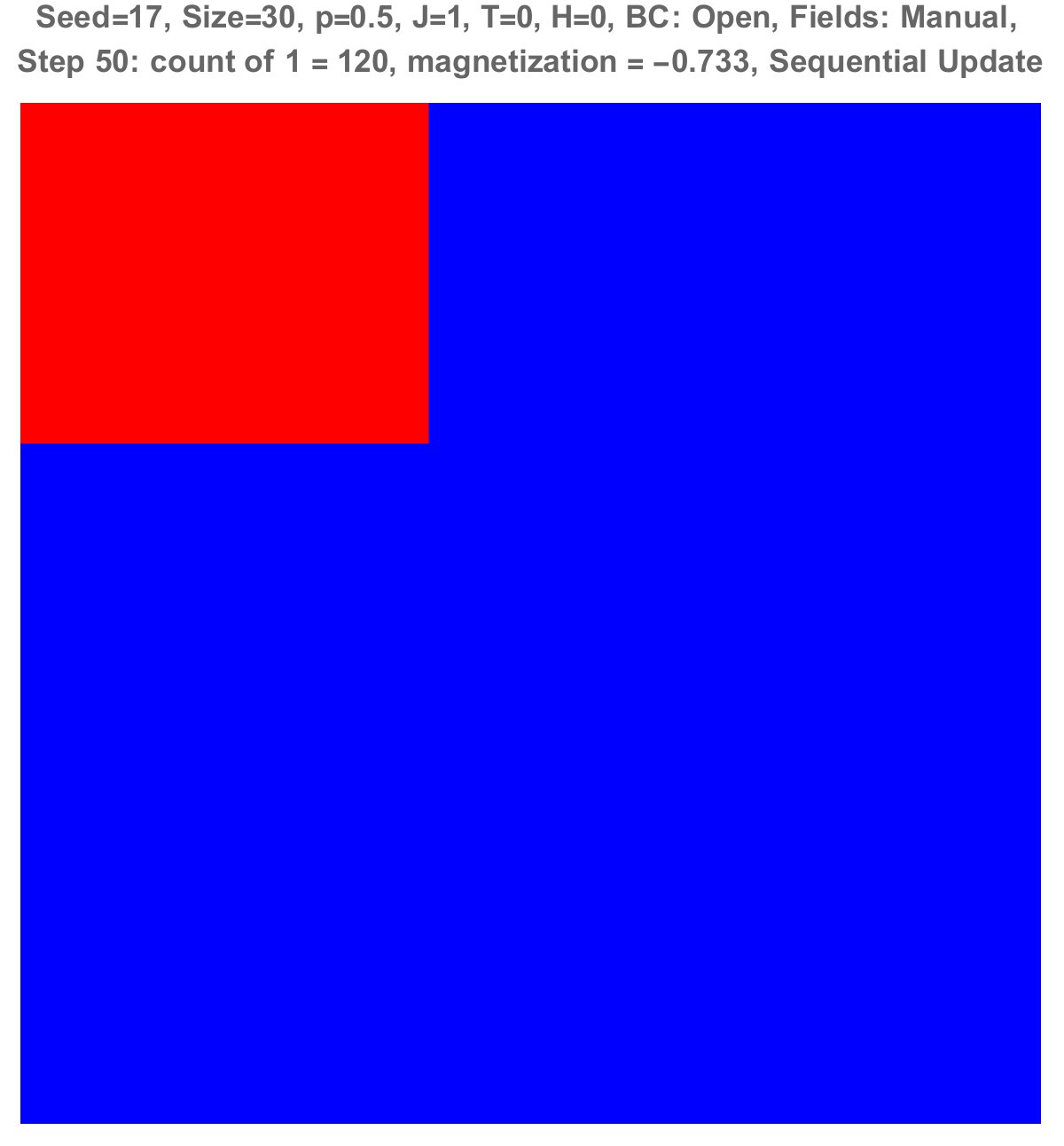}}
\\
\subfigure[]{\includegraphics[width=0.26\textwidth]{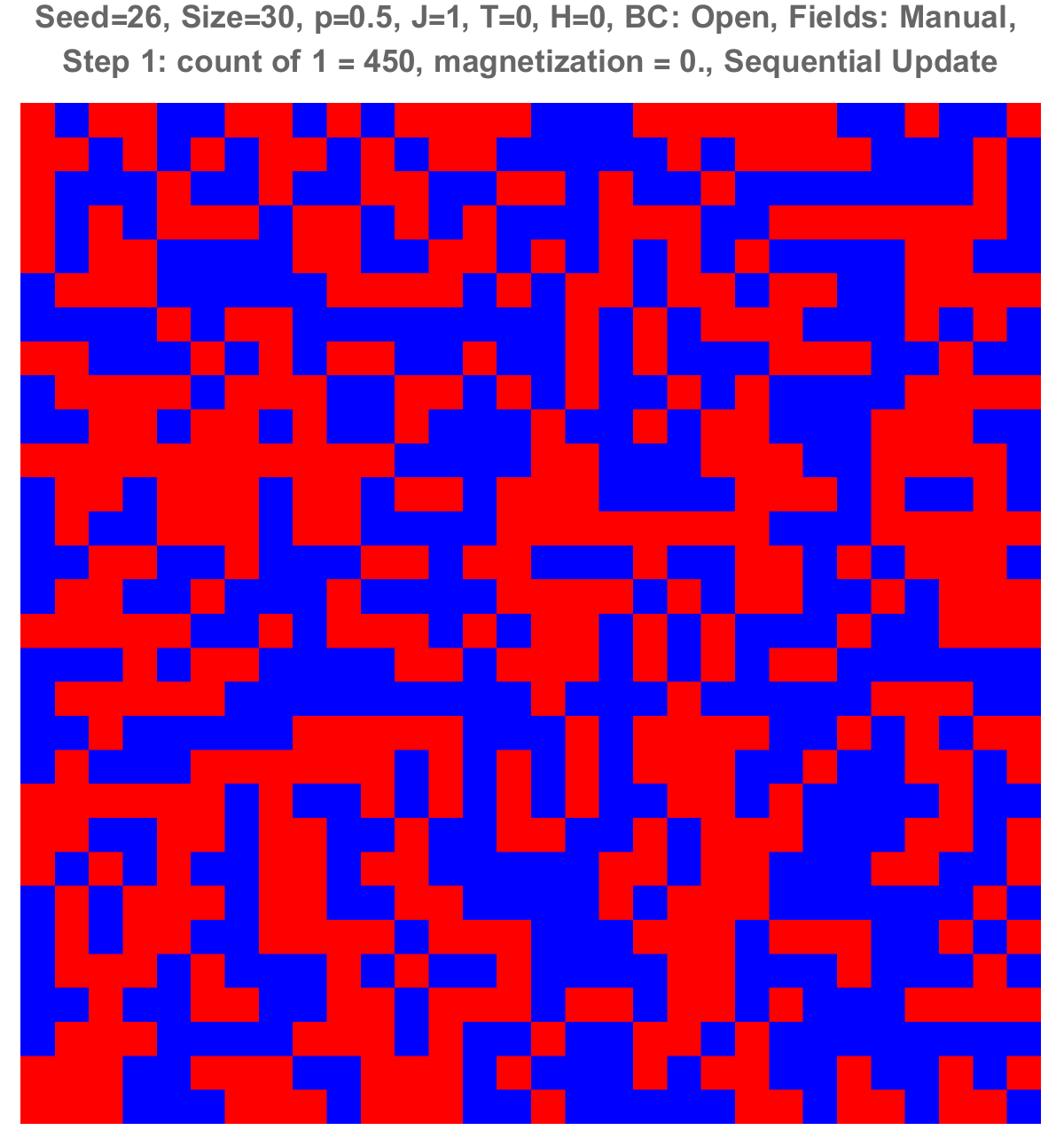}}
\subfigure[]{\includegraphics[width=0.26\textwidth]{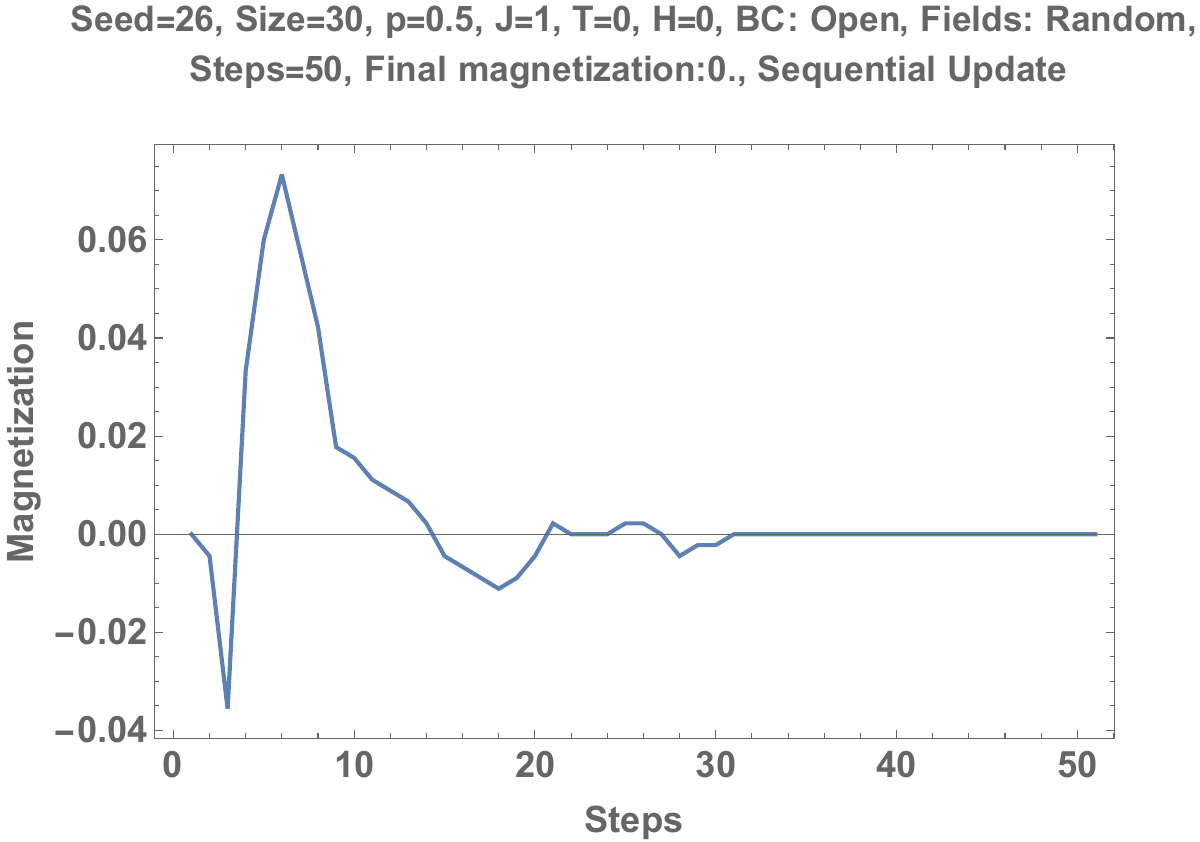}}
\subfigure[]{\includegraphics[width=0.22\textwidth]{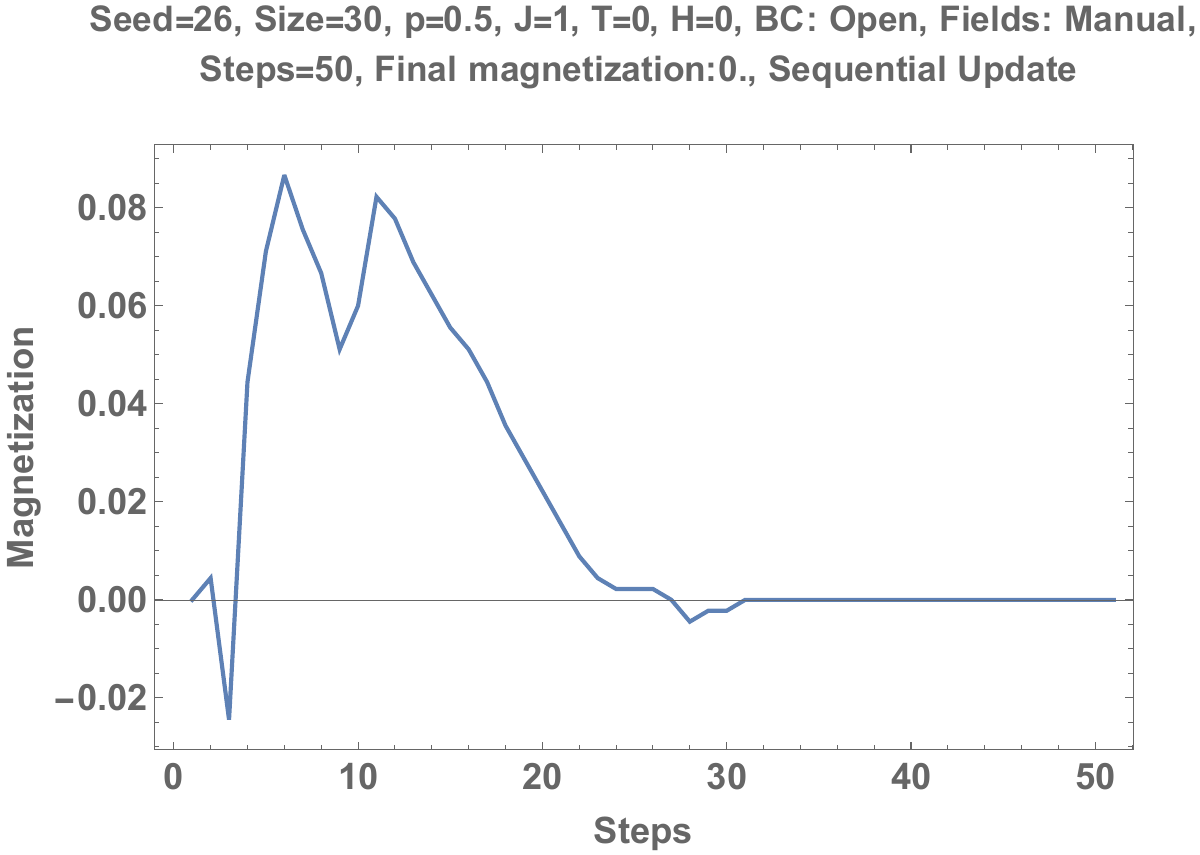}}
\subfigure[]{\includegraphics[width=0.22\textwidth]{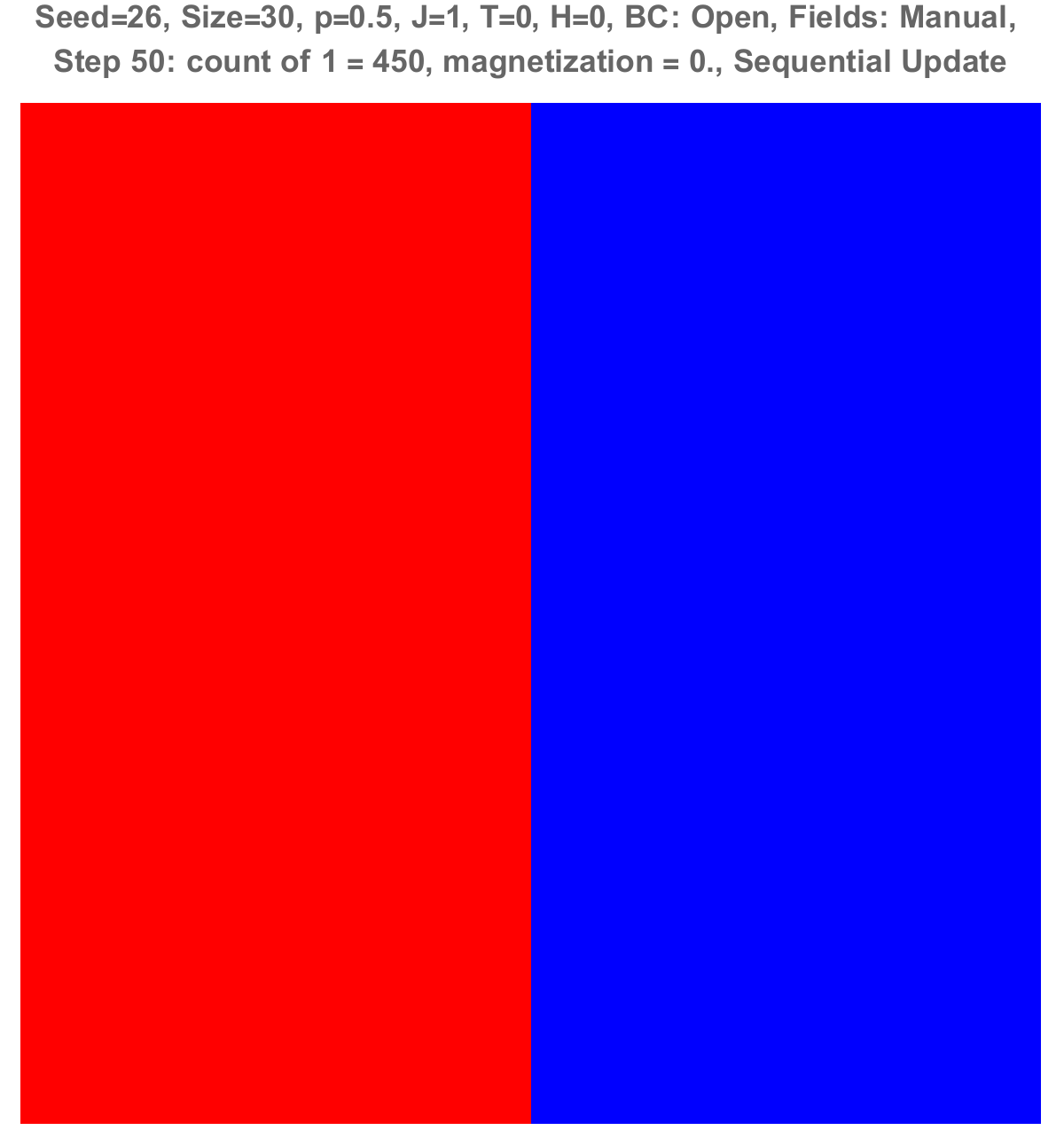}}
\\
\subfigure[]{\includegraphics[width=0.22\textwidth]{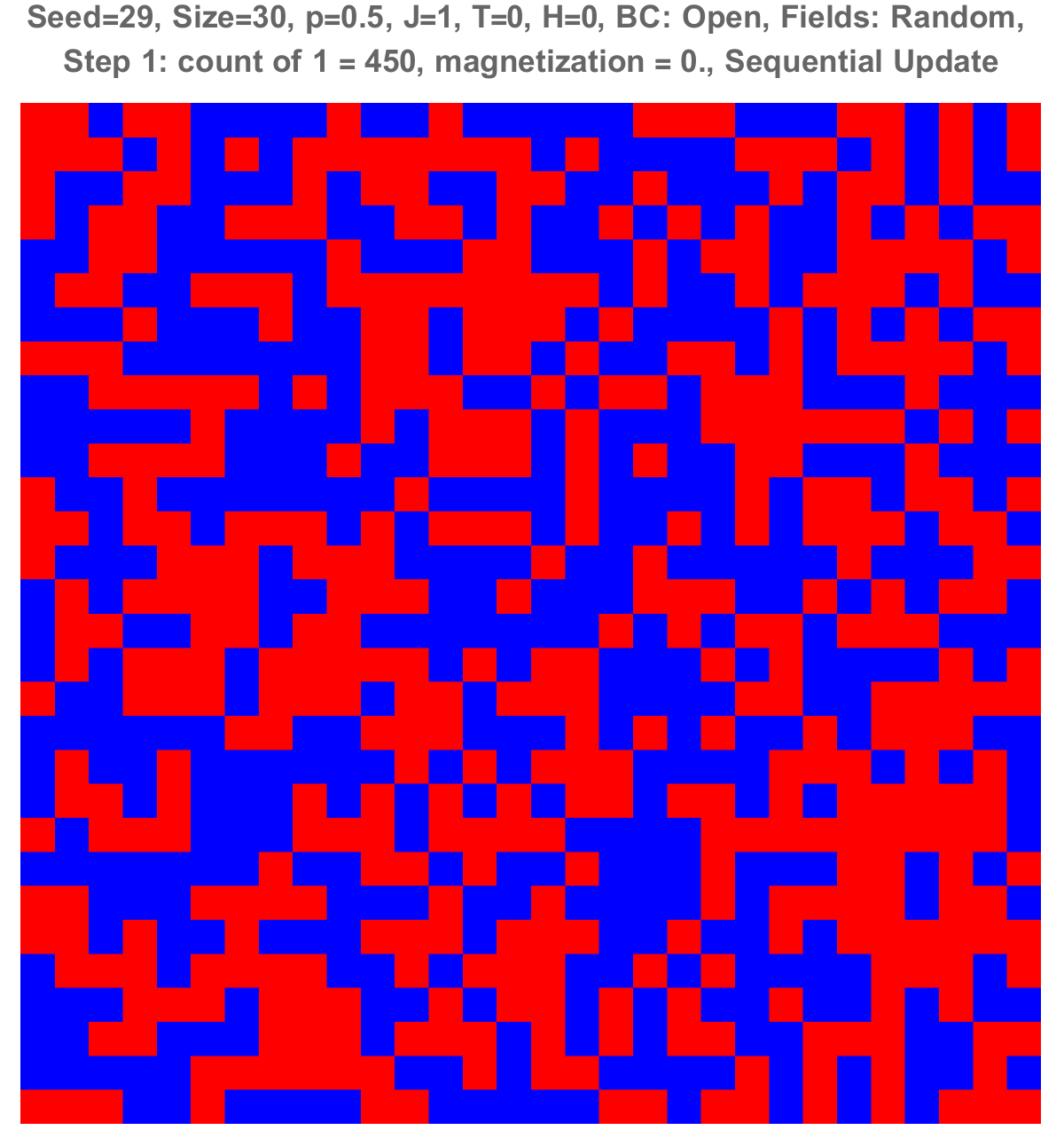}}
\subfigure[]{\includegraphics[width=0.26\textwidth]{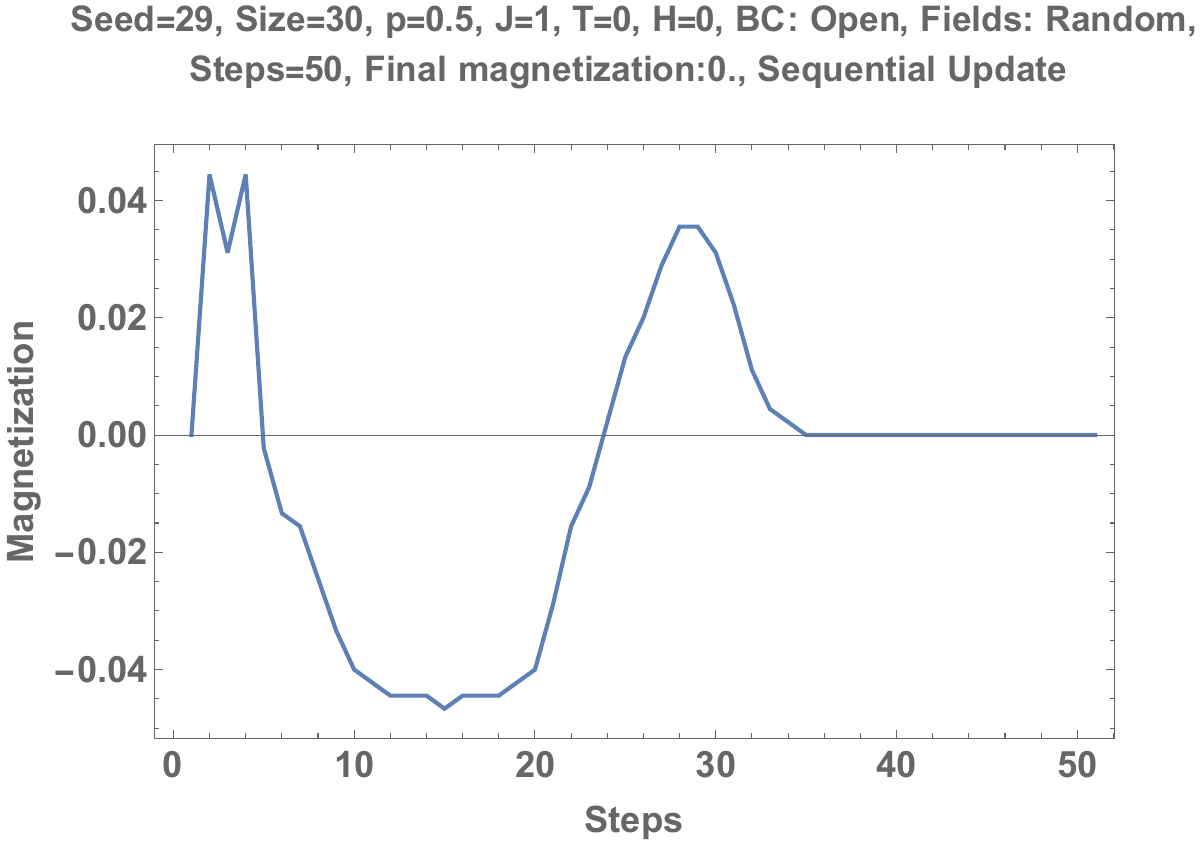}}
\subfigure[]{\includegraphics[width=0.26\textwidth]{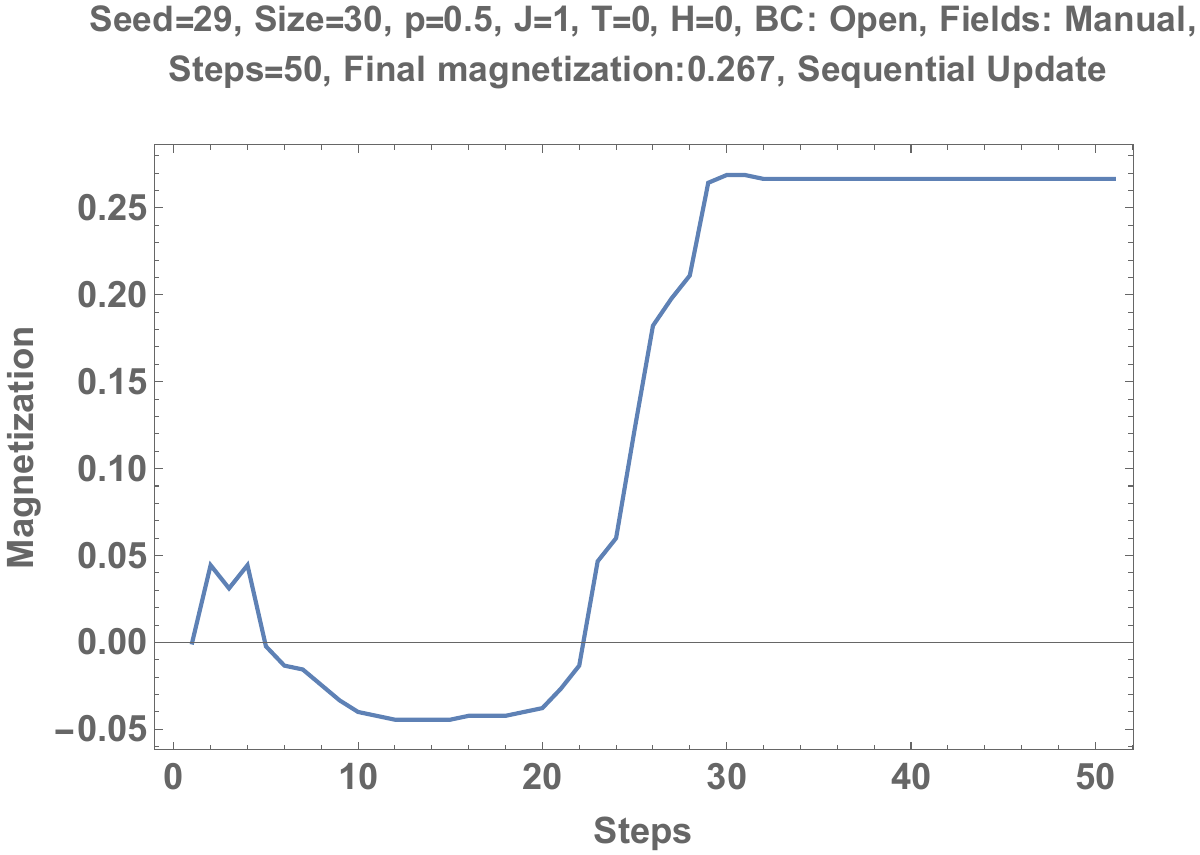}}
\subfigure[]{\includegraphics[width=0.22\textwidth]{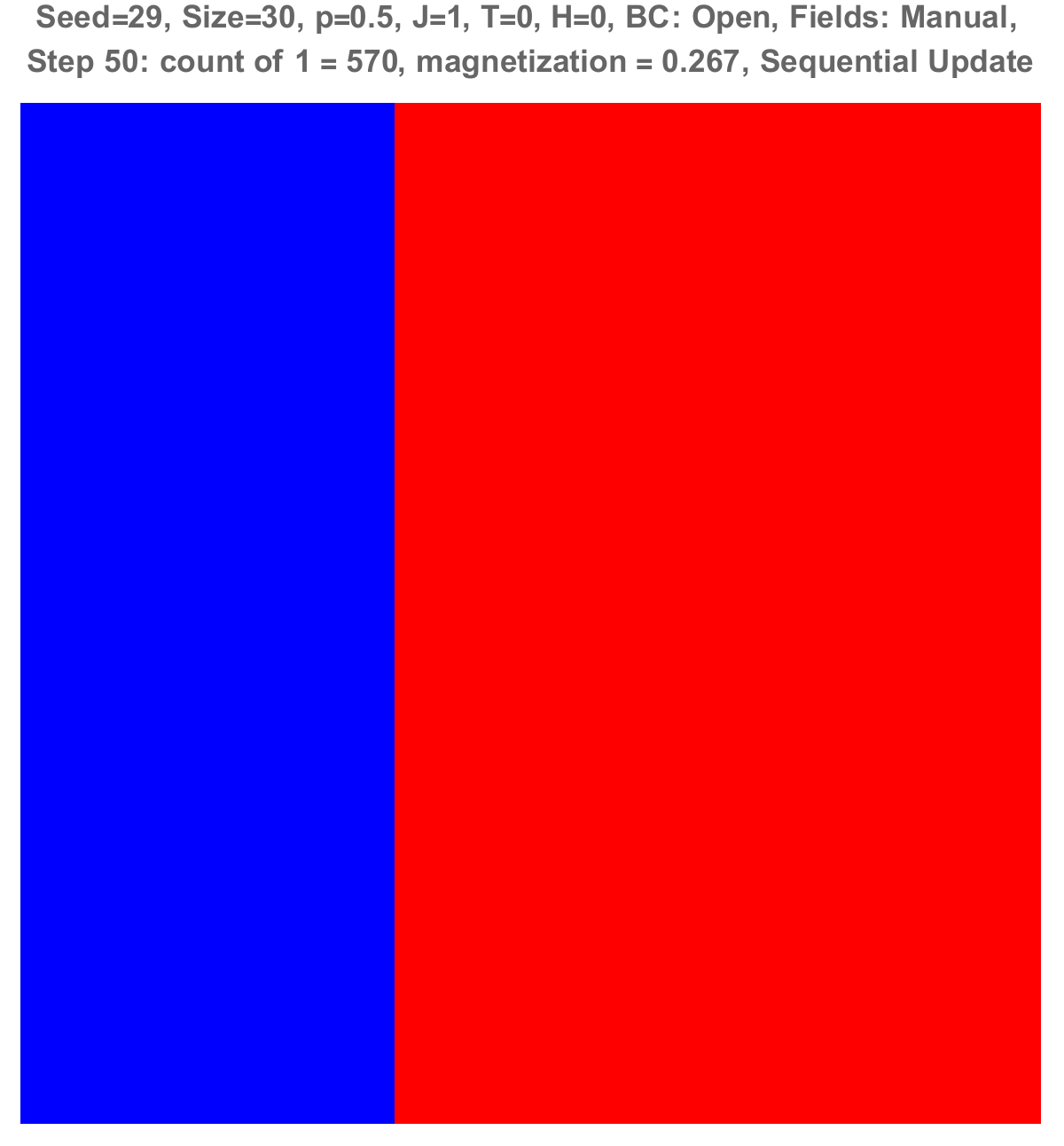}}
\end{figure}

\newpage 

\noindent\captionof{figure}{Subparts (a, b, c, d)  show that one single red local field located at a specific site of a distribution of equal proportions of red and blue colors (Seed = 16, $p=0.50$) distorts the related spontaneous blue symmetry breaking towards a red total symmetry breaking. Subpart (e) shows a different initial distribution of $p=0$ choices (Seed = 17) while subpart (f) exhibits the associated dynamics of symmetry breaking towards blue unanimity. Adding the subpart (c) local red field modifies the dynamics with a stabilization of a minority red domains as seen in subparts (g, h). Subpart (i) shows another different initial distribution of $p=0$ choices (Seed = 26) leading to a stabilization of two equal red and blue domains as seen in subpart (j). Here, applying the red local field modifies the dynamics but preserves the final coexistence of two equal domains as shown in subparts (k, l). Subpart (m) shows another initial distribution $p=0$ leading again to a stabilization of two equal red and blue domains (subpart (n)). However, now the red local field produces a majority of red choices as seen from subparts (o, p).}
\label{ff}

\subsubsection{Figure (\ref{gg})}

Figure (\ref{ff}) has shown the substantial impact that one single red local field well positioned can have on the dynamics of interactions among initial distributions of red and blue choices with $p=0.50$. Figure (\ref{gg}) revisits the effect starting form initial configurations with $p \neq 0.50$.

Subparts (a, b, c, d), (e, f, g, h), (i, j, k, l) keep the single red local field located at the same position. Zero effect is seen when starting from $p=0$ as could be expected (subparts (a, b, c, d)).

Subpart (e) show an initial configuration with $p=0.44$. Subparts (f, g) shows the dynamics of interactions with respectively zero and one local field. Very little effect is observed with the same final state (subpart (h)) as with $p=0$.

However, with a one percent increase in red initial choices from $p=0.44$ to $p=0.45$ (subpart (i)), the single red local field is found to have quite a strong impact with a final majority domain with 630 red choices as seen in subparts (j, k, l).

Last row show the effect of stil one single red local field but located at a different position on the grid as seen in subpart (m) with previous coalition and subpart (n) with the new location. Subpart (o) shows the associated dynamics with the same initial distribution (Seed = 61, $p=0.45$) and subpart (p) with a different initial distribution (Seed = 62, $p=0.45$). While the new location of the red single field increases the support for red choices in the final outcome as seen comparing subparts (o) and (k), it has zero effect on the second distribution as shown in subpart (p).

\begin{figure}
\centering
\vspace{-3cm}
\subfigure[]{\includegraphics[width=0.22\textwidth]{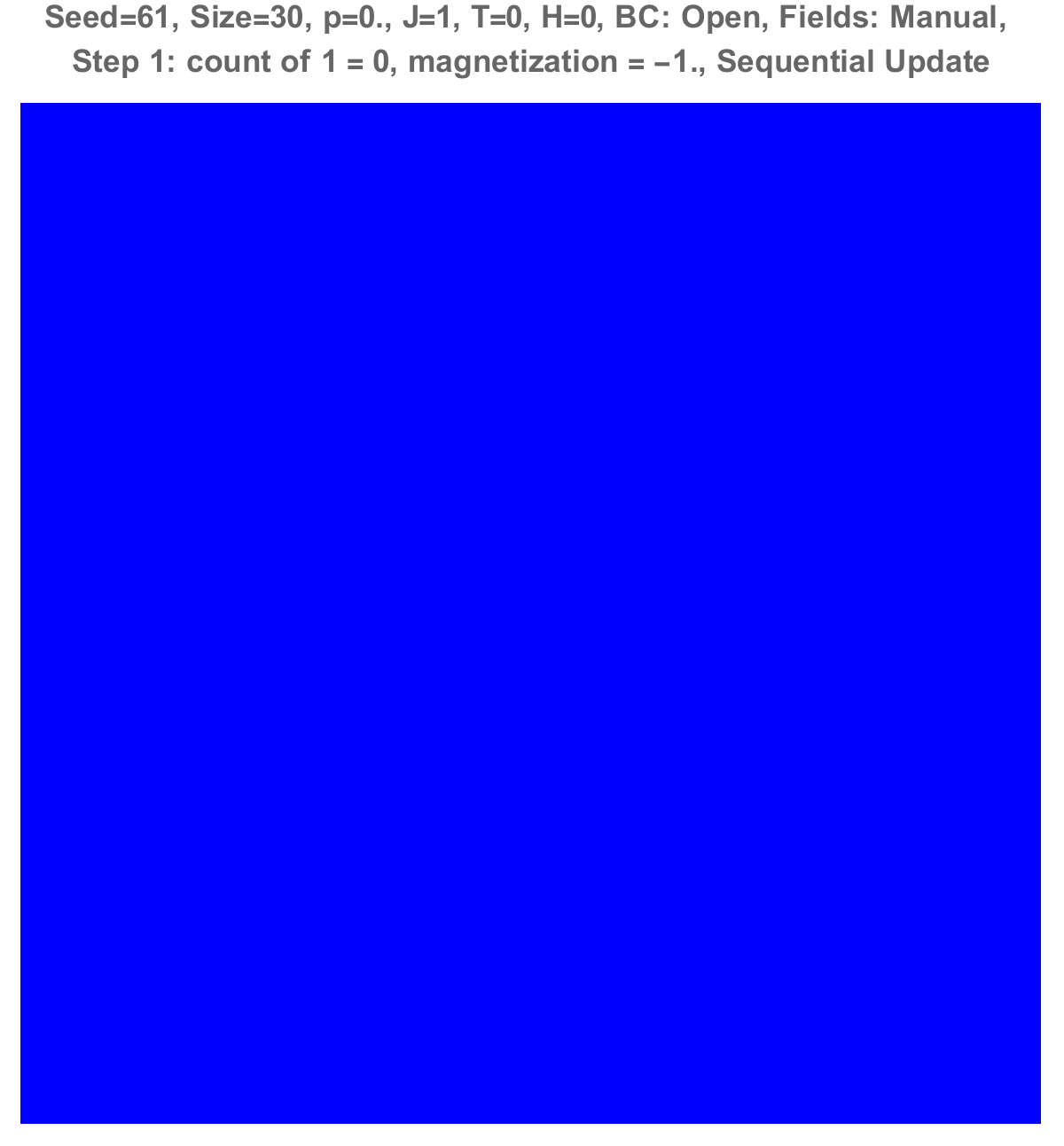}}
\subfigure[]{\includegraphics[width=0.26\textwidth]{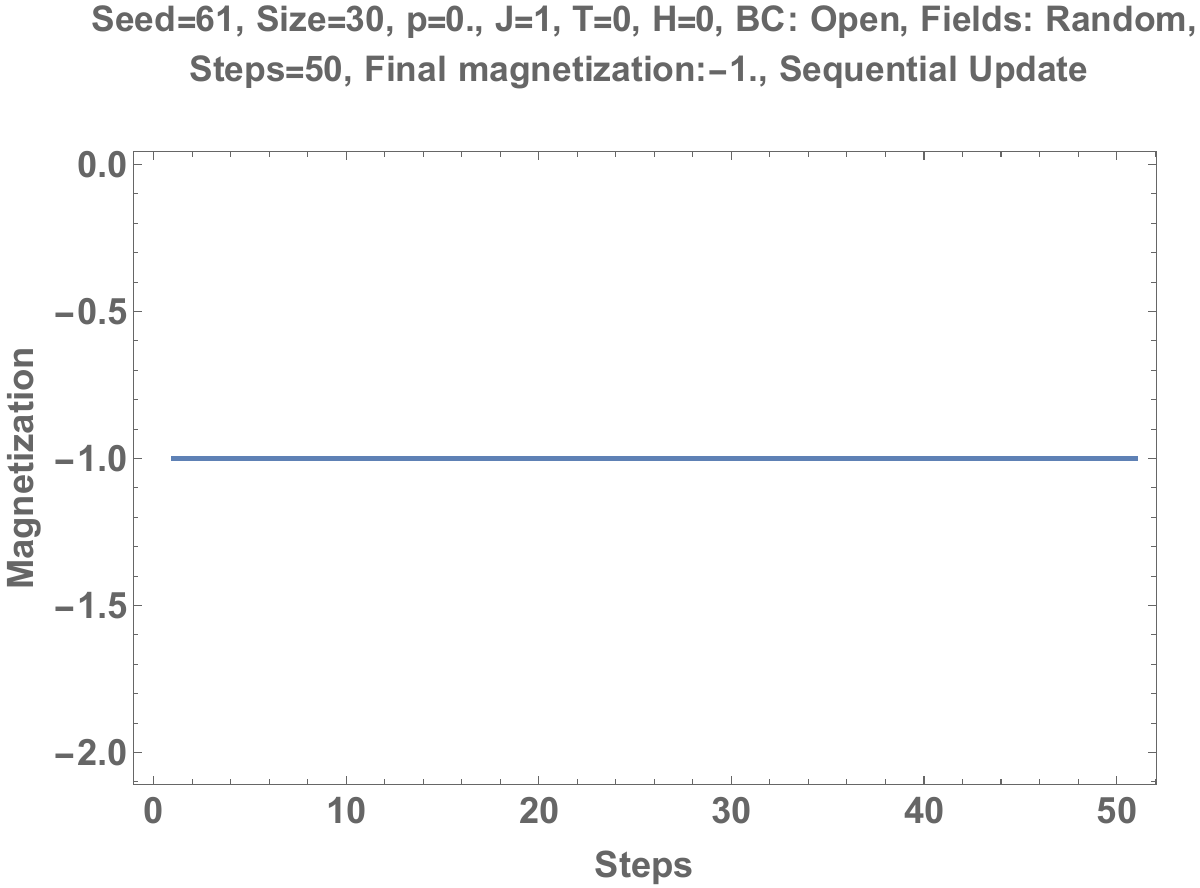}}
\subfigure[]{\includegraphics[width=0.26\textwidth]{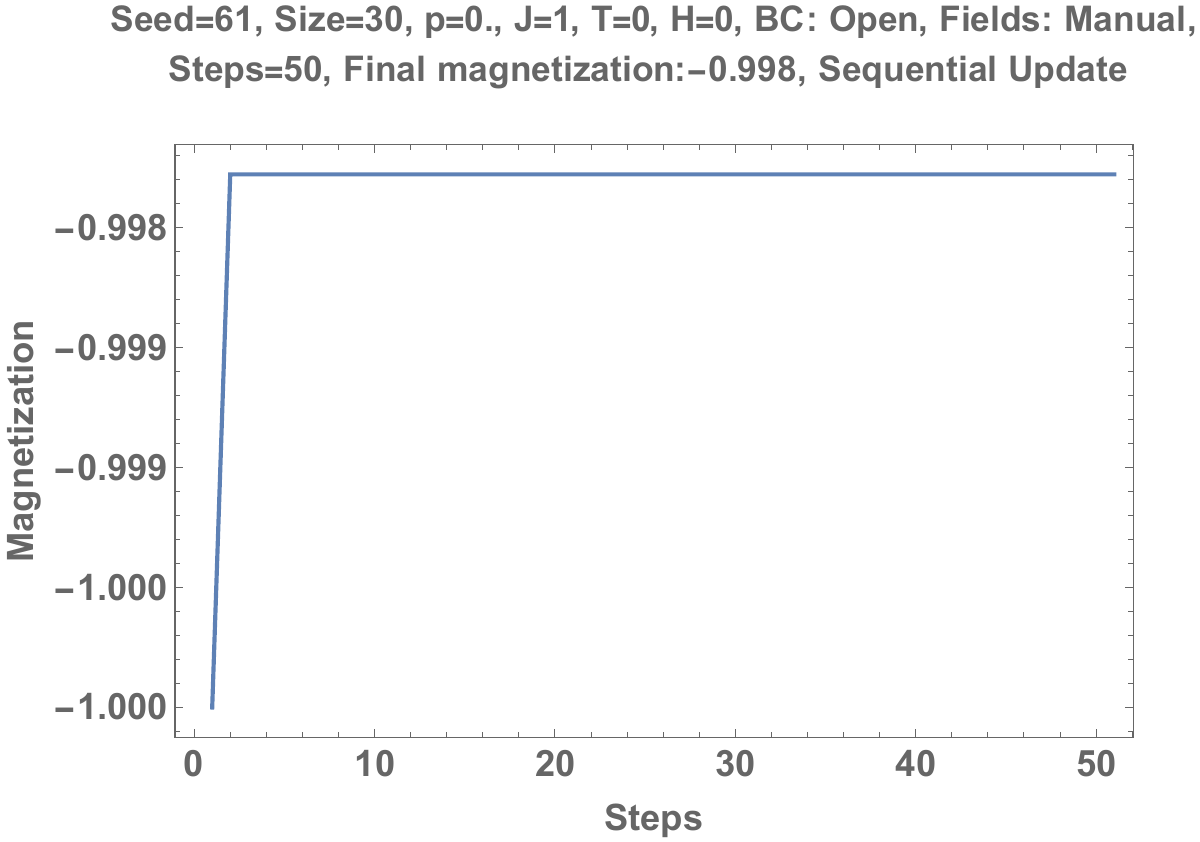}}
\subfigure[]{\includegraphics[width=0.22\textwidth]{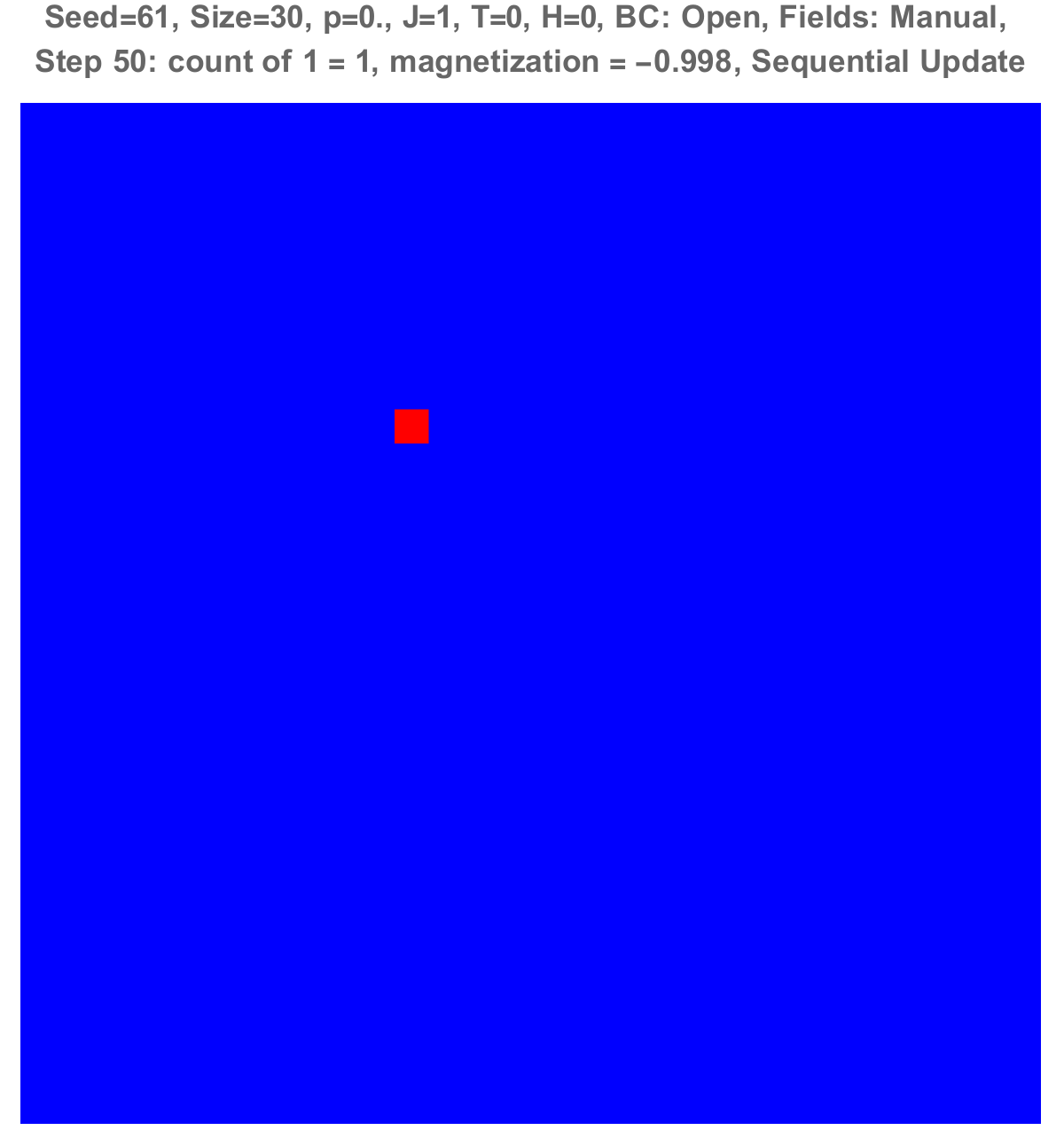}}
\\
\subfigure[]{\includegraphics[width=0.22\textwidth]{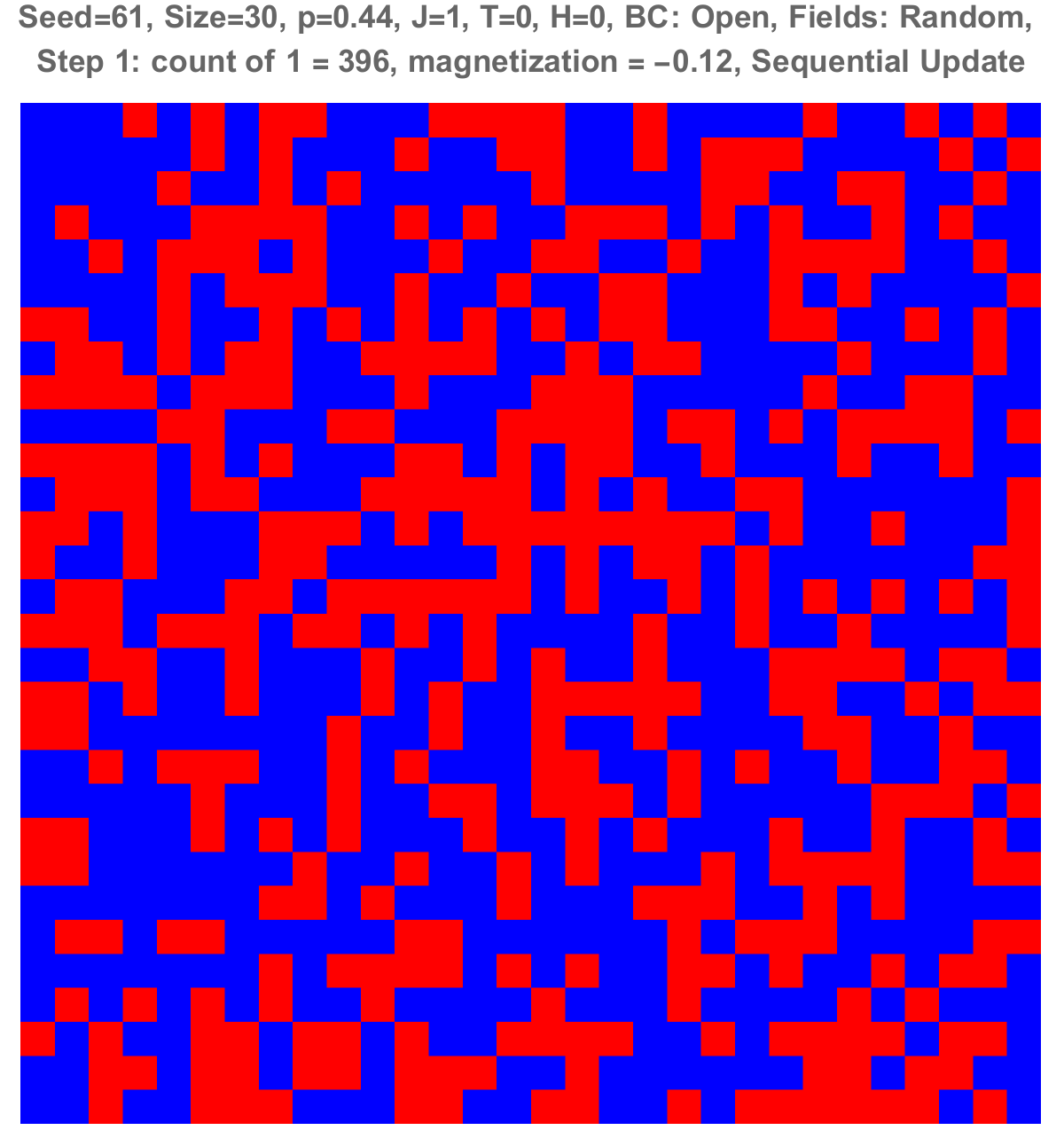}}
\subfigure[]{\includegraphics[width=0.26\textwidth]{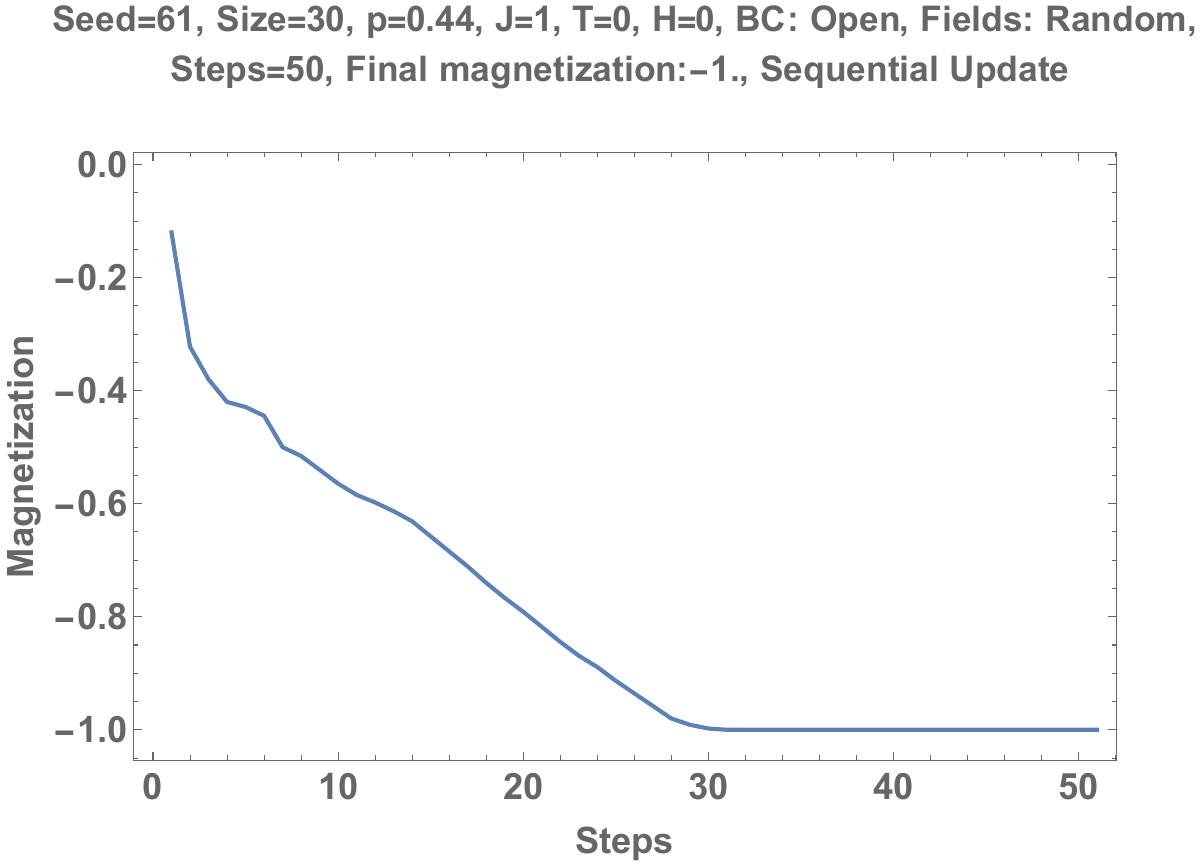}}
\subfigure[]{\includegraphics[width=0.26\textwidth]{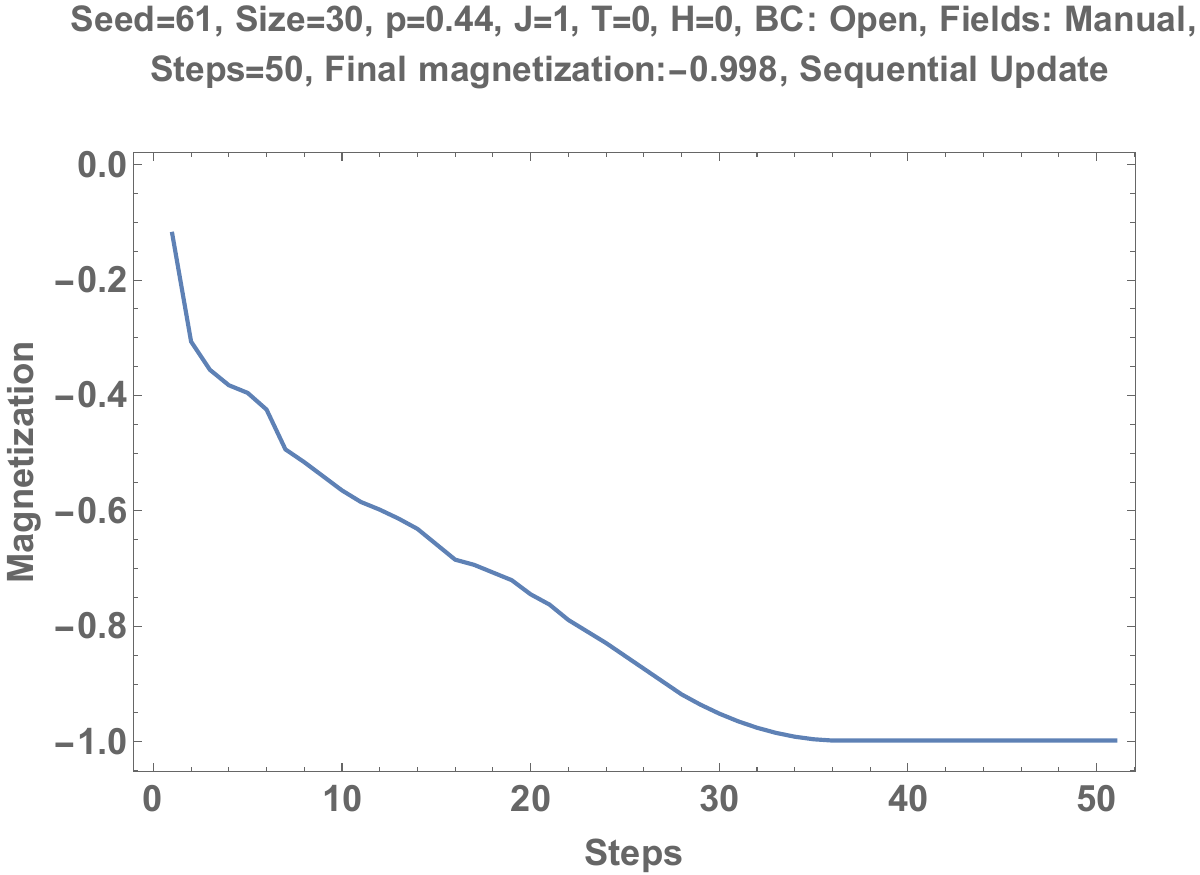}}
\subfigure[]{\includegraphics[width=0.22\textwidth]{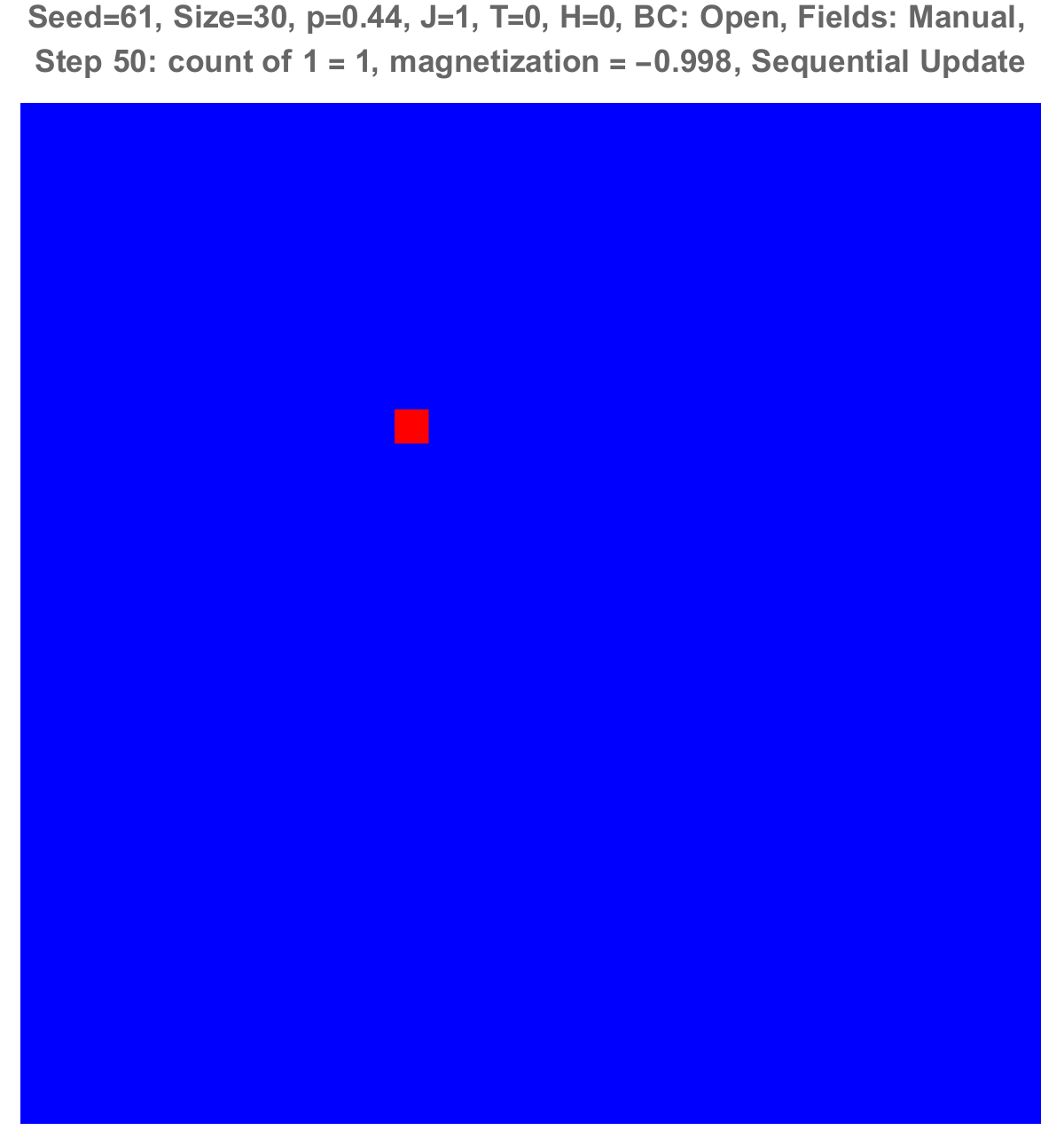}}
\\
\subfigure[]{\includegraphics[width=0.22\textwidth]{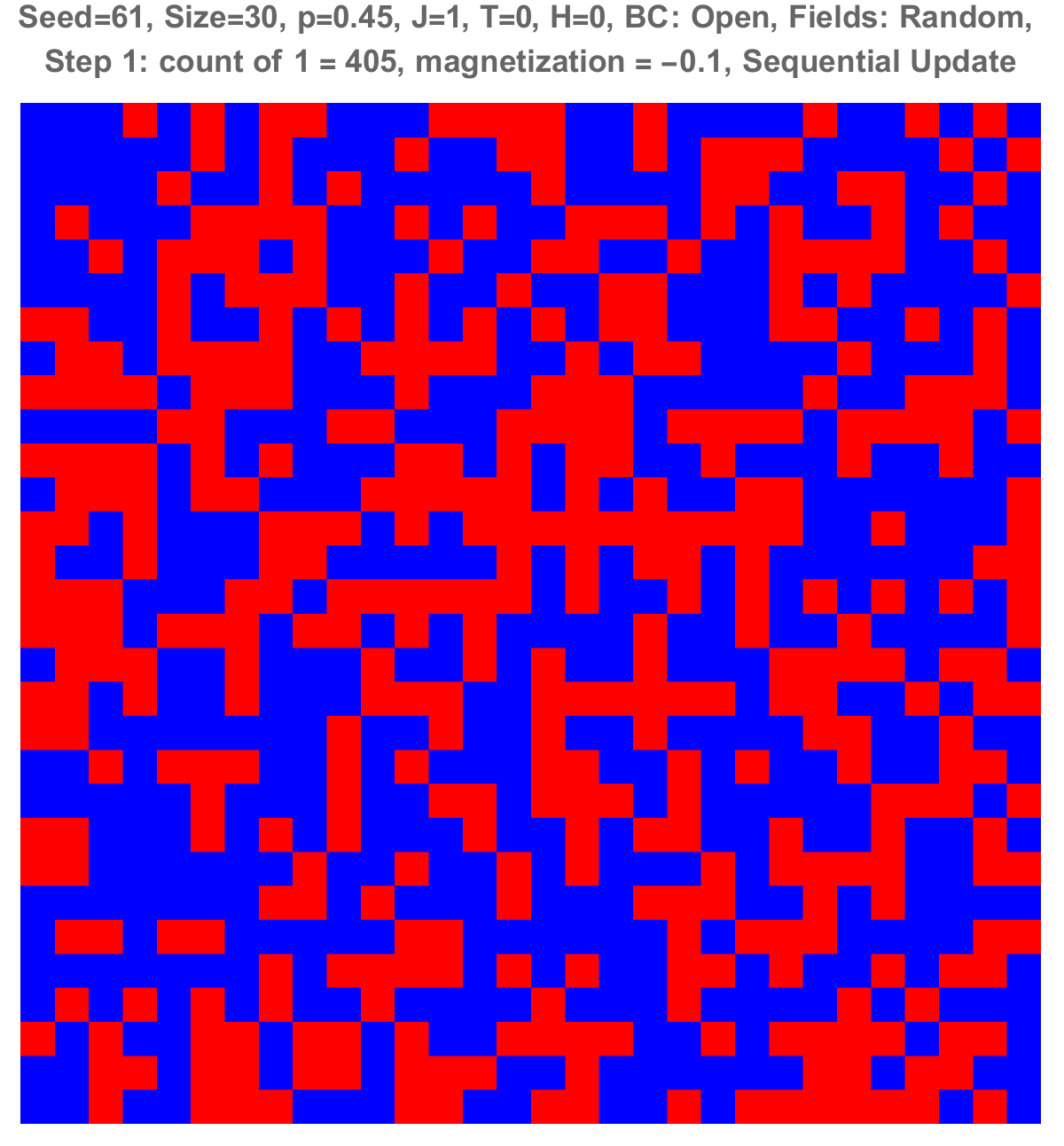}}
\subfigure[]{\includegraphics[width=0.26\textwidth]{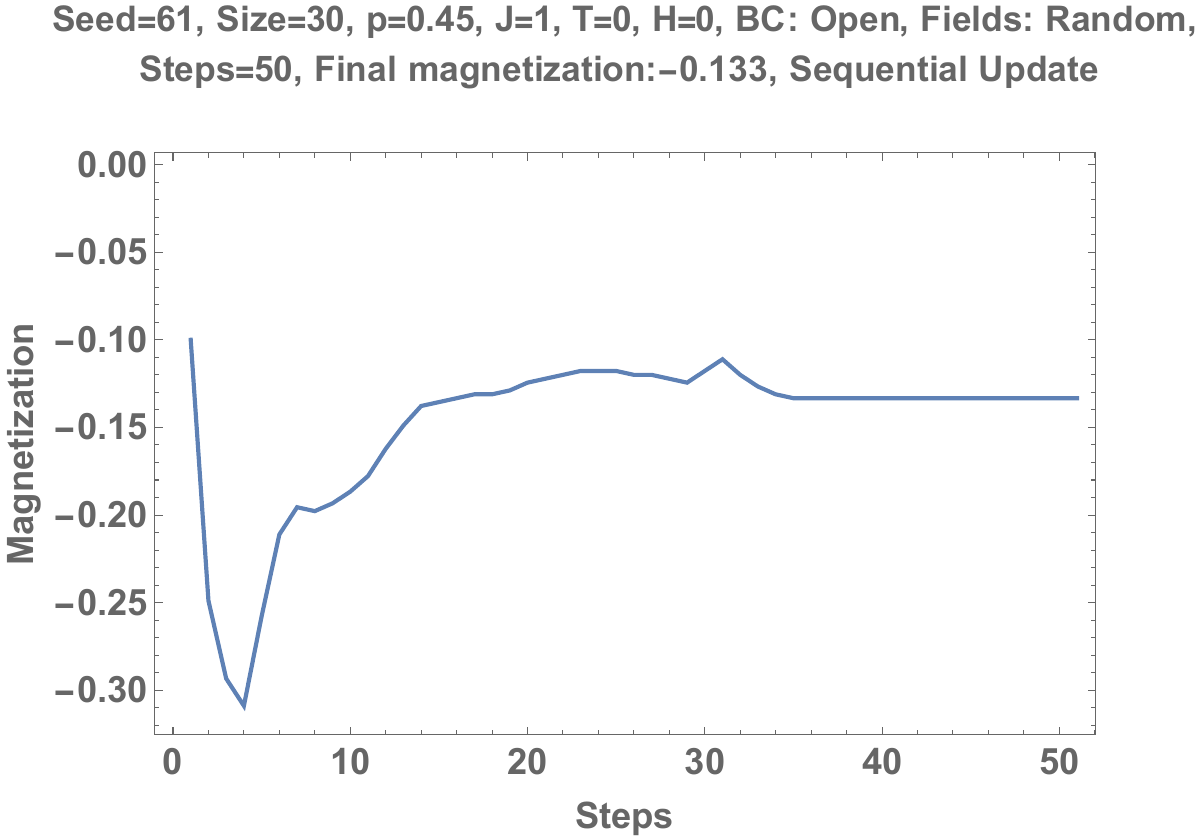}}
\subfigure[]{\includegraphics[width=0.26\textwidth]{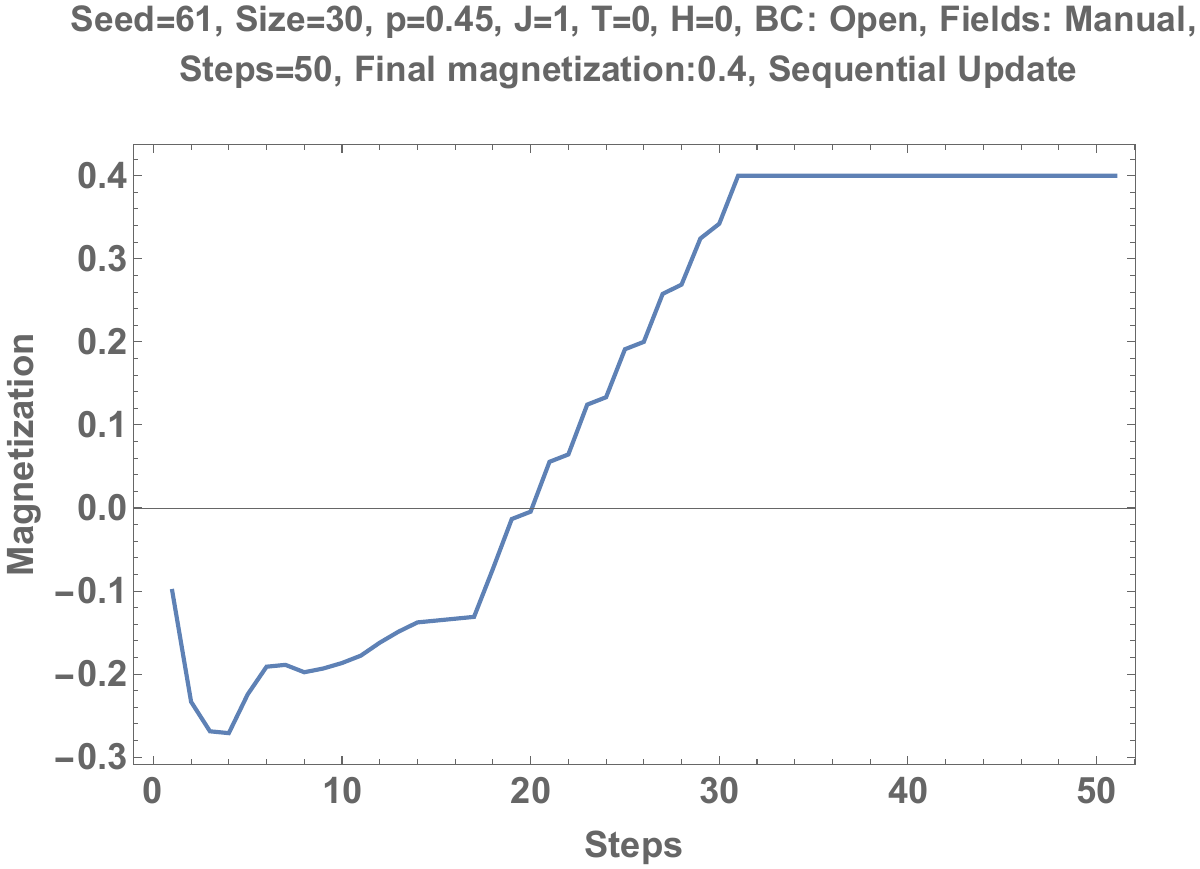}}
\subfigure[]{\includegraphics[width=0.22\textwidth]{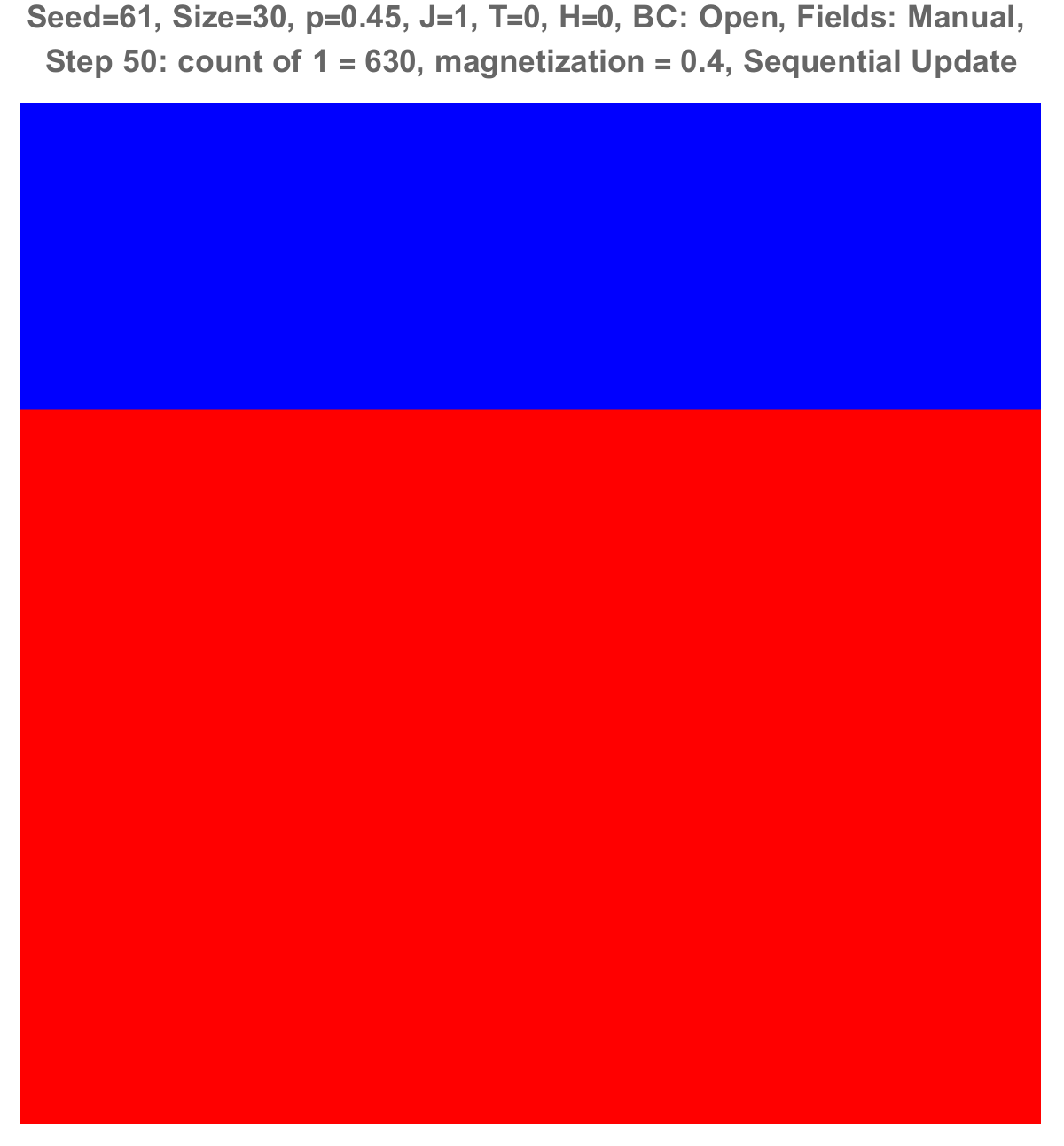}}
\\
\subfigure[]{\includegraphics[width=0.22\textwidth]{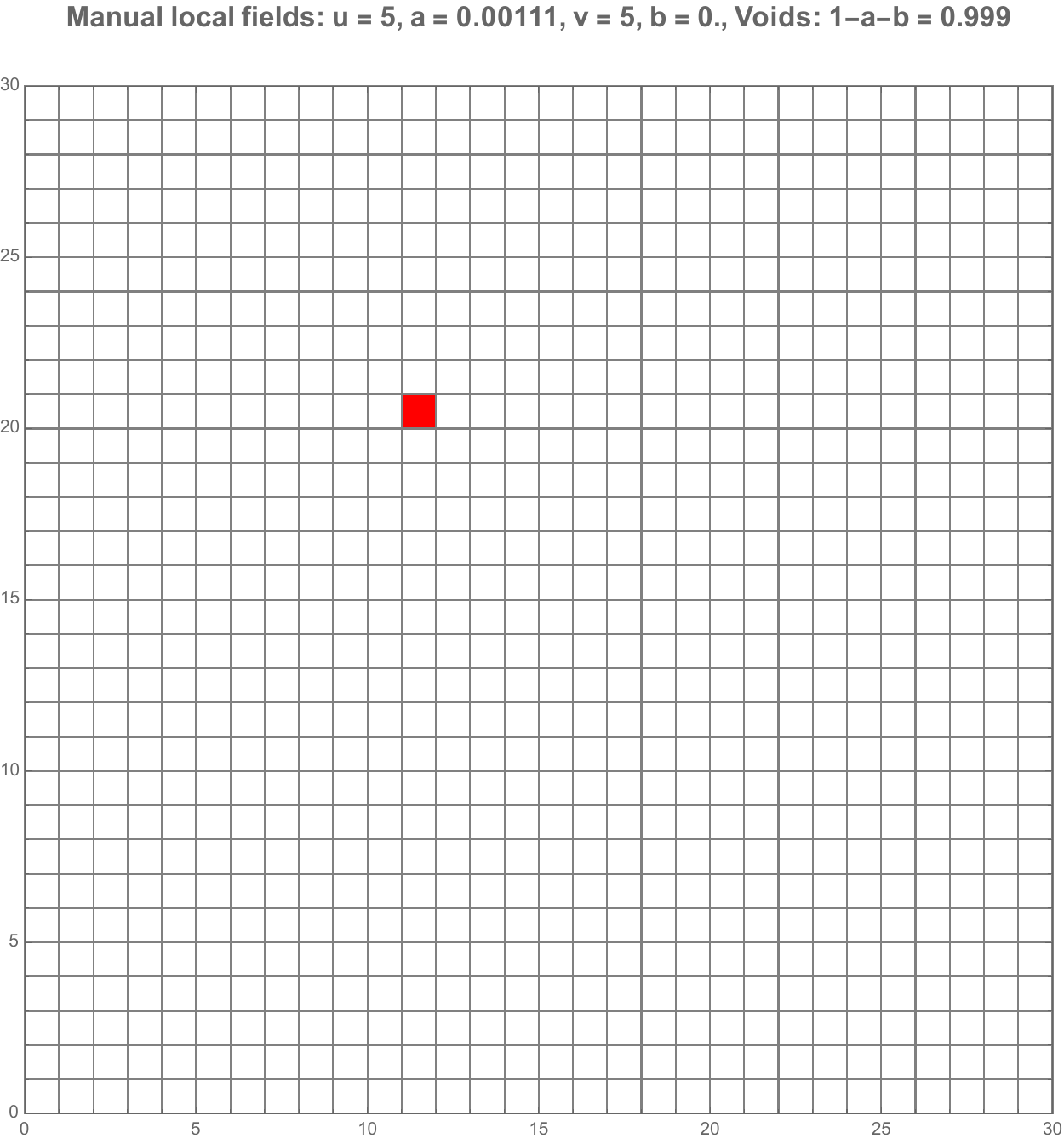}}
\subfigure[]{\includegraphics[width=0.22\textwidth]{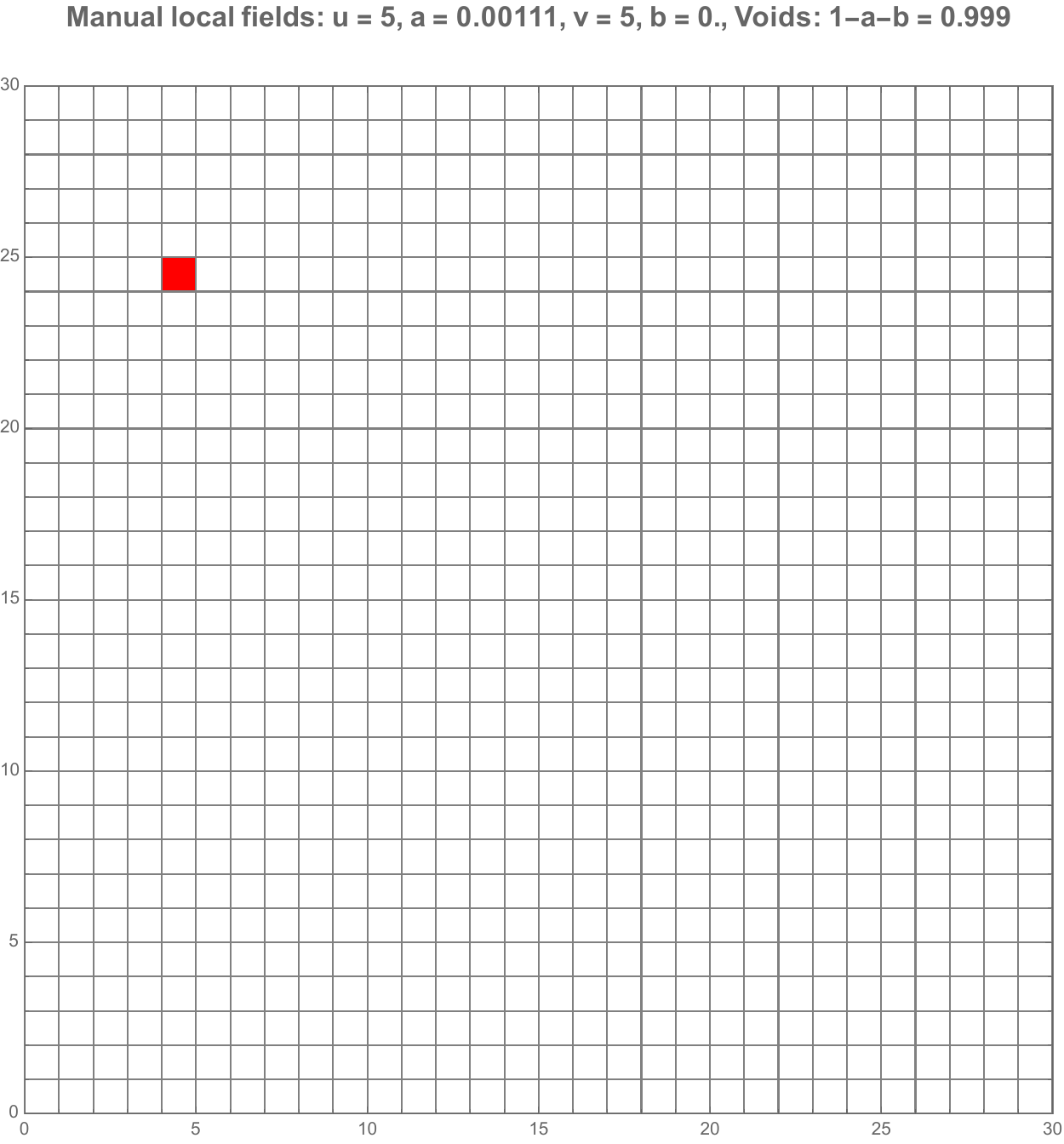}}
\subfigure[]{\includegraphics[width=0.26\textwidth]{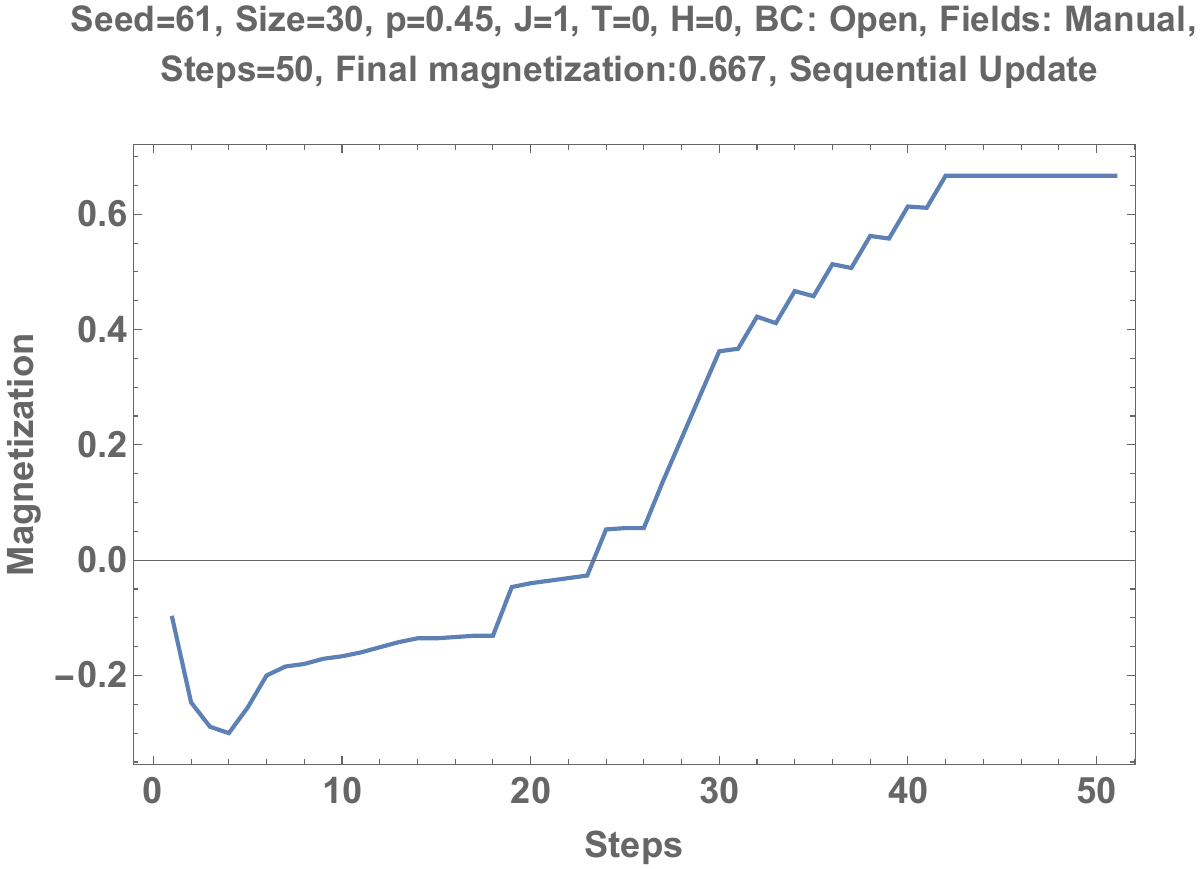}}
\subfigure[]{\includegraphics[width=0.26\textwidth]{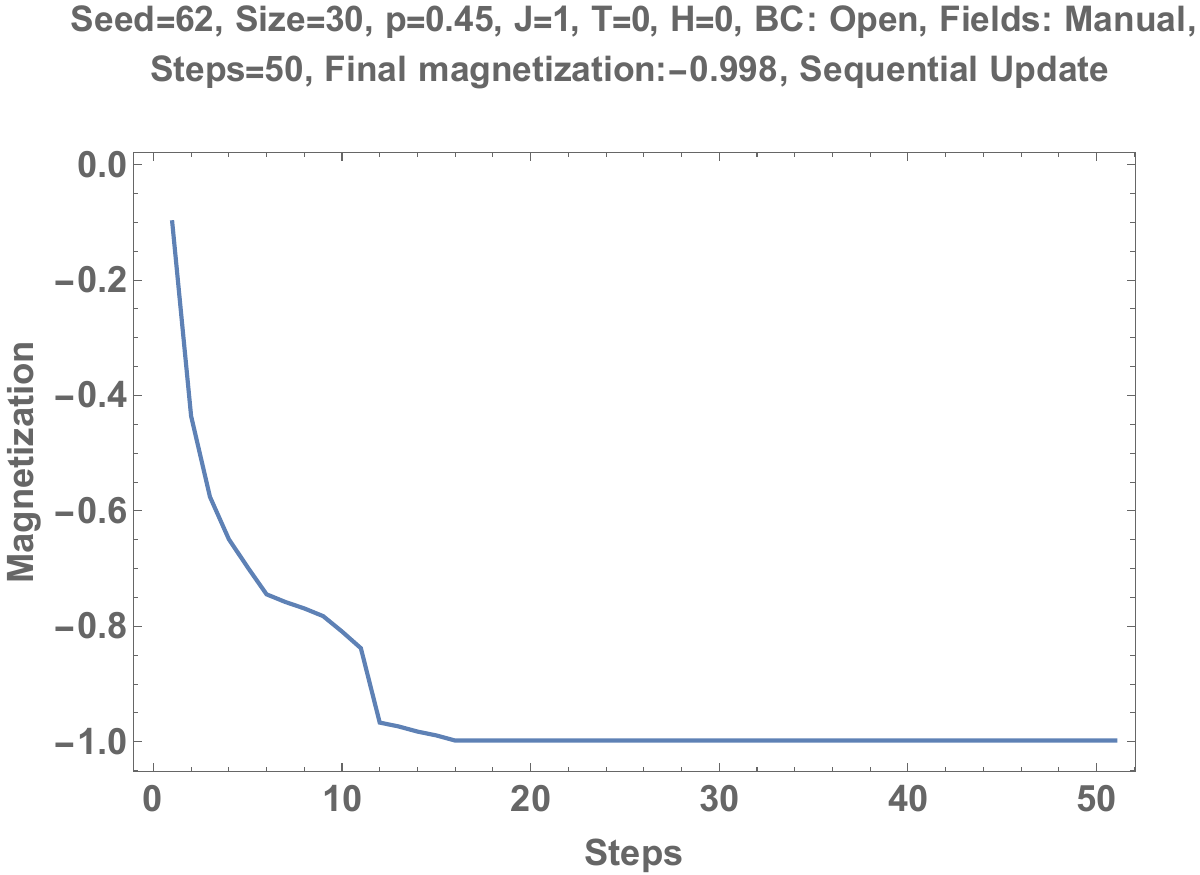}}

\end{figure}

\newpage 

\noindent\captionof{figure}{Subparts (a, b, c, d), (e, f, g, h), (i, j, k, l) have a single red local field located at the same position as in Figure (\ref{gg}). Zero effect is seen when starting from $p=0$ as could be expected (subparts (a, b, c, d)). Subpart (e) show an initial configuration with $p=0.44$ and the related dynamics in subparts (f, g) respectively zero and one local field. Very little effect is observed with the same final state (subpart (h)) as with $p=0$. With one percent increase in red initial choices from $p=0.44$ to $p=0.45$ (subpart (i)), the single red local field is found to have quite a strong impact with a final majority domain with 630 red choices as seen in subparts (j, k, l). Last row show the effect of stil one single red local field but located at a different position on the grid as seen in subpart (m) with previous coalition and subpart (n) with the new location. Subpart (o) shows the associated dynamics with the same initial distribution (Seed = 61, $p=0.45$) and subpart (p) with a different initial distribution (Seed = 62, $p=0.45$). While the new location of the red single field increases the support for red choices in the final outcome as seen comparing subparts (o) and (k), it has zero effect on the second distribution as shown in subpart (p).}
\label{gg}


\subsubsection{Figure (\ref{hh})}

To end the Monte Carlo exploration, which went through using a sequential update I consider here a few cases using a random update to demonstrate that similar qualitative results are obtained dismissing the possibility that they could have been artefacts of the sequential update. However, for the same initial conditions sequential and random updates do yield different quantitative final states.

Subpart (a) shows an initial distribution of red and blue choices with $p=0.45$ as in subparts (i, j, k, l, m, n, o, p) of Figure (\ref{gg}). However, actual distributions are different with Seed = 34 here and Seed = 62 in subpart (p) while Seed = 61 in Figure (\ref{gg}). Subpart (b) shows the dynamics of interactions leading to an overwhelming majority of blue choices.

Unlike previous cases with one single red local field, six red local fields are applied at specific sites (subpart (c)) to reverse the final outcome with a huge majority of red choices after 200 MC steps as seen from subparts (d, e). 

In contrast, erasing one red local field (subpart(f)) restores a blue majority as shown in subparts (g, h).

Yet, applying a sequential update to the distribution of subpart (a) without local fields yields a similar outcome as illustrated with subpart (i) in contrast to subpart (b). However, 30 MC steps are sufficient to get full blue symmetry breaking against 150 with random update.

Yet, applying the six local fields of subpart (c) does not yield a red majority as seen from subparts (j, k).

\begin{figure}
\centering
\vspace{-3cm}
\subfigure[]{\includegraphics[width=0.32\textwidth]{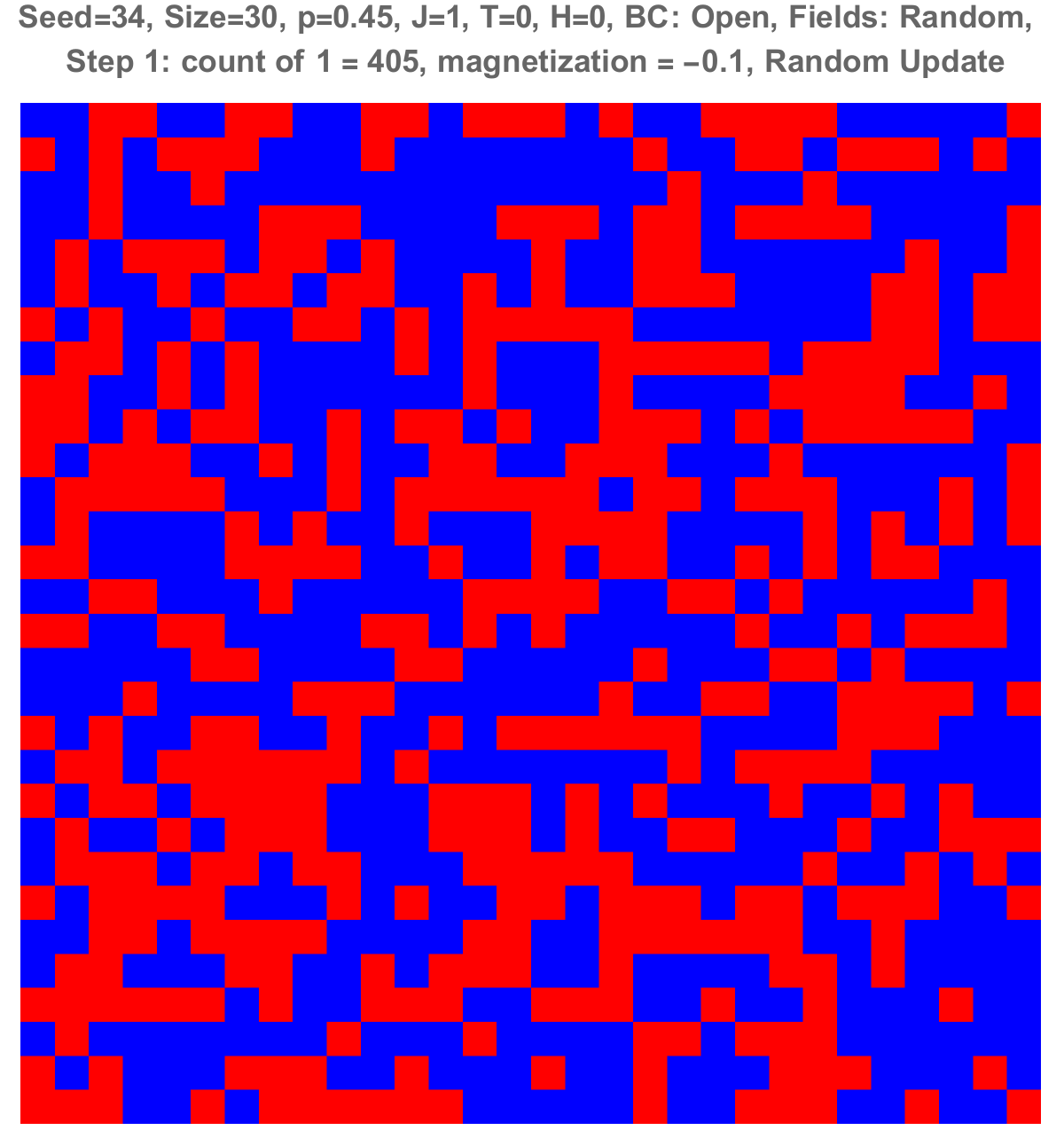}}
\subfigure[]{\includegraphics[width=0.45\textwidth]{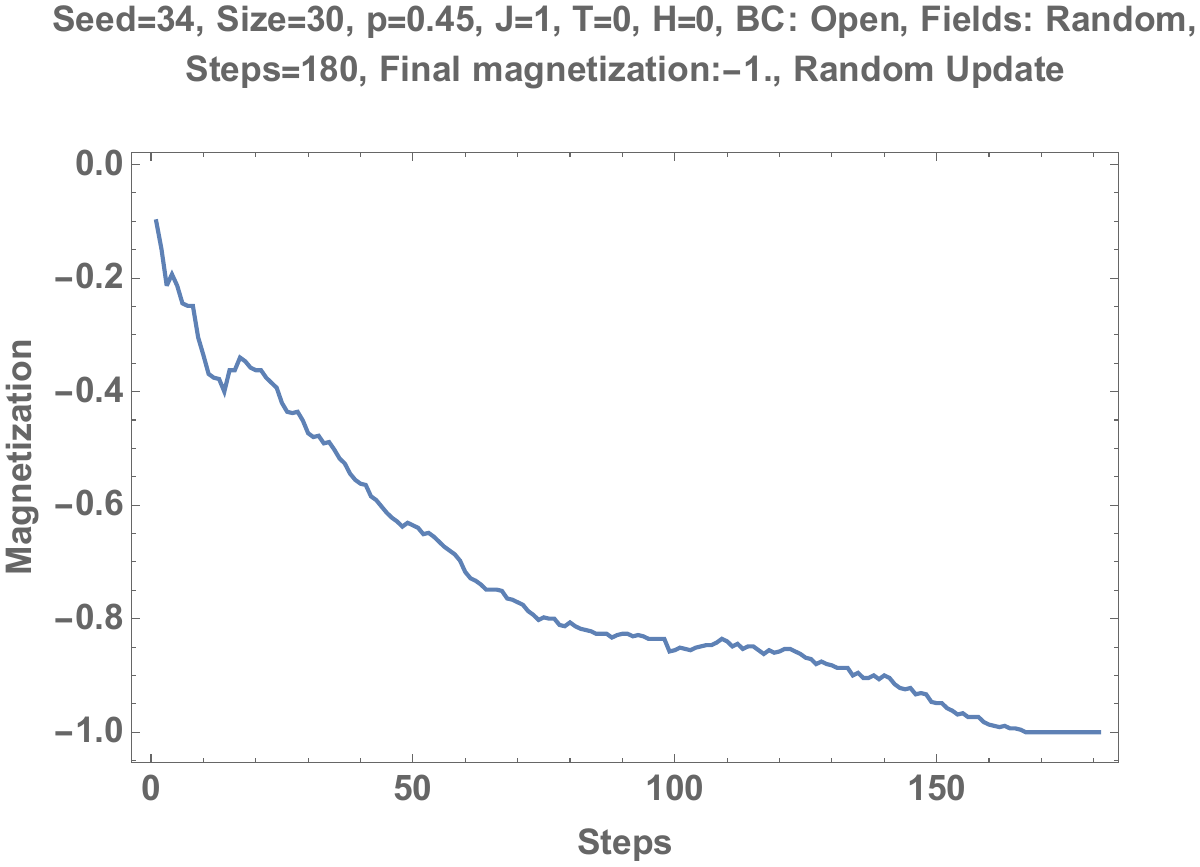}}
\\
\subfigure[]{\includegraphics[width=0.32\textwidth]{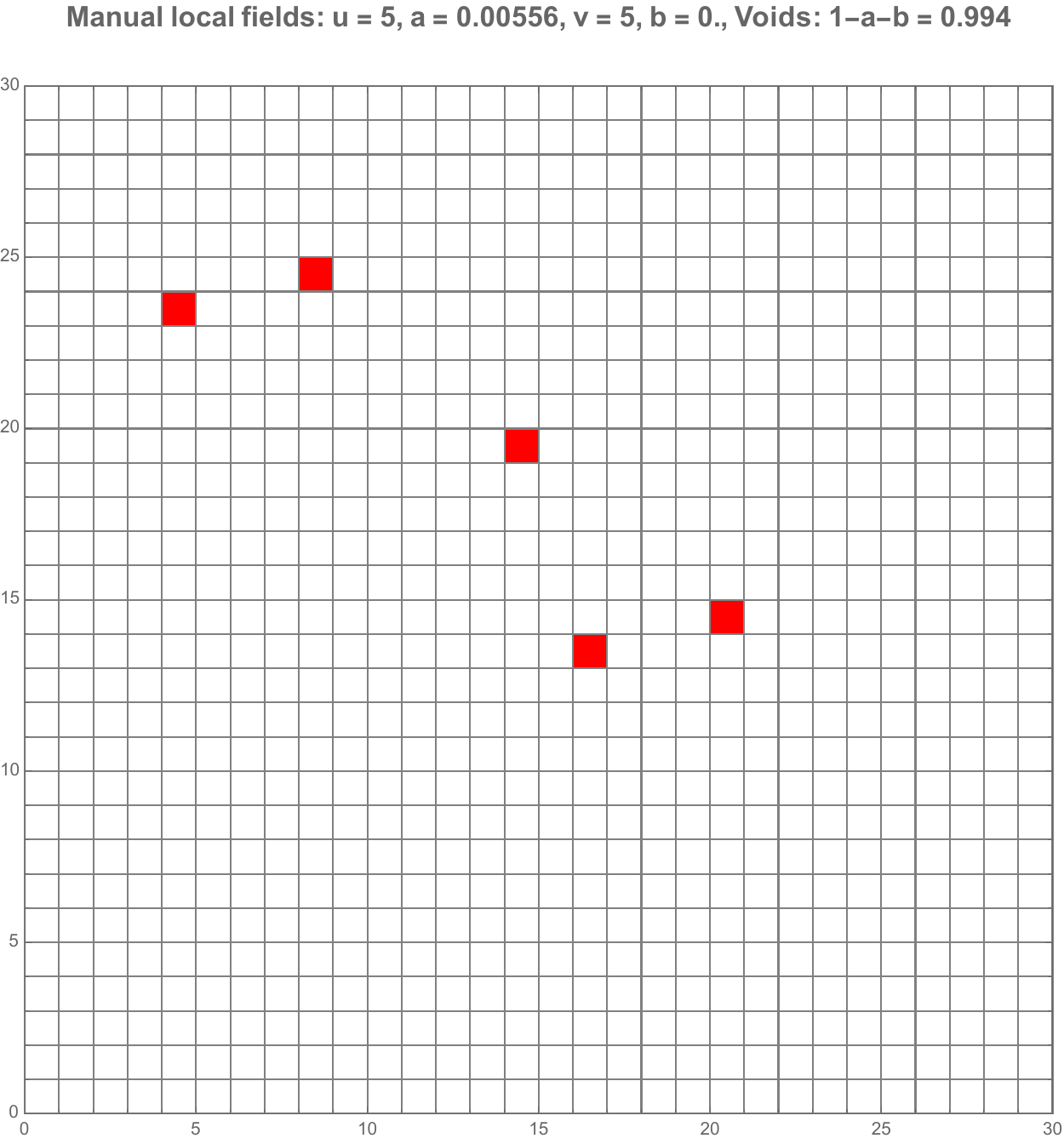}}
\subfigure[]{\includegraphics[width=0.32\textwidth]{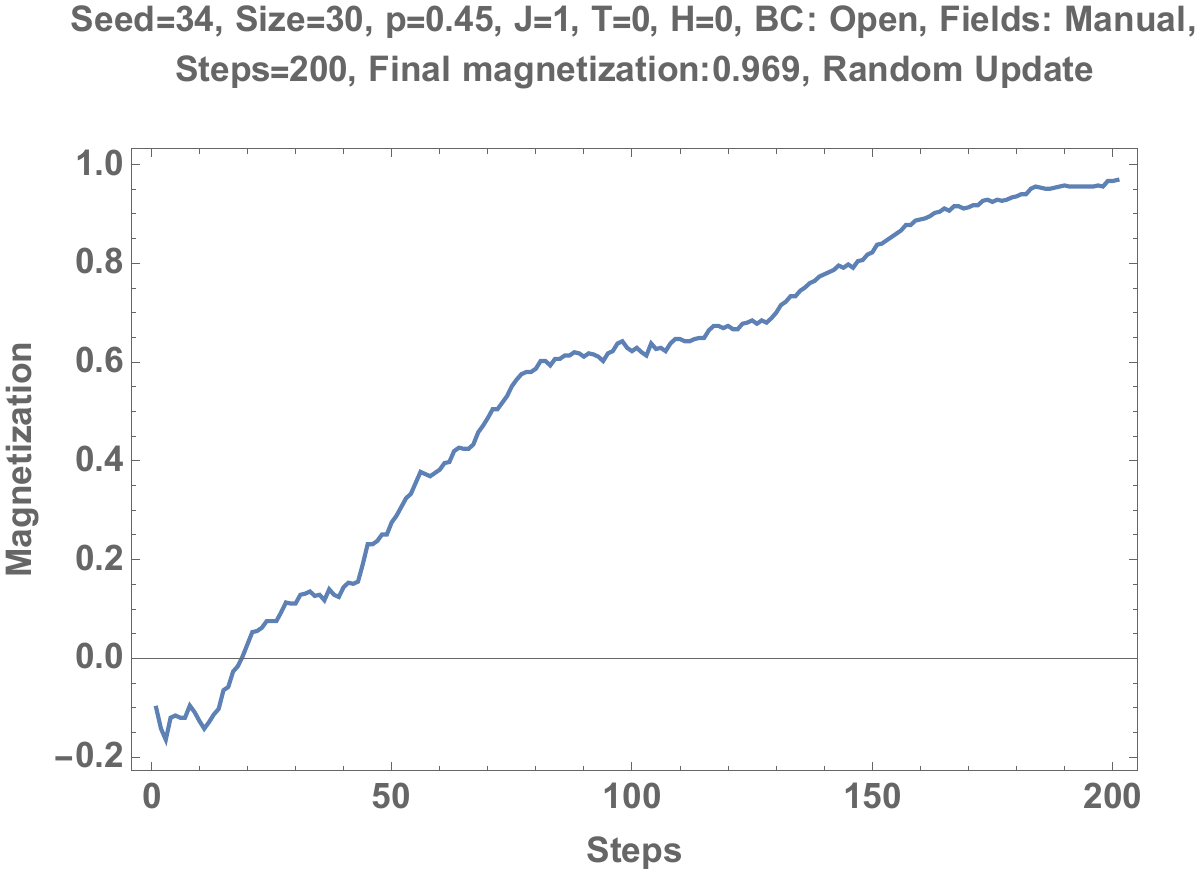}}
\subfigure[]{\includegraphics[width=0.32\textwidth]{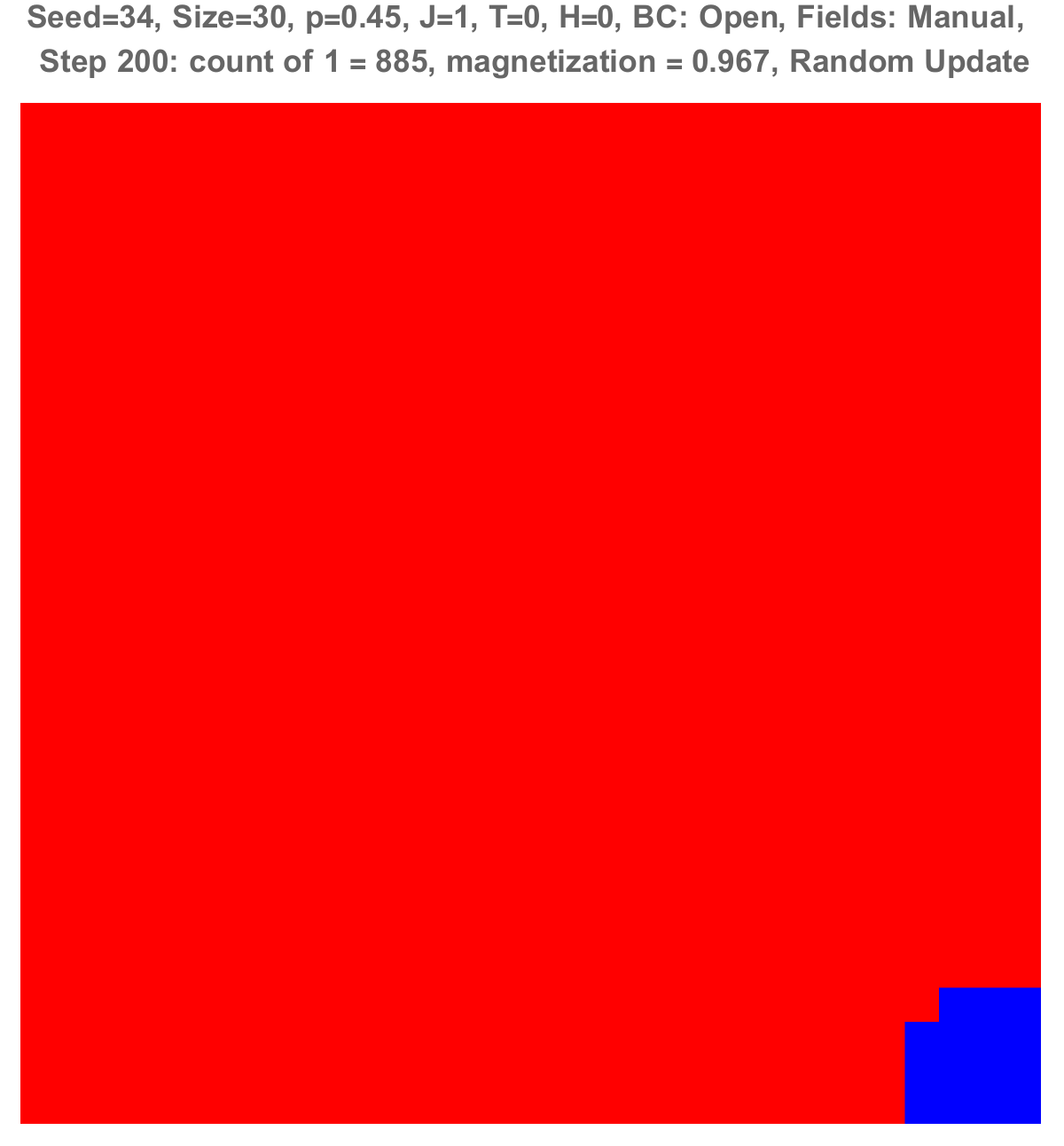}}
\\
\subfigure[]{\includegraphics[width=0.32\textwidth]{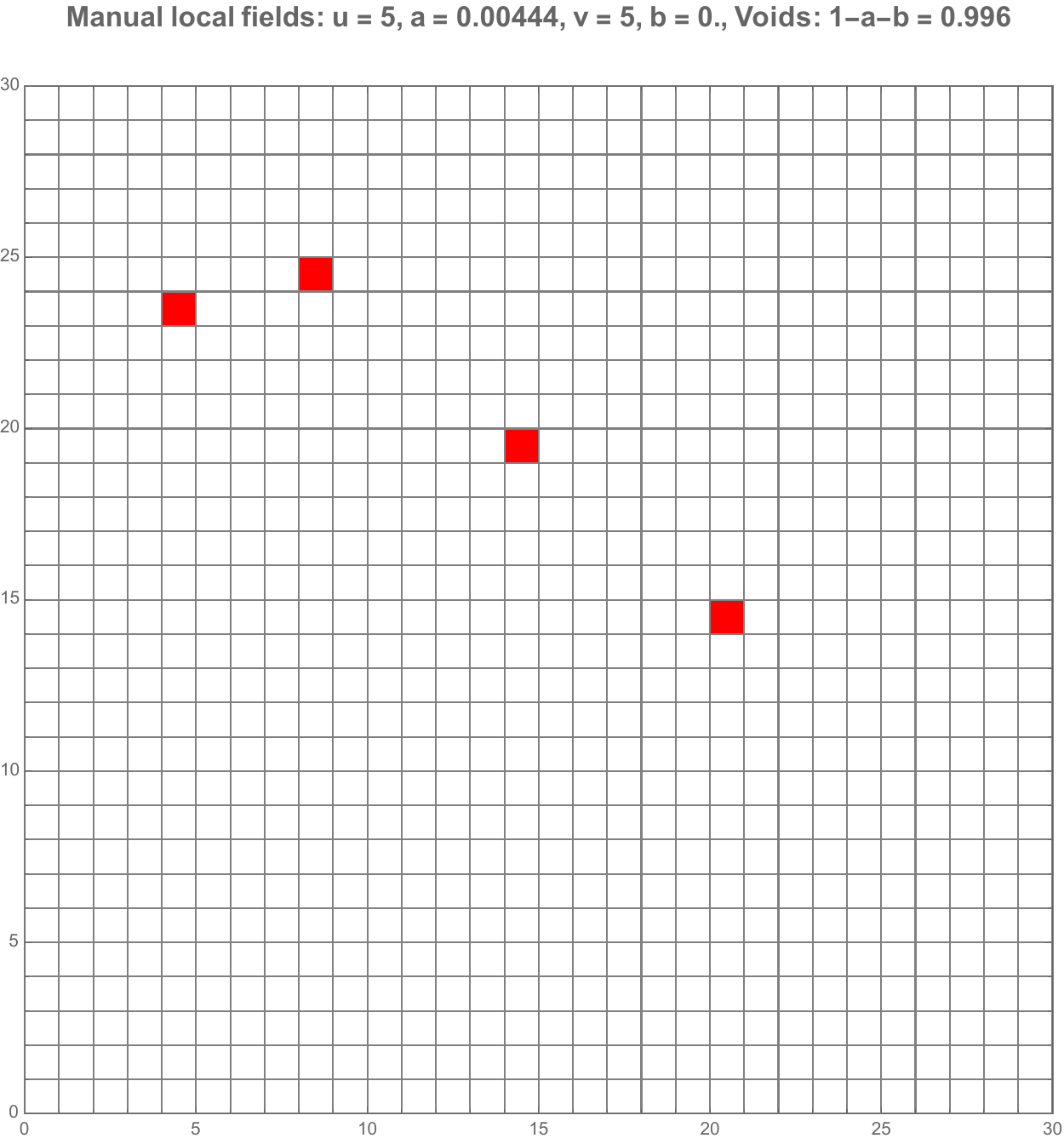}}
\subfigure[]{\includegraphics[width=0.32\textwidth]{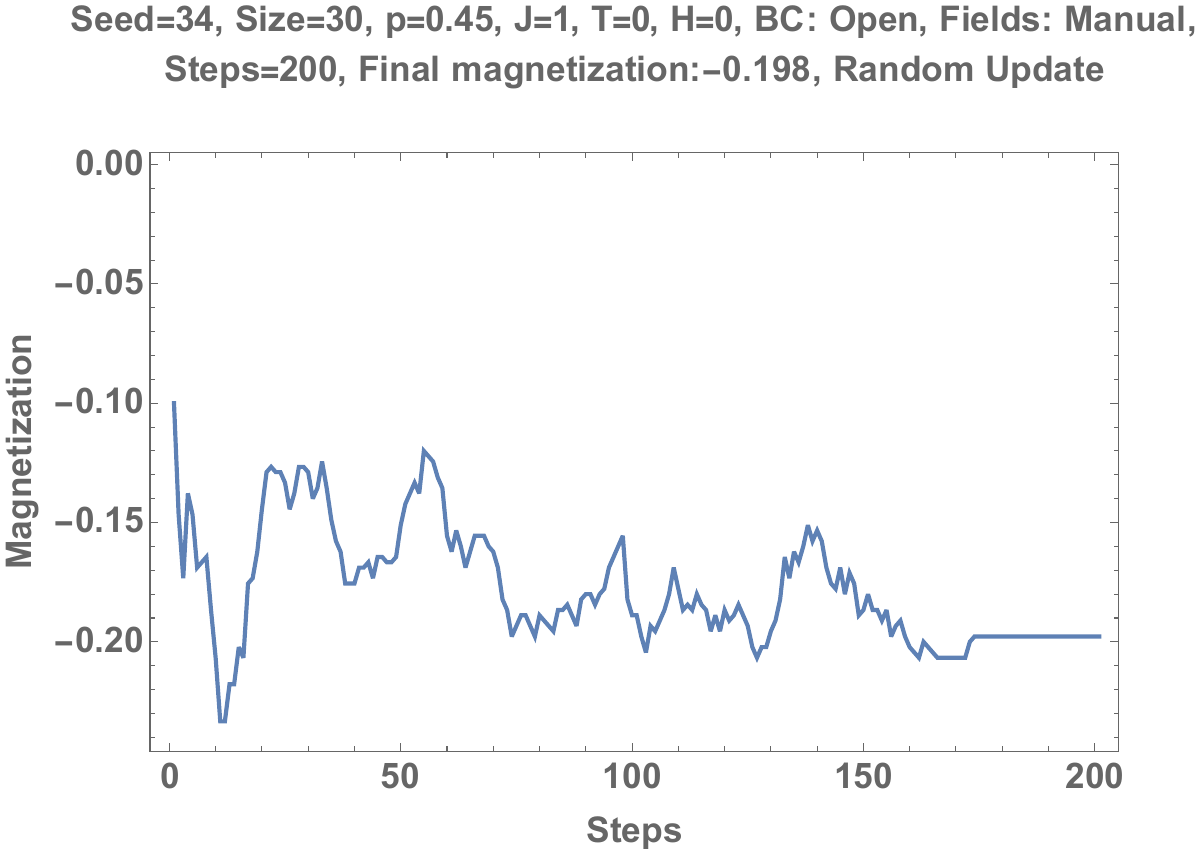}}
\subfigure[]{\includegraphics[width=0.32\textwidth]{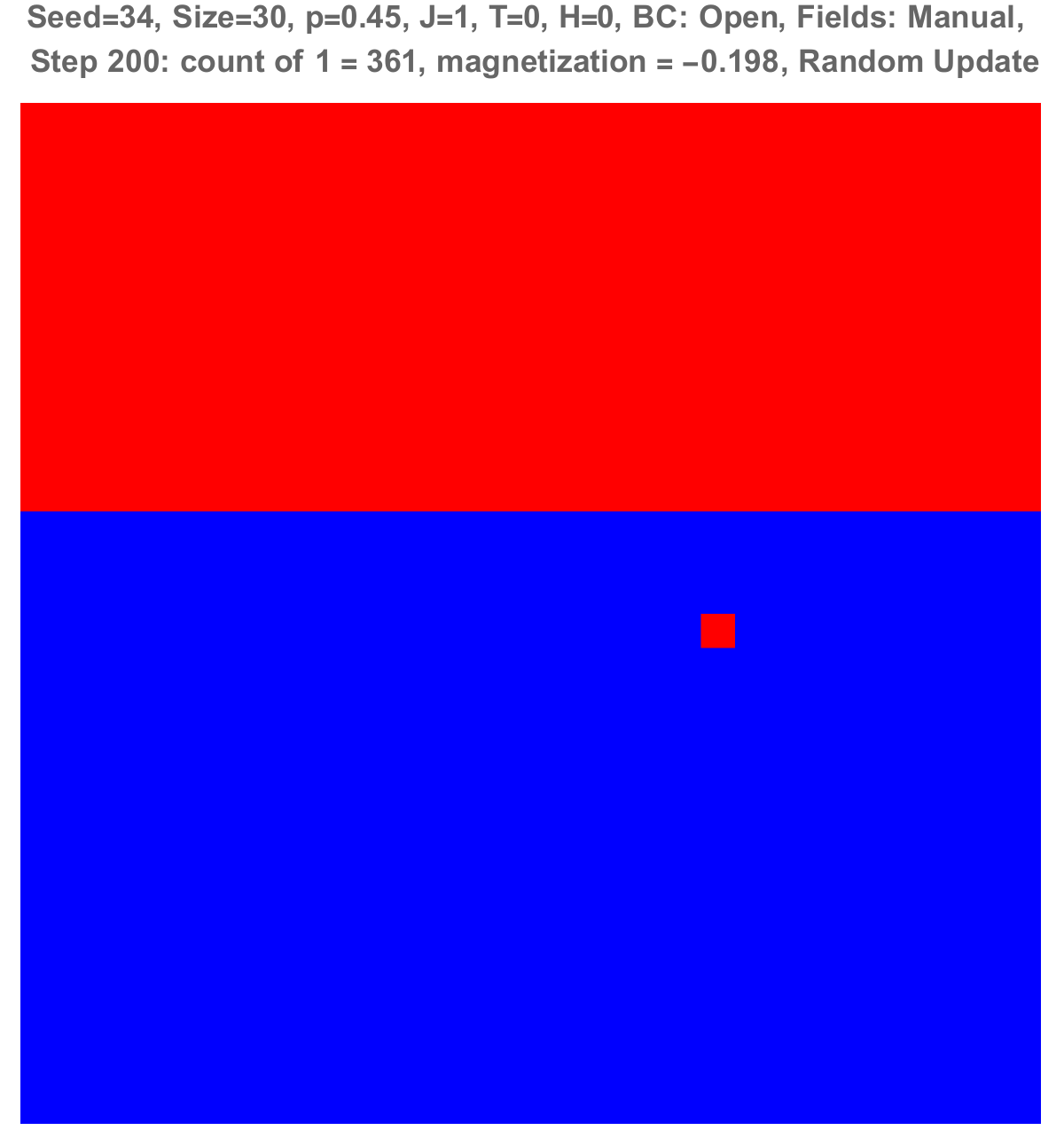}}
\\
\subfigure[]{\includegraphics[width=0.32\textwidth]{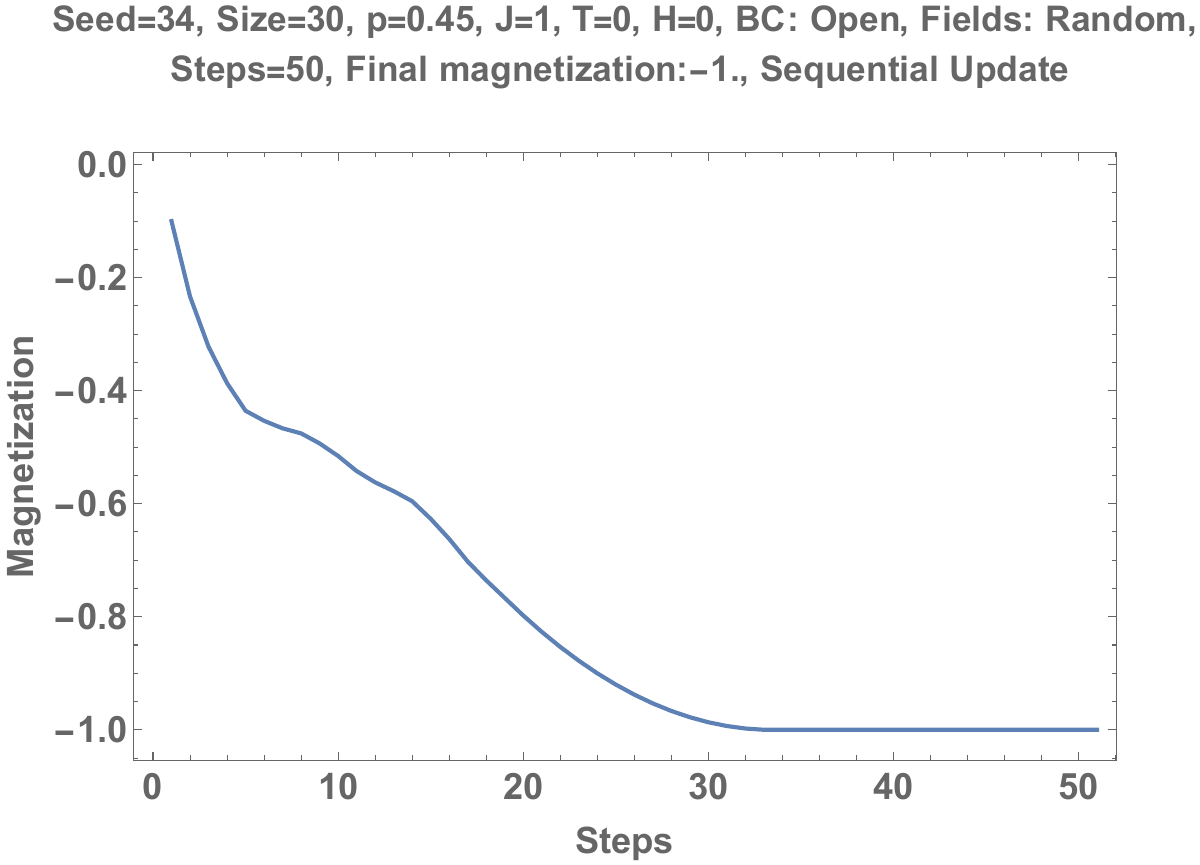}}
\subfigure[]{\includegraphics[width=0.32\textwidth]{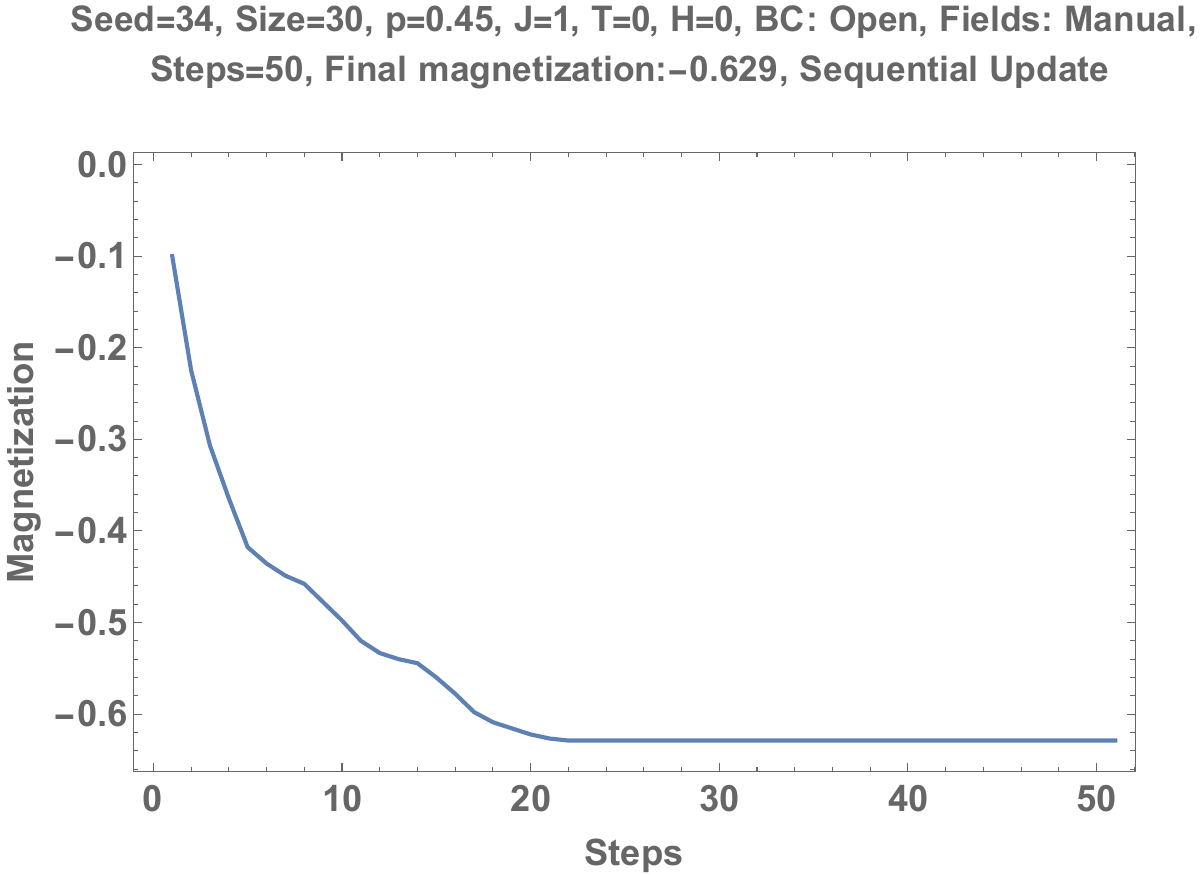}}
\subfigure[]{\includegraphics[width=0.32\textwidth]{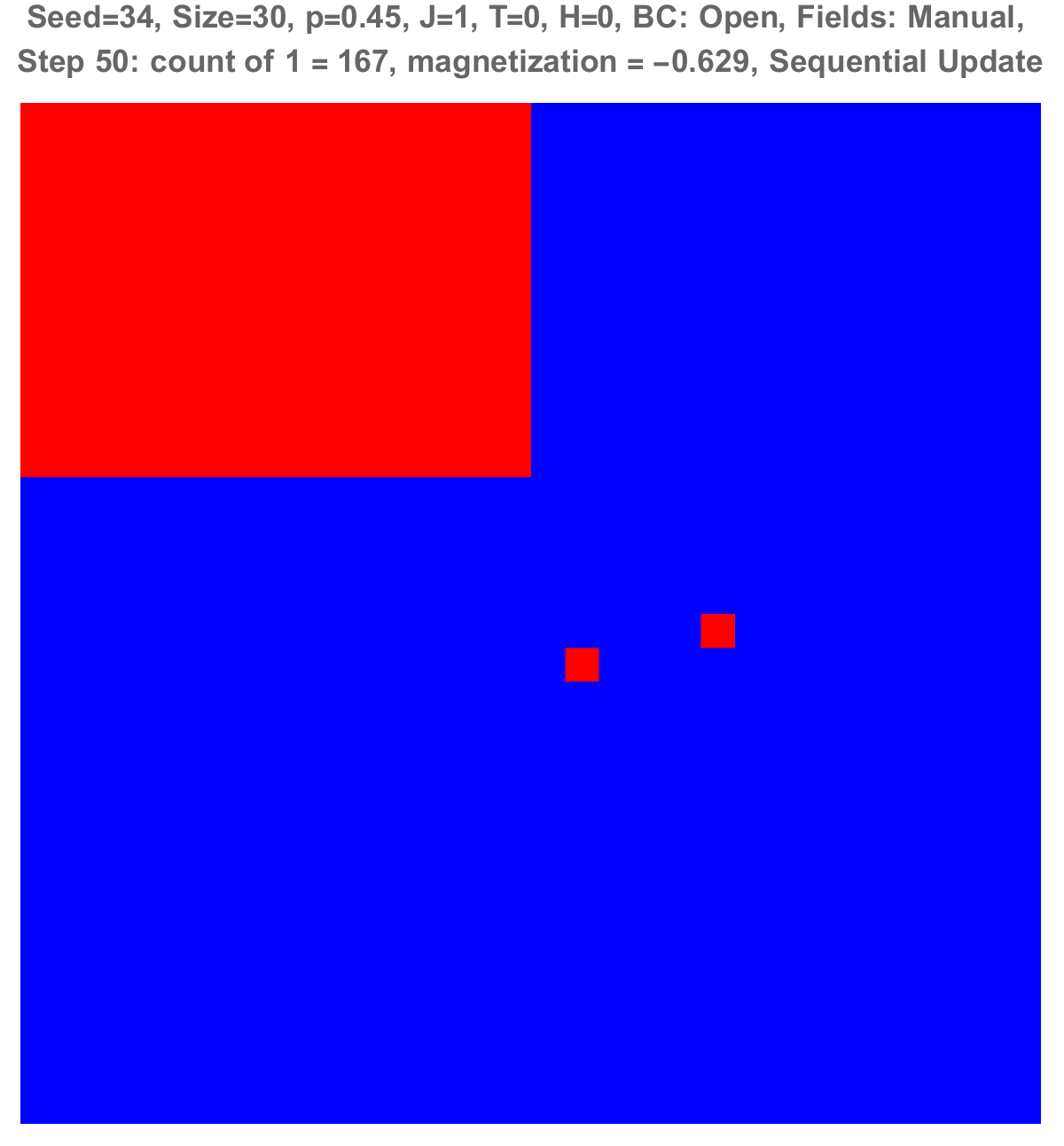}}

\end{figure}

\newpage 

\noindent\captionof{figure}{Subpart (a) shows an initial distribution of red and blue choices with $p=0.45$ with Seed = 34  Subpart (b) shows the dynamics of interactions leading to an overwhelming majority of blue choices. Six red local fields located at specific sites (subpart (c)) reverse the final outcome with a huge majority of red choices after 200 MC steps as seen from subparts (d, e). In contrast, erasing one red local field (subpart(f)) restores a blue majority as shown in subparts (g, h). Applying a sequential update to the distribution of subpart (a) without local fields yields more quickly a similar outcome as illustrated with subparts (i, b). Applying the six local fields of subpart (c) does not yield a red majority as seen from subparts (j, k).}
\label{hh}

\section{Conclusion}

In this paper I have shown how spontaneous symmetry breaking is vulnerable to even minimal distortions of randomness with disproportionate consequences. A handful of local fields, when positioned at tipping sites, were found to be decisive in steering the entire system. This dual nature, strength in effect but fragility in position, underscores the system's sensitivity to localized quenched local fields in the making of symmetry  breaking.

The tipping sites when activated by a local field behave like structural weak points. They are indistinguishable fragile from others, yet capable of redirecting the evolution of the whole system. Much like a crack in a material \cite{f1} or a mutation in a biological network \cite{f2}, these localized vulnerabilities magnify small perturbations into large-scale outcomes, highlighting both the strength and fragility inherent to related dynamics in such configurations.

The distinction between spontaneous and distorted polarization provides a useful lens for interpreting this vulnerability. In the model, spontaneous polarization arises from internal interactions without a preferred direction, reflecting the natural symmetry-breaking tendency of the system whose direction is randomly selected. Indeed, it is a function of the actual distribution of initial local choices and the update scheme being implemented.

Distorted polarization, by contrast, emerges when  invisible manipulation implemented by external local fields or internal local biases are activated at these fragile sites, producing directed stable configurations. This dynamic highlights how minimal interventions can lock a system into trajectories that obscure its underlying capacity for autonomous self-organization. Interventions do not need to be strong. They only need to exploit existing geometrical vulnerabilities to override the system's capacity for spontaneous random symmetry breaking.

The results expose a fundamental tension in Ising-like models of opinion dynamics. In the absence of local bias, the system can undergo spontaneous symmetry breaking, leading to large-scale consensus or polarization generated by stochastic fluctuations and local coupling alone. Yet, when a handful local fields are applied, whether external (media influence, institutional pressure) or internal (individual biases, beliefs, prejudices), these vulnerable tipping sites, when activated with local fields, amplify their effect, producing directed polarization that reduces the system's intrinsic variability. 

This mechanism resonates with real-world social systems, where opinion formation reflects both endogenous dynamics and exogenous constraints. Recognizing how fragility at local scales enables disproportionate global control is essential for interpreting consensus and division, in both physical models and socio-political contexts.

In the end, consensus is not always self-organization. It may be symmetrically broken under a hidden constraint deliberately shaped and strategically imposed by one of the competing sides. Those findings may open a new critical view of the nature of social media. Although obtained within a two-dimensional Ising-like model, the results invite a critical reassessment of the so-called  ``democratic'' character of social media. They may shed new light on hidden limitations within social media platforms, which subtly guide and sometimes even dictate the direction of ostensibly ``spontaneous'' democratic collective choices.

\subsection*{Acknowledgments}


\end{document}